\documentclass{emulateapj}
\usepackage{graphicx}

\slugcomment{Accepted for publication in the Astrophysical Journal, November, 2011. doi - 10.1088/0004-637X/745/1/79.}
\shorttitle{Collision Scaling Laws}
\shortauthors{Leinhardt and Stewart}

\usepackage{natbib}
\makeatletter
\def\gsim{\compoundrel>\over\sim}

\def\compoundrel#1\over#2{\mathpalette\compoundreL{{#1}\over{#2}}}
\def\compoundreL#1#2{\compoundREL#1#2}
\def\compoundREL#1#2\over#3{\mathrel
      {\vcenter{\hbox{$\m@th\buildrel{#1#2}\over{#1#3}$}}}}
\makeatother
\begin{document} 

\title{Collisions between Gravity-Dominated Bodies: 1. Outcome Regimes and Scaling Laws}

\author{Zo\"e M. Leinhardt}
\affil{School of Physics, University of Bristol\\
	H.H. Wills Physics Laboratory, Tyndall Avenue, Bristol BS8 1TL, U.K.}
\email{Zoe.Leinhardt@bristol.ac.uk}

\author{Sarah T. Stewart}
\affil{Department of Earth and Planetary Sciences, Harvard University, \\
  20 Oxford St., Cambridge, MA 02138, U.S.A.}
\email{sstewart@eps.harvard.edu}

\begin{abstract}
  Collisions are the core agent of planet formation. In this work, we
  derive an analytic description of the dynamical outcome for any
  collision between gravity-dominated bodies. We conduct
  high-resolution simulations of collisions between planetesimals; the
  results are used to isolate the effects of different impact
  parameters on collision outcome. During growth from planetesimals to
  planets, collision outcomes span multiple regimes: cratering,
  merging, disruption, super-catastrophic disruption, and hit-and-run
  events.  We derive equations (scaling laws) to demarcate the
  transition between collision regimes and to describe the size and
  velocity distributions of the post-collision bodies.  The scaling
  laws are used to calculate maps of collision outcomes as a function
  of mass ratio, impact angle, and impact velocity, and we discuss the
  implications of the probability of each collision regime during
  planet formation.

  Collision outcomes are described in terms of the impact conditions
  and the catastrophic disruption criteria, $Q^*_{RD}$ -- the specific
  energy required to disperse half the total colliding mass.  All
  planet formation and collisional evolution studies have assumed that
  catastrophic disruption follows pure energy scaling; however, we
  find that catastrophic disruption follows nearly pure momentum
  scaling. As a result, $Q^*_{RD}$ is strongly dependent on the impact
  velocity and projectile-to-target mass ratio in addition to the
  total mass and impact angle. To account for the impact angle, we
  derive the interacting mass fraction of the projectile; the outcome
  of a collision is dependent on the kinetic energy of the interacting
  mass rather than the kinetic energy of the total mass.  We also
  introduce a new material parameter, $c^*$, that defines the
  catastrophic disruption criteria between equal-mass bodies in units
  of the specific gravitational binding energy. For a diverse range of
  planetesimal compositions and internal structures, $c^*$ has a value
  of $5\pm2$; whereas for strengthless planets, we find
  $c^*=1.9\pm0.3$.  We refer to the catastrophic disruption criteria
  for equal-mass bodies as the principal disruption curve, which is
  used as the reference value in the calculation of $Q^*_{RD}$ for any
  collision scenario. The analytic collision model presented in this
  work will significantly improve the physics of collisions in
  numerical simulations of planet formation and collisional evolution.
\end{abstract}


\section{Introduction}

Planet formation is common and the number and diversity of planets
found increases almost daily
\citep[e.g.,][]{Borucki:2011,Howard:2011}. As a result, planet
formation theory is a rapidly evolving area of research. At present,
observations principally provide snapshots of either early
protoplanetary disks or stable planetary systems. Little direct
information is available to connect these two stages of planet
formation, therefore, numerical simulations are used to infer the
details of possible intermediate stages.  However, the diversity of
extrasolar planetary systems continues to surprise observers and
theorists alike.

A complete model of planet formation has eluded the astrophysics
community because of both incomplete physics in numerical simulations
and computational constraints. In order to make the problem of planet
formation more tractable, the process is often divided into separate
stages, which are then tackled in isolation. This method has had some
success. For example, $N$-body simulations show that large
($\sim100$~km) planetesimals may grow into protoplanets of about a
lunar mass on million year time scales
\citep[e.g.,][]{Kokubo:2002}. Other simulations, focusing on later
stages of planet formation, created a variety of stable planetary
systems from initial distributions of protoplanet size bodies
\citep[e.g.,][]{Chambers:2001,Agnor:1999}. Recently, the distribution
of stable planets has been investigated in population synthesis models
\citep[e.g.,][]{Mordasini:2009,Ida:2010,Schlaufman:2010,Alibert:2011}. However,
given the complexity of planet formation, it is unsurprising that the
predictions from the first population synthesis models have been
overturned by the rapidly growing catalog of exoplanets
\citep{Howard:2010,Howard:2011}.  Hence, the diversity of the
extrasolar planets is still unexplained.

At the heart of the standard core-accretion model of planet formation
is the growth of planetestimals. The evolution of planetesimals is
dominated by a series of individual collisions with other
planetesimals \citep[e.g.,][]{Beauge:1990,Lissauer:1993}. The outcome
of each collision depends on the specific impact conditions: target
size, projectile size, impact parameter, impact velocity, and some
internal properties of the target and projectile, such as composition
and strength.  In the past, direct global simulations of planetesimal
evolution have assumed very simplified collision models. In $N$-body
simulations, terrestrial planet embryos were shown to grow easily from
an annulus of large planetesimals if the only outcome of collisions is
merging \citep[e.g.,][]{Kokubo:2002}. However, the computational
demands of such numerical methods did not permit for the tracking of
the very large numbers of bodies necessary to be able to include
direct calculations of the erosion of planetesimals.

Statistical methods are required to describe the full population of
bodies from dust size to planets. For example, \citet{Kenyon:2009}
conducted simulations that included fragmentation but still relied
upon a simple collision model. Specifically, their simulations did not
fully account for the effects of the mass ratio, impact velocity, or
impact angle on the collision outcome. In order to overcome these
simplifications some previous studies have employed a multi-scale
approach that includes direct simulations of collision outcomes within
a top-level simulation of planet growth
\citep{Leinhardt:2005,Leinhardt:2009a,Genda:2011}.  However,
multi-scale calculations significantly increase the computational
requirements. In addition, the numerical methods employed in previous
studies were only valid for a specific impact velocity regime. In the
case of \citet{Leinhardt:2005} and \citet{Leinhardt:2009a}, the
collision model assumed subsonic collisions and could not be extended
past oligarchic growth. In the case of \citet{Genda:2011}, the
technique assumed strengthless bodies and cannot be used in the early
phases of planetesimal growth.
     
A general description of collision outcomes that spans the growth from
dust to planets is required to build a self-consistent model for
planet formation. In previous work, the description of collision
outcomes drew upon a combination of laboratory experiments and limited
numerical simulations of collisions between two planetary-scale bodies
\citep[see review by][]{Holsapple:2002}. Collision outcomes themselves
are quite diverse, and several distinct collision regimes are
encountered during planet formation: cratering, merging/accretion,
fragmentation/erosion, and hit-and-run encounters.

Individual collision regimes have been described in quite varying
detail. In the laboratory, the erosive regimes (cratering and
disruption) have been studied most comprehensively
\citep{Holsapple:1993,Holsapple:2002}; however, even these regimes
lack a complete description of the dependence on all impact parameters
(particularly mass ratio and impact angle). Recently, numerical
studies of collisions between self-gravitating bodies of similar size
have identified new types of collision outcomes including hit-and-run
and mantle-stripping events
\citep{Agnor:2004,Asphaug:2006,Marcus:2009,Marcus:2010,Leinhardt:2010,
  Asphaug:2010,Kokubo:2010,Benz:2007,Genda:2011b}. Up to this point
our understanding of these new regimes has not been sufficient to
implement the diversity of collision outcomes in planet formation
codes. In addition, the transitions between regimes are not clearly
demarcated in the literature.

In the work reported here, we present a complete description of
collision outcomes for gravity-dominated bodies. Using a combination
of published hydrocode and new and published $N$-body gravity code
simulation results, we derive analytic equations to demarcate the
transitions between collision regimes and the size and velocity
distribution of the post-collision bodies.  We describe how these
scaling laws can be used to increase the accuracy of numerical
simulations of collisional evolution without sacrificing
efficiency. In a companion paper \citep{Stewart:2011}, we apply these
scaling laws to the end stage of terrestrial planet formation by
analyzing the range of collision outcomes from recent $N$-body
simulations.

This paper is organized as follows: \S \ref{sec:nummethod} summarizes
the numerical method for the new $N$-body simulations. \S
\ref{sec:resultsdisruption} derives a general catastrophic disruption
scaling law. Then, we develop general scaling laws for the size and
velocity distribution of fragments in the disruption regime. \S
\ref{sec:resultsother} defines the super-catastrophic and hit-and-run
regimes. \S \ref{sec:transitions} presents the transition boundaries
between collision outcome regimes from our numerical simulations and
our analytic model.  \S \ref{sec:discussion} discusses the range of
applicability of our results, areas needing future work, and the
implications of the scaling laws on aspects of planet formation.  The
Appendix summarizes the implementation of the scaling laws in
numerical simulations of planet formation and collisional
evolution. Table \ref{tab:vars} presents the definitions of variables
and annotations used in this work.

\section{Numerical Method} \label{sec:nummethod}

In this section, we describe the numerical method used in the new
impact simulations presented in this work. Simulations of relatively
slow subsonic impacts were conducted using a $N$-body code with
finite-sized spherical particles, PKDGRAV \citep{Stadel:2001}, which
has been extensively used to study the dynamics of collisions between
small bodies
\citep[e.g.,][]{Leinhardt:2000,Michel:2001,Leinhardt:2002,
  Durda:2004,Leinhardt:2005,Leinhardt:2009,Leinhardt:2010}.

Both the target and projectile are assumed to be rubble piles:
gravitational aggregates with no bulk tensile strength
\citep{Richardson:2002}. The rubble pile particles are bound
together purely by self-gravity. The particles themselves are
indestructible and have a fixed mass and radius (for cases without
merging). The equations of motion of the particles are governed by
gravity and inelastic collisions. The amount of energy lost in each
particle-particle collision is parameterized through the normal and
tangential coefficients of restitution. The rubble piles are created
by placing particles randomly in a spherical cloud and allowing the
cloud to gravitationally collapse with highly inelastic particle
collisions. Randomizing the internal structure of the rubble piles
avoids spurious collision results due to crystalline structure of
hexagonal close packing
\citep[see][]{Leinhardt:2000,Leinhardt:2002}. The crystalline
structure can cause large uncertainties in collision outcomes for
super-catastrophic events.

All simulations had a target with radius of 10 km, mass of $4.2\times
10^{15}$ kg, bulk density of 1000 kg m$^{-3}$, and escape velocity of
7.5 m s$^{-1}$. The current study includes four projectile-to-target
mass ratios ($\gamma$), four impact angles ($\theta$), and a range of
impact velocities spanning merging to super-catastrophic disruption.
These results for a single size target body are used to derive scaling
laws for any size body in the gravity regime. The target and
projectile are initially separated by the sum of their respective
radii to ensure that the impact angle of the impact is unchanged from
the initial trajectory.

In order to resolve the size distribution after the collisions, each
body needs a relatively high number of particles ($N_{\rm targ} \sim
10^4$, $N_{\rm p} \sim 250 - 10^4$ depending on the mass ratio). However,
large numbers of particles are also time consuming to integrate,
especially in a rubble pile configuration where there is a high
frequency of particle-particle collisions. Each simulation here uses
high resolution with inelastic particle collisions to resolve the
initial impact. Once the velocity field is well established, the
particles are allowed to merge with one another. Thus, our method has
both accuracy and efficiency, resolving the size distribution to small
fragments and completing the simulations as quickly as possible.

\begin{figure}
 \includegraphics[scale=0.4]{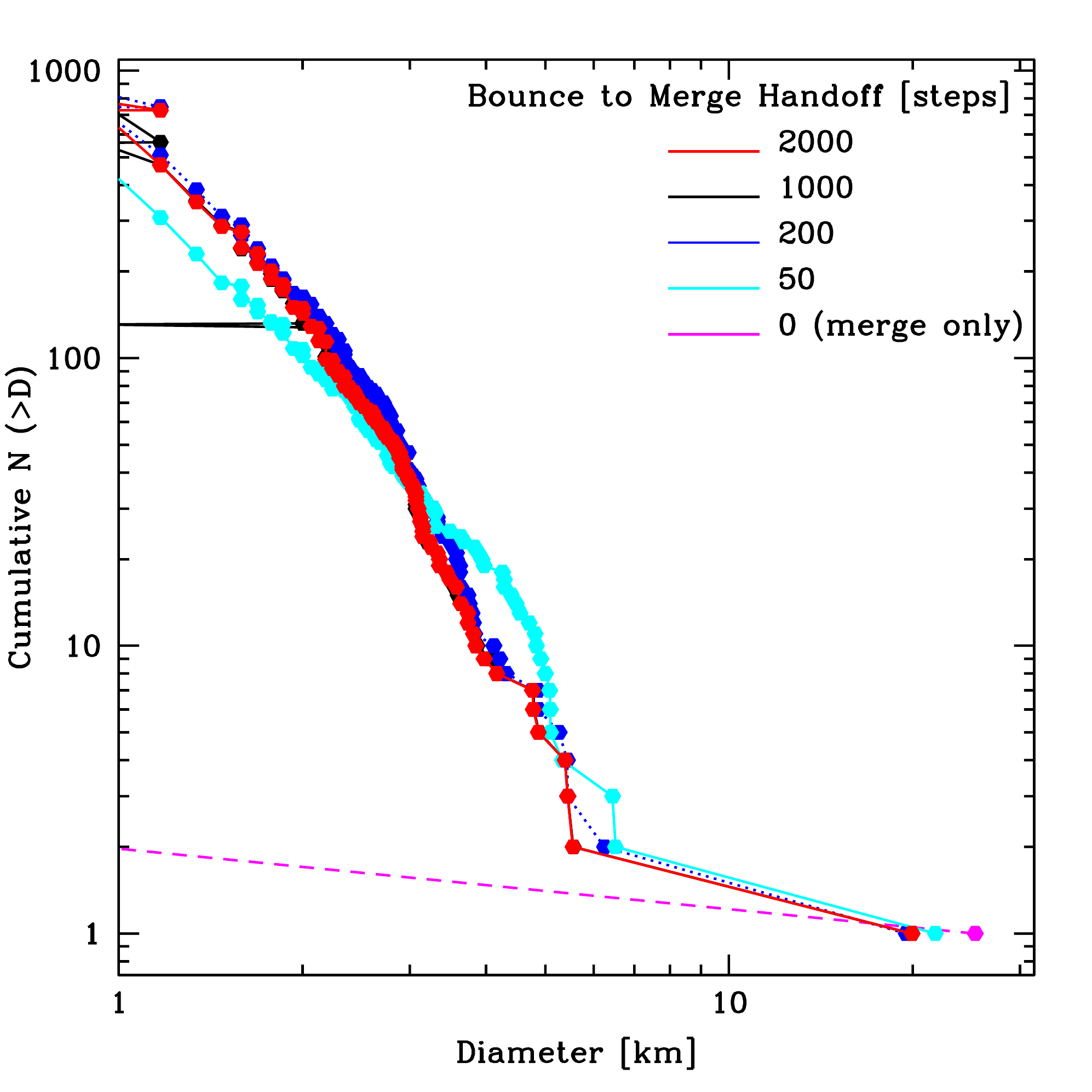}
 \caption{Cumulative size distribution of collisional debris after the
   catastrophic impact between two 20-km diameter bodies. Line colors
   represent different handoff times from inelastic bouncing to
   perfect merging for the outcome of collisions between pairs of
   PKDGRAV particles. Each step corresponds to 1 minute in simulation
   time. The same initial impact is used for all distributions shown:
   $\gamma = 1$, $V_i = 30$ m s$^{-1}$, $\theta = 0$, $N_{\rm targ} = N_{\rm p} =
   1\times10^4$.
   \label{fig:bouncetomerge}}
\end{figure}

We considered the possible influence of the time of the transition
from inelastic bouncing to perfect
merging. Figure~\ref{fig:bouncetomerge} presents a test case using a
head-on catastrophic impact between equal-sized objects (mass ratio
$\gamma = 1$). Each body has $\sim 10^4$ particles; in the inelastic
bouncing phase, each particle has a normal coefficient of restitution,
$\epsilon_n = 0.5$, and a tangential coefficient of restitution,
$\epsilon_t = 1$, consistent with field observations and friction
experiments on rocky materials \citep[e.g.,][]{Chau:2002}. At a
certain time, the colliding particles are allowed to merge, producing
one particle with the same mass and bulk density as the two original
particles. If merging is turned on too early, the mass of the largest
remnant is overestimated (green and cyan lines) due to a geometric
effect known as runaway merging \citep{Leinhardt:2009}. The results of
this numerical test show that the mass distribution is stable if
merging is turned on after 50 steps, where 1 step is one minute in
simulation time in the frame of the particles.  However, we choose to
be conservative and merge after 250 steps of inelastic bouncing in all
of the new simulations presented in this paper. All simulations were
run for at least 0.2 years, at which point the size distribution had
stabilized and clumps of rubble pile fragments were easily
identifiable.

Previous studies using PKDGRAV did not have the numerical resolution
to determine an accurate size or velocity distribution of the
collisional remnants \citep[e.g., $N_{\rm targ} \sim
10^3$,][]{Leinhardt:2000}. In this work, we present more extensive
simulations at an order of magnitude higher resolution ($N_{\rm targ}
\sim 10^4$). Note that $N = 10^4$ is high resolution for $N$-body
simulations of colliding rubble-pile bodies with bouncing
particles. We conducted several resolution tests and find that the
random error on the mass of the largest remnant is a few percent of
the total system mass. Hence, the super-catastrophic impacts, where
the largest remnant mass is a few percent of the total system mass,
have the highest error. We achieve excellent reproducibility of the
slope of the size and velocity distributions with the nominal
resolution compared to the higher resolution tests.

Note that $N$-body simulations are inherently higher resolution
compared to smoothed particle hydrodynamics (SPH) simulations. Our
simulations resolve over a decade in fragment size, comparable to SPH
simulations using an order of magnitude more particles
\citep{Durda:2004}. Fragments of radius 0.5 km are considered the
smallest usable fragments in these simulations.

In the following sections, we also include published results of
subsonic and supersonic collisions from previous work
\citep{Leinhardt:2009,Agnor:2004,Agnor:2004a,Marcus:2009,Marcus:2010,
  Durda:2004,Durda:2007,Jutzi:2010,Benz:1999,Benz:2000,Stewart:2009,
  Korycansky:2009,Nesvorny:2006,Benz:2007}. Studies of supersonic
collisions utilize shock physics codes, which include the effects of
irreversible shock deformation. For computational efficiency, the
shock code is generally used to calculate only the early stages of an
impact event; after a few times the shock wave crossing time, the
amplitude of the shock decays to the point where further
deformation is negligible. After the hydrocode step, the gravitational
reaccumulation stage of disruptive events has been calculated directly
using PKDGRAV \citep[e.g.,][]{Leinhardt:2009,Durda:2004,Michel:2002}
or indirectly by iteratively solving for the mass bound to the largest
fragment \citep[e.g.,][]{Benz:1999,Benz:2000,Marcus:2009}.

\section{Results: The Disruption Regime} \label{sec:resultsdisruption}

In our model, the boundaries between collision outcome regimes are
defined using the catastrophic disruption criteria, the specific
energy required to gravitationally disperse half the total mass,
because it provides a convenient means of calculating the mass of the
largest remnant. Our definition of the disruption regime refers to
collisions in which the energy of the event results in mass loss
(fragmentation) between about 10\% and 90\% of the total mass. More
quantitatively, the disruption regime is defined as collision
that result in the largest remnant having a linear dependence on the
specific impact energy. The rationale for this definition will become
apparent in \S \ref{sec:univlaw}.

This section focuses on deriving the dynamical outcome (the mass and
velocity distribution of post-collision fragments) in the disruption
regime.  Other collision regimes are discussed in \S
\ref{sec:resultsother}. Before discussing the results of our new
numerical simulations, we briefly review the catastrophic disruption
criteria, as it is a fundamental part of our story.

In the literature on planetary collisions, $Q$ traditionally denotes
the specific energy of the impact (kinetic energy of the
projectile/target mass) and $Q^*$ indicates the catastrophic
disruption criteria, where the largest remnant has half the target
mass.  Upon recognition that gravitational dispersal was important,
$Q^*_S$ and $Q^*_D$ denoted the criteria for shattering in the
strength regime and dispersal in the gravity regime, respectively. All
of the previous definitions for $Q^*$ assumed that the projectile
mass, $M_{\rm p}$, was much smaller than the target mass, $M_{\rm
  targ}$; however, in several phases of planet formation it is
expected that $M_{\rm p} \sim M_{\rm targ}$. Therefore, in previous
work, we developed a disruption criteria in the center of mass
reference frame in order to study collisions between comparably sized
bodies \citep{Stewart:2009}. The subscript $R$ was added in the
modification of the specific energy definition to denote reduced mass.
The center of mass specific impact energy is given by
\begin{eqnarray}
Q_R &=& (0.5 M_{\rm p}{V^2}_{\rm p} + 0.5 M_{\rm targ}{V^2}_{\rm targ})/M_{\rm tot}, \nonumber \\
      &=& 0.5  \mu V_i^2 / M_{\rm tot}, \label{eqn:qr}
\end{eqnarray}
where $M_{\rm tot}=M_{\rm p}+M_{\rm targ}$, $\mu$ is the reduced mass $M_{\rm p}
M_{\rm targ}/M_{\rm tot}$, $V_i$ is the impact velocity, and $V_{\rm p}$ and
$V_{\rm targ}$ are the speed of the projectile and target with respect to
the center of mass, respectively. At exactly the 
catastrophic disruption threshold, 
\begin{eqnarray} \label{eqn:vstardef}
Q^*_{RD} &=& 0.5  \mu {V^*}^2 / M_{\rm tot},
\end{eqnarray}
where we explicitly define $V^*$ to be the critical impact velocity
required to disperse half of the total mass for a specific impact
scenario (total mass and mass ratio).

The catastrophic disruption criteria is a strong function of size with
two components: a strength regime where the critical specific energy
decreases with increasing size and a gravity regime where the critical
specific energy increases with increasing size. The transition between
regimes occurs between a few 100-m and few-km radius, depending on the
strength of the bodies \citep[see Figure~2,][]{Stewart:2009}. A general
formula for $Q^*_{RD}$ as a function of size was derived by
\citet{Housen:1990} using $\pi$-scaling theory,
\begin{eqnarray} \label{eqn:qstarred}
Q^*_{RD}=q_s \left ( S/\rho_1 \right )^{3\bar \mu(\phi+3)/(2\phi+3)}
R_{C1}^{9\bar \mu/(3-2\phi)} V^{*(2-3\bar \mu)} + \nonumber \\
 q_g \left (\rho_1 G
\right )^{3 \bar \mu/2} R_{C1}^{3 \bar \mu} V^{*(2-3\bar \mu)},
\end{eqnarray}
where the first term represents the strength regime and the second the
gravity regime. $R_{C1}$ is the spherical radius of the combined
projectile and target masses at a density of $\rho_1 \equiv
1000$~kg~m$^{-3}$. The variable $R_{C1}$ was introduced by
\citet{Stewart:2009} in order to fit and compare the disruption
criteria for collisions with different projectile-to-target mass
ratios and to account for bodies with different bulk densities (e.g.,
rock and ice). $G$ is the gravitational constant; $q_s$ and $q_g$ are
dimensionless coefficients with values near 1.  $S$ is a measure of
the material strength in units of Pa s$^{3/(\phi+3)}$, and the
remaining variables, $\phi$ and $\bar \mu$, are dimensionless material
constants. $\phi$ is a measure of the strain-rate dependence of the
material strength with values ranging from 6 to 9
\citep[e.g.,][]{Housen:1990,Housen:1999}.  $\bar \mu$ is a measure of
how energy and momentum from the projectile are coupled to the target;
$\bar \mu$ is constrained to fall between 1/3 for pure momentum
scaling and 2/3 for pure energy scaling \citep{Holsapple:1987}.  Note
that the form of equation \ref{eqn:qstarred} assumes that the
projectile and target have the same density.

In the strength regime, the largest post-collision remnant is a
mechanically intact fragment. The catastrophic disruption criteria
decreases with increasing target size because more flaws grow and
coalesce during the longer loading duration in larger impact events
\citep[e.g.,][]{Housen:1999}.  In the gravity regime, disruption
requires both fracturing and gravitational dispersal
\citep{Melosh:1997,Benz:1999}; hence the disruption criteria increases
with increasing target size. In this regime, the largest remnant is a
gravitational aggregate composed of smaller intact fragments. In both
regimes, the disruption criteria increases with impact velocity
because more of the impact kinetic energy is dissipated by shock
deformation at higher velocities \citep{Housen:1990}. This work
focuses on the gravity regime; the strength regime will be the
subject of future studies.

Both equations \ref{eqn:vstardef} and \ref{eqn:qstarred} are satisfied
by collisions at exactly the catastrophic disruption threshold.  The
general formula for catastrophic disruption given by equation
\ref{eqn:qstarred} describes a family of curves that depend on size,
impact velocity, and material parameters ($q_g$ and $\bar \mu$). Most
previous work fit the material parameters in equation
\ref{eqn:qstarred} to planetary bodies of a particular composition
under various assumptions (e.g., a fixed impact velocity).  Next
(\S \ref{sec:minteract}--\ref{sec:princ}), we present a
general method to calculate the values for the disruption energy and
critical impact velocity for specific impact scenarios and materials.

\subsection{Derivation of a general catastrophic disruption law in the
gravity regime} \label{sec:univ}

\subsubsection{The Universal Law} \label{sec:univlaw}

\begin{figure}
  \includegraphics[scale=0.2]{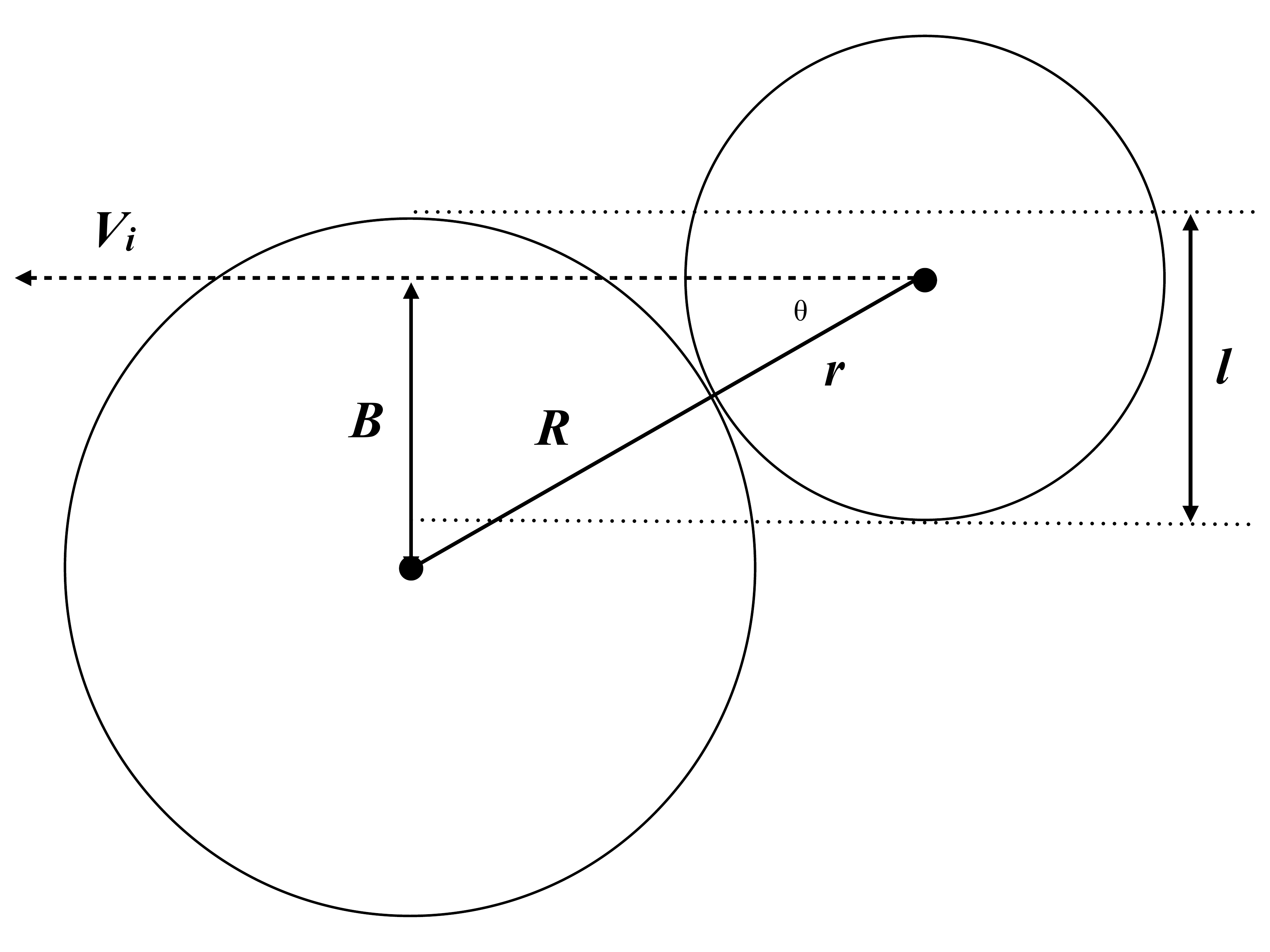}
  \caption{Schematic of the collision geometry. The target is
    stationary and the projectile is moving from right to left with
    speed $V_i$. The impact angle, $\theta$, is defined at the time of
    first contact as the angle between the line connecting the centers
    of the two bodies and the normal to the projectile velocity
    vector. The impact parameter is $b=\sin
    \theta$. \label{fig:cartoon}}
\end{figure}

\begin{figure}
 \includegraphics[scale=0.40]{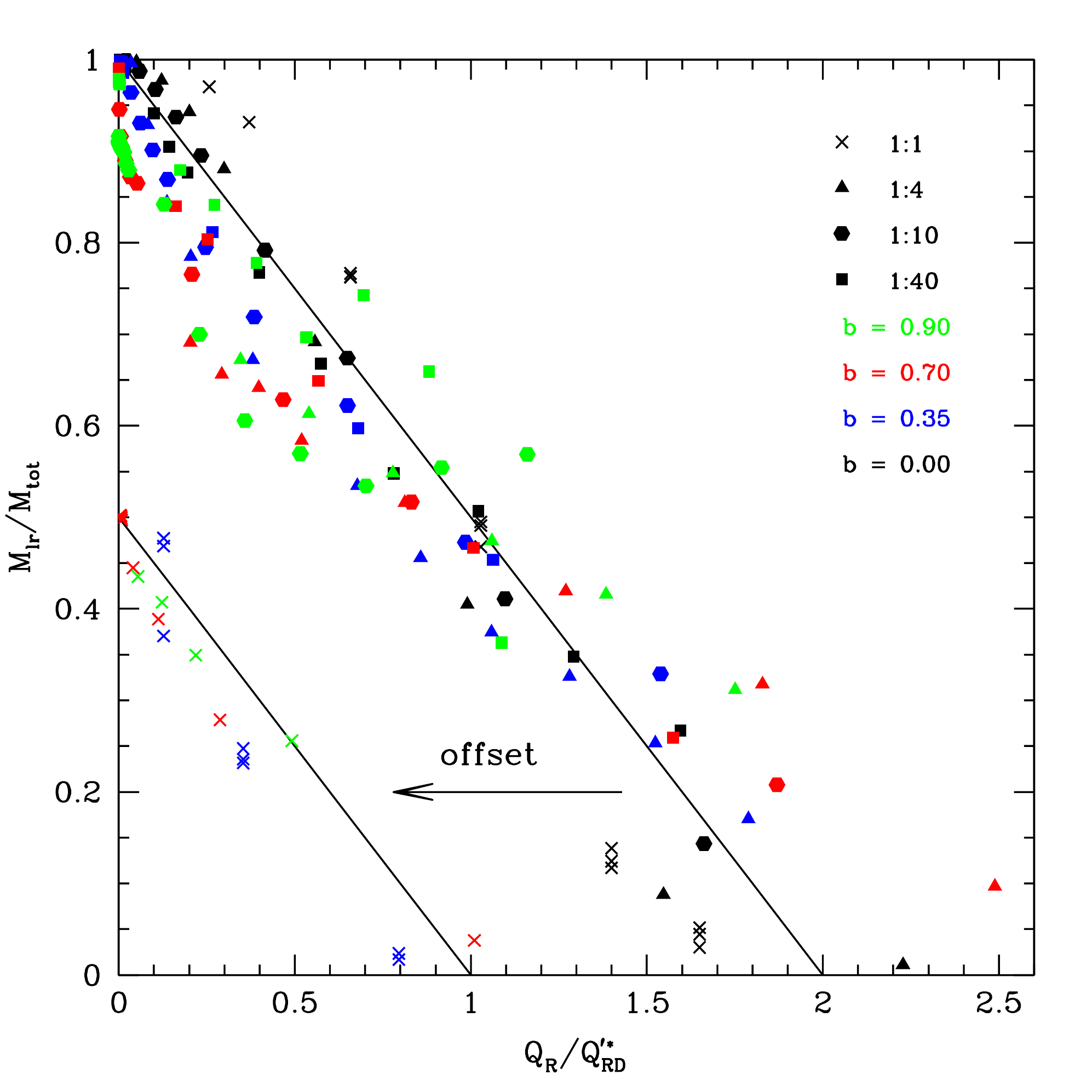}
 \caption{Normalized mass of the largest post-collision remnant versus
   normalized impact energy for all collisions in the disruption
   regime. The impact energy is scaled by the empirical catastrophic
   disruption criteria $Q^{\prime *}_{RD}$ (Table~\ref{tab:pkd}). The
   solid lines are the universal law for the mass of the largest
   remnant (equation \ref{eqn:univlaw}); see text for discussion of
   1:1 oblique impacts. The symbol denotes the projectile-to-target
   mass ratio, and the color denotes the impact
   parameter. \label{fig:univ}}
\end{figure}

In previous work, using simulations of head-on impacts, we determined
the value of $Q^*_{RD}$ for a particular pair of planetary bodies by
fitting the mass of the largest post-collision remnant, $M_{\rm lr}$,
as a function of the specific impact energy, $Q_R$. The simulations
held the projectile-to-target mass ratio fixed and varied the impact
velocity. For a wide range of target masses, projectile-to-target mass
ratios, and critical impact velocities, we found that the mass of the
largest remnant is approximated by a single linear relation,
\begin{equation} \label{eqn:univlaw}
M_{\rm lr}/M_{\rm tot} = -0.5(Q_R/Q^*_{RD} -1) + 0.5,
\end{equation}
where $Q^*_{RD}$ was fitted to be the specific energy such that
$M_{\rm lr}=0.5M_{\rm tot}$ \citep{Stewart:2009,Leinhardt:2009}.  We
found that a single slope agreed well with results from both
laboratory experiments and numerical simulations.  Furthermore, the
dimensional analysis by \citet{Housen:1990} supports the linearity of
the largest remnant mass with impact energy near the catastrophic
disruption threshold.  Hence, we refer to equation~\ref{eqn:univlaw}
as the ``universal law'' for the mass of the largest remnant.

However, the most likely collision between two planetary bodies is not
a head-on collision; a $45^\circ$ impact angle is most probable
\citep{Shoemaker:1962}.  The impact parameter is given by $b = \sin
\theta$, where $\theta$ is the angle between the centers of the bodies
and the velocity vector at the time of contact
(Figure~\ref{fig:cartoon}). The impact parameter has a significant
effect on the collision outcome because the energy of the projectile
may not completely intersect the target when the impact is
oblique. For example, in the collision geometry shown in
Figure~\ref{fig:cartoon}, the top of the projectile does not directly
hit the target (above the dotted line). As a result, a portion of the
projectile may shear off and only the kinetic energy of the
interacting fraction of the projectile will be involved in disrupting
the target. Thus, a higher specific impact energy is required to reach
the catastrophic disruption threshold for an oblique impact.

The new PKDGRAV simulations conducted for this study were used to
develop a generalized catastrophic disruption law, as previous work
did not independently vary critical parameters. Table \ref{tab:pkd}
presents the subset of the simulations discussed in detail below; for
a complete listing see Table \ref{tab:pkdall} in the Appendix.  The
simulations are grouped by impact scenario: fixed mass ratio and
impact angle. The value for the catastrophic disruption criteria,
$Q^{\prime *}_{RD}$, is found by fitting a line to the mass of the
largest remnant as a function of increasing impact energy in each
group. The prime notation in the catastrophic disruption criteria
indicates an impact condition that may be oblique ($b>0$).

With the new data, we first consider how impact angle influences the
universal law for the mass of the largest remnant. Figure
\ref{fig:univ} presents the normalized mass of the largest remnant
versus normalized specific impact energy. Our previous simulations
(all at $b=0$) are shown on the same universal law in
\citet{Stewart:2009}.  Note that comparable mass collisions with $b>0$
need to be considered carefully (offset for emphasis in
Figure~\ref{fig:univ}). Such collisions transition from merging to an
inelastic bouncing regime (called hit-and-run, discussed in \S
\ref{sec:minteract}) before reaching the disruption regime. As a
result, the mass of the largest remnant has a discontinuity between
$M_{\rm tot}$ and $M_{\rm targ}$ with increasing impact energy. So, in
the case of equal-mass collisions\footnote{A robust fit requires
  several points between $0.1M_{\rm tot}$ and $M_{\rm targ}$. A
  similar procedure was applied to fit the $Q^{\prime *}_{RD}$ for the
  $b=0.5$, $\gamma=1$ and $\gamma=0.5$ simulations from
  \citet{Marcus:2010} that are shown in Figure~\ref{fig:qsrd}.}, only
the fragments with $M_{\rm lr}<M_{\rm targ}$ are fit by a line of
slope -0.5.

Our new results demonstrate that the same universal law for the mass
of the largest remnant found for head-on collisions can be generalized to any
impact angle:
 \begin{equation} \label{eqn:univlawangle}
M_{\rm lr}/M_{\rm tot} = -0.5(Q_R/Q^{\prime \star}_{RD} -1) + 0.5.
\end{equation}
In detail, the mass of the largest fragment for a specific subset of
simulations may deviate slightly from the universal law
(Figure~\ref{fig:univ}).  Note that the deviations vary between subsets,
with some results systematically sloped more steeply and others sloped
more shallowly. The deviations in $M_{\rm lr}/M_{\rm tot}$ from equation
\ref{eqn:univlawangle} are about 10\% for near-normal impacts ($b=0.00$ and $b=0.35$) and somewhat
larger and more varied for highly oblique impacts.  

Overall, the universal law provides an excellent representation for
mass of the largest remnant for all disruptive collisions in the
gravity regime.  As a result, we have chosen to use the range of
impact energies that satisfy the universal law for the mass of the
largest remnant as the technical definition of the disruption regime.
At higher specific impact energies, the linear universal law breaks
down in a transition to the super-catastrophic regime
($Q_R/Q^*_{RD}\ge1.8$, see \S \ref{sec:super}). At lower specific
impact energies, the outcomes are merging or cratering (\S
\ref{sec:transitions}). Using our definition, the disruption regime
encompasses less than a factor of two in specific impact energy. The
outcomes in the disruption regime span partial accretion of the
projectile onto the target to partial erosion of the target body.

Note that the derived values for $Q^{\prime *}_{RD}$ are strong
functions of both the mass ratio and the impact parameter (Table
\ref{tab:pkd}). The catastrophic disruption energy rises with smaller
projectiles and larger impact parameters. \citet{Benz:1999}
investigated the effect of impact parameter on the disruption
criteria; however, their study fixed the impact velocity and varied
the mass ratio of the bodies. Hence, the individual roles of the
impact parameter and mass ratio cannot be discerned from their
data. In the next two sections, the influence of each factor is
isolated and quantified.

\begin{table*}
  \caption{Summary of parameters and results from selected PKDGRAV simulations (for full list of simulations see Appendix).\label{tab:pkd}}
\begin{tabular}{lccccccccccc}
  $\underline{M_{\rm p}}$ & $b$ &  $V_i$ & $\underline{M_{\rm lr}}$ & $\underline{M_{\rm slr}}$ & $\beta$ & $Q_{R}$ & $Q^{\prime *}_{RD}$ & $\alpha$ & $M_{\rm lr}/M_{\rm tot}$ & $M_{\rm slr}/M_{\rm tot}$ \\
$M_{\rm targ}$ & -- & m/s & $M_{\rm tot}$ & $M_{\rm tot}$ & -- & J/kg & J/kg & -- & Predicted & Predicted\\
\hline\hline
1.00 & 0.00 & 24 & 0.76 & 0.004 & 4.0 &  $7.2 \times 10^1$  &  &  &  0.67 & 0.01 \\
1.00$^*$ & 0.00 & 30 & 0.50 & 0.01 & 3.2 &  $1.1 \times 10^2$  &  $1.1 \times 10^2$  &1 & 0.49 & 0.01\\
1.00 & 0.00 & 35 & 0.12 & 0.05 & 2.9 &  $1.5 \times 10^2$  &  & &0.31 & 0.02\\
\hline
1.00 & 0.35 & 17 & 0.48 & 0.46 & 4.8  &  $3.6 \times 10^1$  &  & & 0.44$^\dag$ & 0.44$^\dag$ \\
1.00 & 0.35 & 30 & 0.23 & 0.20 & 3.1 &  $1.1 \times 10^2$  & $5.4 \times 10^1$ & 0.72 & 0.33$^\dag$ & 0.33$^\dag$ \\
1.00 & 0.35 & 45 & 0.02 & 0.01 & 3.7 &  $2.6 \times 10^2$  & &  & 0.11$^\dag$ & 0.11$^\dag$\\
\hline
1.00 & 0.70 & 12 & 0.50 & 0.49 & --  & $1.8 \times 10^1$ & & & 0.50$^\dag$ & 0.50$^\dag$ \\
1.00$^*$ & 0.70 & 80 & 0.28 & 0.27 & 4.1 &  $8.0 \times 10^2$  & $4.7 \times 10^2$ & 0.22 & 0.36$^\dag$ & 0.36$^\dag$\\
1.00 & 0.70 & 150 & 0.04 & 0.01 & 3.2 &  $2.8 \times 10^3$  & & & super-cat\\
\hline
1.00 & 0.90 & 20 & 0.50 & 0.50 & --  & $5.0 \times 10^1$ &    & & 0.50$^\dag$ & 0.50$^\dag$\\
1.00 & 0.90 & 400 & 0.35 & 0.34 & 4.6 &  $2.0 \times 10^4$  & $1.5 \times 10^4$& 0.03 & 0.39$^\dag$ & 0.39$^\dag$\\
1.00 & 0.90 & 600 & 0.25 & 0.25 & 3.2 &  $4.5 \times 10^4$  &  & & 0.25$^\dag$& 0.25$^\dag$\\
\hline
0.25 & 0.00 & 30 & 0.69 & 0.01 & 3.7 & $7.2 \times 10^1$ & & & 0.72 & 0.01\\
0.25$^{\dag *}$ & 0.00 & 40 & 0.40 & 0.02 & 4.4 & $1.3 \times 10^2$ & $1.3 \times 10^2$ & 1 & 0.51 & 0.01 \\
0.25 & 0.00 & 50 & 0.09 & 0.02 & 3.4 & $2.0 \times 10^2$ &  &  & 0.23 & 0.02\\
\hline
0.25 & 0.35 & 30 & 0.67 & 0.01 & 4.5 & $7.2 \times 10^1$ & & & 0.81 & 0.005\\
0.25$^*$ & 0.35 & 40 & 0.53 & 0.01 & 4.1 & $1.3 \times 10^2$ & $1.9 \times 10^2$ & 0.93 & 0.66 & 0.01\\
0.25 & 0.35 & 60 & 0.25 & 0.01 & 3.8 & $2.9 \times 10^2$ & & & 0.24 & 0.02\\
\hline
0.25 & 0.70 & 50 & 0.69 & 0.01 & 2.8 & $2.0 \times 10^2$ & & & 0.90 & 0.003\\
0.25 & 0.70 & 100 & 0.52 & 0.004 & 3.3 & $8.0 \times 10^2 $& $9.9 \times 10^2$ & 0.33 & 0.60 & 0.01 \\
0.25 & 0.70 & 150 & 0.32 & 0.01 & 2.5 & $1.8 \times 10^3$ &  & & 0.09 & 0.02 \\
\hline
0.25 & 0.90 & 120 & 0.77 & 0.14 &  3.17  & $1.2 \times 10^3$ &     & & 0.94$^{**}$ & 0.002$^{**}$\\
0.25 & 0.90 & 350 & 0.47 & 0.003 & 4.40 & $9.9 \times 10^3$ & $9.3 \times 10^3$ & 0.05 & 0.47& 0.01\\
0.25 & 0.90 & 450 & 0.31 & 0.01 & 3.28 & $1.6 \times 10^4$ &   & & 0.13 & 0.02 \\
\hline
0.10 & 0.00 & 40 & 0.79 & 0.001 & 4.9 & $6.7 \times 10^1$ &   &  & 0.79 & 0.005\\
0.10 & 0.00 & 65 & 0.41 & 0.01 & 3.7 & $1.8 \times 10^2$ &  $1.6 \times 10^2$ & 1 & 0.45 & 0.01\\
0.10 & 0.00 & 80 & 0.14 & 0.03 & 3.4 & $2.7 \times 10^2$ & & & 0.17 & 0.02\\
\hline
0.10 & 0.35 & 40 & 0.79 & 0.002 & 3.7 & $6.7 \times 10^1$ & & &0.88 & 0.003 \\
0.10 & 0.35 & 80 & 0.47 & 0.01 & 4.5 & $2.7 \times 10^2$ & $2.7 \times 10^2$ & $1^{\dag \dag}$ & 0.51 & 0.01\\
0.10$^{\dag *}$ & 0.35 & 100 & 0.33 & 0.01 & 3.6 & $4.2 \times 10^2$ & & & 0.23 & 0.02\\
\hline
0.10$^*$ & 0.70 & 100 & 0.77 & 0.002 &  3.6  & $4.2 \times 10^2$ &  & & 0.90 & 0.003\\
0.10 & 0.70 & 200 & 0.52 & 0.004 & 3.8 & $1.7 \times 10^3$ & $2.0 \times 10^3$ & 0.46 & 0.59 & 0.01\\
0.10 & 0.70 & 300 & 0.21 & 0.01 & 3.3 & $3.7 \times 10^3$ & & & 0.07 & 0.02\\
\hline
0.10 & 0.90 & 400 & 0.70 & 0.001 &  5.2  & $6.7 \times 10^3$ & & & 0.89 & 0.003 \\
0.10 & 0.90 & 700 & 0.53 & 0.002 & 4.8 & $2.0 \times 10^4$ & $2.9 \times 10^4$  & 0.07 & 0.65 & 0.01\\
0.10 & 0.90 & 900 & 0.57 & 0.01 & 4.3 & $3.4 \times 10^4$ & & & 0.42 & 0.02\\
\hline
0.025 & 0.00 & 100 & 0.77 & 0.001 & -- & $1.2 \times 10^2$    &  & & 0.91 & 0.002\\
0.025 & 0.00 & 140 & 0.55 & 0.01 & 4.45 & $2.3 \times 10^2$   &  $6.4 \times 10^2$ & 1 & 0.82 & 0.01\\
0.025 & 0.00 & 160 & 0.51 & 0.01 & 4.10 & $3.1 \times 10^2$     &  & & 0.76 & 0.01\\
\hline
0.025 & 0.35 & 160 & 0.60 & 0.01 & 3.97 & $3.1 \times 10^2$   &  & & 0.79 & 0.01\\
0.025 & 0.35 & 200 & 0.45 & 0.01 & 4.78 & $4.8 \times 10^2$ & $7.2 \times 10^2$ & $1^{\dag \dag}$ & 0.67 & 0.01\\
0.025 & 0.35 & 300 & 0.07 & 0.05 & 3.25 & $1.1 \times 10^3$ & & & 0.25 & 0.02\\
\hline
0.025$^{\dag *}$ & 0.70 & 300 & 0.65 & 0.002 &  5.08  & $1.1 \times 10^3$&  & & 0.73 & 0.01\\
0.025 & 0.70 & 400 & 0.47 & 0.01 & 3.59 & $1.9 \times 10^3$ & $2.0 \times 10^3$ & 0.74 & 0.52 & 0.01\\
0.025 & 0.70 & 500 & 0.26 & 0.01 & 3.23 & $3.0 \times 10^3$ & & & 0.26 & 0.02\\
\hline
0.025* & 0.90 & 800 & 0.74 & 0.001 & -- & $7.7 \times 10^3$ &    & & 0.65 & 0.01\\
0.025${^\dag *}$ & 0.90 & 900 & 0.66 & 0.002 & 4.06 & $9.7 \times 10^3$ &  $1.1 \times 10^4$ & 0.12 & 0.56 & 0.01\\
0.025 & 0.90 & 1000  &  0.36   & 0.003   &   3.77   &   $1.2 \times 10^4$ &    &  & 0.46 & 0.01\\
\hline
\end{tabular}

\footnotesize{$\frac{M_{\rm p}}{M_{\rm targ}}$ -- mass of projectile normalized
  by mass of target; $b$ -- impact parameter; $V_i$ -- projectile
  impact velocity; $\frac{M_{\rm lr}}{M_{\rm tot}}$ -- mass of largest remnant
  normalized by total mass; $M_{\rm slr}$ -- mass of the second largest
  remnant; $\beta$ -- slope of cummulative size distribution; $Q_{R}$
  -- center of mass specific energy; $Q^{\prime *}_{RD}$ -- empirical
  critical center of mass specific energy for catastrophic disruption
  and gravitational dispersal derived from the simulations. In all
  cases, the target contained $\sim 1\times 10^4$ particles,
  $M_{\rm targ}=4.2\times 10^{15}$ kg, $R_{\rm targ}=10^4$ m; $^*$ indicates
  models shown {\bf in blue} in Figure~\ref{fig:cumr}; $^\dag$ indicates $N_{\rm lr} = 2$
  and $N_{\rm slr} = 4$; $^{**}$ erosive hit-and-run regime, the
  disruption regime model does not apply; $^{\dag \dag}$ indicates
  an $\alpha$ for which $b > 0$ but $l < R$ thus $\alpha = 1$; -- not
  enough material to fit a power law; $^{\dag *}$ indicates models shown in Figure~\ref{fig:vwrtlr}. }
\end{table*}

\subsubsection{Dependence of disruption on impact angle and derivation
  of the interacting mass} \label{sec:minteract}

In order to describe the dependence of catastrophic disruption on
impact angle, we introduce two geometrical collision groups
(Figure~\ref{fig:cartoon}): {\it non-grazing} -- most of the projectile
interacts with the target, and {\it grazing} -- less than half the
projectile interacts with the target. Following \citet{Asphaug:2010},
the critical impact parameter,
\begin{equation} \label{eqn:bcrit}
b_{\rm crit} =\left( \frac{R}{R+r} \right),
\end{equation}
is reached when the center of the projectile (radius $r$) is tangent to the surface
of the target (radius $R$). Grazing impacts are defined to occur when $b >
b_{\rm crit}$. 

When considering a non-grazing impact scenario with a particular $b$
and $\gamma$, the collision outcome transitions smoothly from merging
to disruption as the impact velocity increases. For grazing impacts,
however, the collision outcome transitions abruptly from merging
($M_{\rm lr} \sim M_{\rm tot}$) to hit-and-run ($M_{\rm lr} \sim
M_{\rm targ}$) and then (less abruptly) to disruption (see \S
\ref{sec:transitions}). Thus, only collision energies that result in
$M_{\rm lr}<M_{\rm targ}$ should be used in the derivation of
$Q^{\prime *}_{RD}$ in the grazing regime, as done for the $\gamma=1$
results shown in Figure~\ref{fig:univ}.

During oblique impact events, a significant fraction of the projectile
may not actually interact with the target, particularly for comparable
mass bodies. For gravity-dominated bodies, the projectile is
decapitated and a portion of the mass misses the target entirely. As a
result, only a fraction of the projectile's total kinetic energy is
deposited in the target, and the impact velocity must increase to
reach the catastrophic disruption threshold.

Using a simple geometric model, we derive the fraction
of the projectile mass that is estimated to be involved in the
collision. First, we define $l$ as the projected length of the
projectile overlapping the target. As shown in Figure~\ref{fig:cartoon}, 
\begin{equation}
l + B = R + r,
\end{equation}
where $B = (R+r)\,\sin\theta$.
Placing the origin at the bottom of the projectile on the center line
and the positive z-axis pointing to the top of the page, the estimated
projectile mass involved in the collision, $m_{\rm interact}$, is
determined by integrating cylinders of height $dz$ and radius $a$ from
$0$ to $l$ along the z-axis,
\begin{equation}
m_{\rm interact} = \rho \int_0^l \pi a^2 dz,
\end{equation}
where $\rho$ is the bulk density of the projectile. The radius of each
cylinder can be defined in terms of the radius of the projectile and
the height from the origin,
\begin{equation}
a^2 = r^2 - (r - z)^2.
\end{equation}
Then,
\begin{equation}
m_{\rm interact} = \rho(\pi r l^2 - (\pi/3) l^3).
\end{equation}
Dividing by the total mass of the projectile, $M_{\rm p}$,
\begin{equation} \label{eqn:alpha}
\frac{m_{\rm interact}}{M_{\rm p}} = \frac{3 r l^2 - l^3}{4 r^3} \equiv \alpha.
\end{equation}
Thus, $\alpha$ is the mass fraction of the projectile estimated to be
involved in the collision (see Table \ref{tab:pkd} for the values of
$\alpha$ in our simulations). The entire projectile interacts with the
target when $R > b (r+R) + r$; then $l<R$ and $\alpha=1$.

In order to account for the effect of impact angle on $Q^{\prime
  *}_{RD}$, we include the kinetic energy of only the interacting
mass. The appropriate reduced mass is then
\begin{equation}
\mu_{\alpha}=\frac{\alpha M_{\rm p} M_{\rm targ}}{\alpha M_{\rm p} + M_{\rm targ}}. \label{eqn:mualpha}
\end{equation}
Now consider the difference between a head-on impact by a projectile
of mass $M_{\rm p}$ at $V^*$ and a head-on impact by a projectile of
mass $\alpha M_{\rm p}$. At the same impact velocity, the impact
energies between the two cases differ by the ratio of the reduced
masses,
\begin{equation} \label{eqn:modqr}
Q^{\prime}_{R} = \frac{\mu}{\mu_{\alpha}} Q_{R}.
\end{equation}
Next, in order to conserve the effective specific impact energy, the
impact velocity must increase with increasing impact angle such that
\begin{equation} \label{eqn:vstarbar}
\bar V^{'*} = \sqrt{ \frac{\mu}{\mu_{\alpha}} V^{*2}}.
\end{equation}
However, the disruption criteria itself depends on the magnitude of
the impact velocity (equation \ref{eqn:qstarred}). In other words,
when the effective projectile mass changes, the change in the impact
velocity required for disruption varies by more than the factor
presented in equation \ref{eqn:vstarbar}.  Combining these two 
effects leads to the relationship between the oblique and head-on
disruption energy for a fixed mass ratio collision,
\begin{eqnarray} 
  Q^{'*}_{RD} &=& \left ( \frac{\mu}{\mu_{\alpha}} Q^*_{RD} \right ) \left ( \frac{\bar V^{'*}}{V^*} \right )^{2-3\bar \mu}, \nonumber \\
                 &=& \left ( \frac{\mu}{\mu_{\alpha}} \right )^{2-3\bar \mu/2} Q^*_{RD}.
\label{eqn:qangle}
\end{eqnarray}
By definition, the critical impact velocity for an oblique impact must
satisfy equation \ref{eqn:vstardef}:
\begin{equation} \label{eqn:vstarprime}
  V^{'*}= \sqrt{\frac{2Q^{'*}_{RD}M_{\rm tot}}{\mu}}.
\end{equation}
The correction for changing the mass ratio is derived in the next
section.

\begin{figure*}
\includegraphics[scale=0.45]{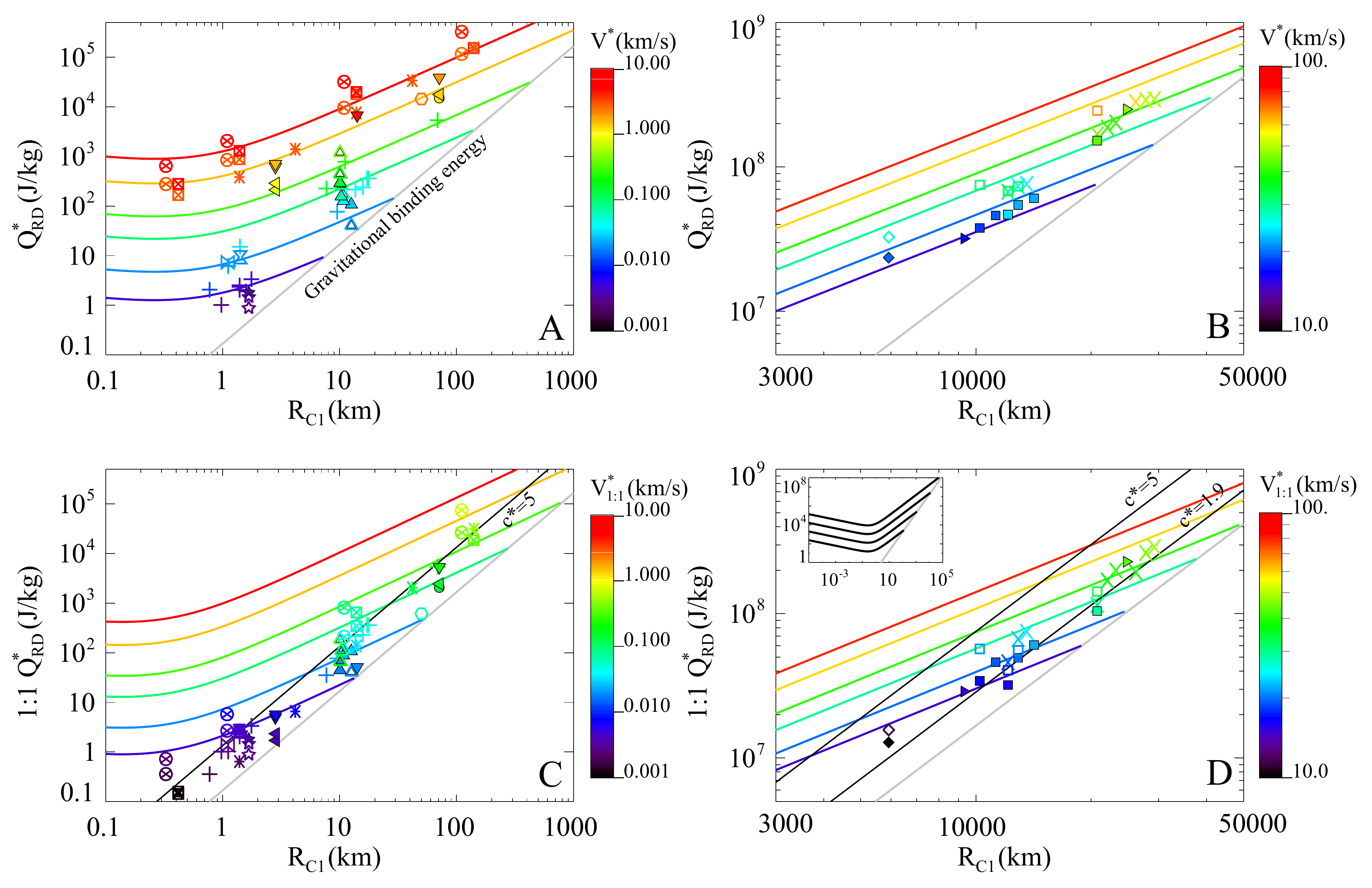}
\caption{A compilation of gravity-regime catastrophic disruption
  simulation results. Symbols denote different target materials
  (Table~\ref{tab:datasources}), and color denotes the critical impact
  velocity, $V^*$. Filled and line symbols are head-on impacts; open
  symbols are oblique impacts.  {\bf A and B}: Data corrected for
  impact angle to equivalent head-on impact using the interacting mass
  (equations \ref{eqn:qheadon} and
  \ref{eqn:vheadon}). Constant-velocity $Q^*_{RD}$ curves (equation
  \ref{eqn:qstarred}) are best fit to all the data with $\bar
  \mu=0.35$ and $q_g=1$.  The fit between the data and model curves is
  very good over almost 5 orders of magnitude in size and 9 orders of
  magnitude in impact energy. Contours for $V^*=.005$, .02, .1, .3,
  1.5 and 5 km~s$^{-1}$ (A \& C) and $V^*=15$, 20, 30, 40, 60, 80
  km~s$^{-1}$ (B \& D).  {\bf C and D}: Data converted to an
  equivalent equal-mass (1:1) disruption criteria using equations
  \ref{eqn:qgamma} and \ref{eqn:vgamma}. The equal-mass data fall on
  lines proportional to $R^2_{C1}$. Fits to the equal-mass data are
  called ``principal disruption curves'' that are defined by $c^*$
  (black lines, equation \ref{eqn:principalq}); $c^*$ represents the
  value for the equal-mass $Q^*_{RD}$ in units of the specific
  gravitational binding energy. Best fit values are $c^*=5$ and $\bar
  \mu=0.37$ for small bodies and $c^*=1.9$ and $\bar \mu=0.36$ for
  hydrodynamic planets.  {\bf Inset}: Full $Q^{*}_{RD}$ curves (0.1,
  1, 10, 100 km s$^{-1}$) showing transition from strength to gravity
  regimes.  \label{fig:qsrd}}
\end{figure*}

Our model for the effect of impact angle is used to derive equivalent
head-on $Q^*_{RD}$ values from our new and previously published
catastrophic disruption data. Using the values for $Q^{'*}_{RD}$ and
$V^{'*}$ fitted to the oblique simulation results, the equivalent
head-on impact disruption criteria are
\begin{eqnarray}
Q^*_{RD} &=& Q^{'*}_{RD} \left ( \frac{\mu}{\mu_{\alpha}} \right
)^{(3\bar \mu/2 - 2)}, \label{eqn:qheadon} \\
V^*     &=& \sqrt{ \frac{2Q^*_{RD}M_{\rm tot}}{\mu} }. \label{eqn:vheadon}
\end{eqnarray}

We considered the catastrophic disruption of a wide variety of
planetary bodies from the studies summarized in Table
\ref{tab:datasources}.  First, we fit the general expression for
$Q^*_{RD}$ (equation \ref{eqn:qstarred}\footnote{In the fitting
  procedure, the strength term is neglected in equation
  \ref{eqn:qstarred}. For the lines plotted in Figure~\ref{fig:qsrd},
  the strength regime parameters are fixed at $\phi=7$, $S=2.4$ Pa
  s$^{0.3}$, and $q_s=1$ based on the work in \citet{Stewart:2009}.})
to the (equivalent) head-on disruption data to derive the values of
$q_g$ and $\bar \mu$ that best describe the entire data set, from
planetesimals to planets. The same value for the material parameter
$\bar \mu$ is used in the angle correction and the fit to equation
\ref{eqn:qstarred}. A small number of data points were excluded from
the global fit, which are discussed in \S \ref{sec:strength}.  The
best fit values for $q_g$ and $\bar \mu$ were found by minimizing the
absolute value of the log of the fractional error, $\delta=|{\rm
  log}(Q^*_{RD,\rm sim}/Q^*_{RD,\rm model})|$.

In some cases, the impact angle correction is significant (e.g., the
impact scenarios with small values of $\alpha$ given in
Table~\ref{tab:pkd}). With the exception of the constant-velocity
results from \citet{Benz:1999} and \citet{Jutzi:2010} and the mixed
velocity data from \citep{Benz:2000}, the disruption data were derived
from simulations conducted with a constant mass ratio and the critical
impact velocity for catastrophic disruption, $V^{'*}$, was found by
fitting to the universal law. For the simulations described in
Table~\ref{tab:pkd}, the model correction for impact angle usually
yields an impact energy within a factor of 2 of the simulation results
for head-on collisions (e.g., within the linear regime for the mass of
the largest remnant). We restricted our fits to cases where
$\alpha>0.5$ to reduce any error contribution from a poor model
correction for highly oblique impacts.

The compiled data and best fit model $Q^*_{RD}$ are presented in
Figure~\ref{fig:qsrd}A and B. The combined data are well fit by
$q_g=1.0$ and $\bar \mu=0.35$ ($\delta=0.14$). Note the good match in
the values for $V^*$ (colors) from the simulations with the lines of
constant $V^*$. Similarly good fits are found for $0.33\le \bar \mu
\le0.36$ and $0.8\le q_g \le 1.2$ with $0.14 < \delta <0.15$.
Amazingly, the compilation of catastrophic disruption data is well fit
by equation \ref{eqn:qstarred} for single values of $q_g$ and $\bar
\mu$ for a wide variety of target compositions and over almost 5
orders of magnitude in size and 9 orders of magnitude in impact
energy.  The critical impact velocities span 1~m~s$^{-1}$ to several
10's km~s$^{-1}$. The best fit value for $\bar \mu$ falls near pure
momentum scaling ($\bar \mu=1/3$).

Upon closer examination, we found that the global fit with equation
\ref{eqn:qstarred} systematically predicts a low disruption energy for
small bodies ($R_{C1}<1000$~km) and a high disruption energy for
planet-sized bodies. Next, we consider separately the data for small
and large bodies. A better fit is found for the small body data in
Figure~\ref{fig:qsrd}A with $0.35 \le \bar \mu \le 0.37$ and $1.4 \le
q_g \le 1.65$ with $0.11 < \delta < 0.12$.  The small body data
includes hydrodynamic to strong bodies and different compositions. The
planet data in Figure~\ref{fig:qsrd}A are best fit with $0.35 \le \bar
\mu \le 0.375$ and $0.85 \le q_g \le 1.0$ with the very small error of
$0.038 < \delta < 0.041$. The planet size data includes three
different target compositions. The data for collisions between small
strong bodies have the largest dispersion; these data will be
discussed in \S \ref{sec:strength}.

\subsubsection{Dependence of disruption on mass
  ratio} \label{sec:massratio}

By fitting such a large collection of data, it is clear that equation
\ref{eqn:qstarred} describes a self-consistent family of possible
$Q^*_{RD}$ values. For a specific impact scenario, the correct value
for $V^*$ at each $R_{C1}$ is ambiguous because $V^*$ depends on both
a material property and the mass ratio.  In studies that hold $V_i$
constant and vary the mass ratio, the derived value for the critical
impact energy only applies for the corresponding critical mass
ratio. As noted in previous work, the critical impact velocity falls
dramatically as the mass ratio approaches 1:1
\citep{Benz:2000,Stewart:2009}. As a result, collisions between
equal-mass bodies require the smallest impact velocity to reach the
catastrophic disruption threshold.

Because a mass ratio of 1:1 defines the lowest disruption energy for a
fixed total mass, we derive the disruption criteria for different mass
ratios with respect to the equal-mass disruption criteria,
$Q^*_{RD,\gamma=1}$. We begin with the equality between the impact
energy and gravity term in the disruption energy (equation
\ref{eqn:qstarred}),
\begin{eqnarray} 
Q_{R} & = & Q^*_{RD}, \nonumber \\
\frac{\mu V^{*2}}{2 M_{\rm tot}} &=& q_g \left (\rho_1 G \right )^{3
  \bar \mu/2} R_{C1}^{3 \bar \mu} V^{*(2-3\bar \mu)}.  
\end{eqnarray} 
Note that
\begin{eqnarray} 
\mu & = & M_{\rm p} M_{\rm targ} / (M_{\rm p} + M_{\rm targ}), \nonumber \\
      & = & \frac{\gamma}{\gamma+1} M_{\rm targ}, 
\end{eqnarray}
and
\begin{equation} 
M_{\rm tot} =  (\gamma+1) M_{\rm targ}.
\end{equation}
Then, substituting for $\mu$ and $M_{\rm tot}$, 
\begin{eqnarray} 
\frac{(\gamma / (\gamma+1)) M_{\rm targ} V^{*2}}{2 (\gamma+1) M_{\rm targ}} &=&
  q_g \left (\rho_1 G \right )^{3 \bar \mu/2} R_{C1}^{3 \bar \mu}
  V^{*(2-3\bar \mu)}, \nonumber \\
V^* &=& \left [ \frac{2(\gamma+1)^2}{\gamma} q_g \left (\rho_1 G
  \right )^{3 \bar \mu/2} R_{C1}^{3 \bar \mu} \right ]^{1/(3 \bar
  \mu)}, \nonumber \\
    &=& \left [ \frac{1}{4} \frac{(\gamma+1)^2}{\gamma} \right ]^{1/(3 \bar
  \mu)}  V^*_{\gamma=1}. \label{eqn:vgamma}
\end{eqnarray}
Then, for the same total mass, the relationship between the equal-mass
disruption energy and any other mass ratio is determined by the
difference in the critical impact velocities,
\begin{eqnarray} 
  Q^*_{RD} &=& Q^*_{RD,\gamma=1} \left ( \frac{V^*}{V^*_{\gamma=1}}
    \right )^{(2-3\bar \mu)}, \nonumber \\
          &=& Q^*_{RD,\gamma=1} \left ( \frac{1}{4} \frac{(\gamma+1)^2}{\gamma}
    \right )^{2/(3\bar \mu) - 1}. \label{eqn:qgamma}
\end{eqnarray}
The equations for $Q^*_{RD,\gamma=1}$ and $V^{*}_{\gamma=1}$ are given
in the next section.

In the compilation of catastrophic disruption data shown in
Figure~\ref{fig:qsrd}C and D, all the $\gamma<1$ data have been
converted to an equivalent equal-mass impact disruption energy and the
colors denote $V^*_{\gamma=1}$. For example, the critical disruption
energy from head-on PKDGRAV simulations with $\gamma=0.03$ are a
factor of three above the disruption energy for $\gamma=1$ in
Figure~\ref{fig:qsrd}A \citep[parallel sets of $+$ from
][]{Stewart:2009}. The data lie on the same line after the correction
in Figure~\ref{fig:qsrd}C. The correction also brings together data
from studies using different numerical methods and vastly different
material properties. For example, the high-velocity $Q^*_{RD}$ for
strong and weak basalt targets ($\blacktriangledown \blacktriangleleft
\otimes \boxtimes$) fall on the same line as the PKDGRAV rubble piles
after the conversion to an equivalent equal-mass impact. Similarly,
studies of the disruption of Mercury \citep{Benz:2007} follow the same
curve as disruption of earth-mass water/rock planets
\citep{Marcus:2010}. The general form for the equal-mass disruption
criteria is derived in the next section.

\subsubsection{The principal disruption curve}\label{sec:princ}

In the previous two sections, we calculated the disruption criteria
for head-on equal-mass collisions by adjusting the critical disruption
energy to account for different impact angles and mass ratios. The
head-on equal-mass data points, derived from the compilation of
numerical simulations, fall along a single curve that we name the
``principal disruption curve'' (black lines in Figure~\ref{fig:qsrd}C
and D).

On the principal disruption curve, the critical impact velocity for
equal-mass head-on impacts, $V^*_{\gamma=1}$, satisfies both equation
\ref{eqn:qr} and the gravity regime term in equation
\ref{eqn:qstarred}:
\begin{eqnarray} 
Q_{R,\gamma=1} & = & Q^*_{RD,\gamma=1} \nonumber \\
\frac{\mu_{\gamma=1} V^{*2}_{\gamma=1}}{2 M_{\rm tot}} &=& q_g \left (\rho_1 G \right )^{3 \bar \mu/2} R_{C1}^{3 \bar \mu} V^{*(2-3\bar \mu)}_{\gamma=1},
\end{eqnarray}
Then, substituting $\mu_{\gamma=1}=M_{\rm targ}/2=M_{\rm tot}/4$,
\begin{eqnarray}
V^{*}_{\gamma=1} & = & \left [8 q_g \left (\rho_1 G \right )^{3 \bar
    \mu/2} R_{C1}^{3 \bar \mu} \right ]^{1/(3 \bar \mu)}, \nonumber \\
              & = & (8 q_g)^{1/(3 \bar \mu)} \left (\rho_1 G \right )^{1/2} R_{C1}. \label{eqn:vstartmp}
\end{eqnarray}
Thus, along a curve with a fixed projectile-to-target mass ratio, the
critical impact velocity has a linear dependence on $R_{C1}$. The
linear dependence of $V^*$ on $R_{C1}$ for a fixed mass ratio was
confirmed by the numerical simulations in \citet{Stewart:2009} ($+$ in
Figure~\ref{fig:qsrd}).

Then, consider the dependence of the catastrophic disruption criteria
on size (equation \ref{eqn:qstarred}) and replace the velocity term
with size,
\begin{eqnarray}
Q^*_{RD,\gamma=1} & \propto & R_{C1}^{3 \bar \mu} V^{*(2-3\bar \mu)},
\nonumber \\
 & \propto & R_{C1}^{3 \bar \mu} R_{C1}^{(2-3\bar \mu)}, \nonumber \\
 & \propto & R_{C1}^2. 
\end{eqnarray}
Thus, the catastrophic disruption criteria scales as radius squared
along any curve with a fixed projectile-to-target mass ratio.

Next, note the proximity of the gravity-regime equal-mass disruption
energy to the specific gravitational binding energy,
\begin{equation}
U=\frac{3GM_{\rm tot}}{5R_{C1}},
\end{equation}
shown as the grey line in Figure~\ref{fig:qsrd}. We define a
dimensionless material parameter, $c^*$, that represents the offset
between the gravitational binding energy and the equal-mass disruption
criteria. Then, the principal disruption curve is given by
\begin{equation} \label{eqn:principalq}
  Q^*_{RD,\gamma=1}= c^* \frac{4}{5} \pi \rho_1 G R_{C1}^2.
\end{equation}
The parameter $c^*$ is a measure of the dissipation of energy within
the target. 

The coefficient $q_g$ is found by substituting equation
\ref{eqn:vstardef} for $Q^*_{RD,\gamma=1}$ in equation
\ref{eqn:principalq} and then equation \ref{eqn:vstartmp} for
$V^{*}_{\gamma=1}$:
\begin{eqnarray} 
\frac{\mu_{\gamma=1} V^{*2}_{\gamma=1}}{2 M_{\rm tot}} & = & c^* \frac{4}{5} \pi \rho_1 G R_{C1}^2, \nonumber \\
(1/8) (8 q_g)^{2/(3 \bar \mu)}  &= & c^* \frac{4}{5} \pi,  \nonumber \\
q_g  &= & \frac{1}{8} \left ( \frac{32 \pi c^*}{5} \right )^{3 \bar
  \mu / 2}. \label{eqn:qg}
\end{eqnarray}
Finally, substituting $q_g$ into equation \ref{eqn:vstartmp} gives 
\begin{equation}
V^{*}_{\gamma=1} = \left ( \frac{32 \pi c^*}{5} \right )^{1/2} (\rho_1 G)^{1/2} R_{C1}. \label{eqn:vequal}
\end{equation}
Hence, the critical velocity along the disruption curve for equal-mass
impacts is solely a function of $R_{C1}$ and $c^*$. 

The principal disruption curve (equation \ref{eqn:principalq}) is a
simple, yet powerful way to compare the impact energies required to
disrupt targets composed of different materials. Each material is
defined by a single parameter $c^*$. In Figure~\ref{fig:qsrd}C and D,
the best fit values are $c^*=5\pm2$ and $\bar \mu=0.37\pm0.01$ for
small bodies with a wide variety of material characteristics and
$c^*=1.9\pm0.3$ and $\bar \mu=0.36\pm0.01$ for the hydrodynamic
planet-size bodies. These simulations span pure hydrodynamic targets
(no strength), rubble piles, ice, and strong rock targets. Hence, for
all the types of bodies encountered during planet formation, $c^*$ is
limited to a small range of values. Note that the difference in $c^*$
between the small and large bodies is not simply because of the
differentiated structure of the large bodies; two pure rock cases
\citep[$\blacktriangleright$,][]{Marcus:2009} fall on the same
$Q^*_{RD,\gamma=1}$ curve. Rather, the large bodies were all studied
using a pure hydrodynamic model, whereas the small bodies were studied
using techniques that incorporated material strength in various
ways. A transition from a higher value for $c^*$ for small bodies to a
lower value for planet-sized bodies is appropriate for planet
formation studies, as discussed in \S\ref{sec:discussion}.

Now it is clear that most of the differences in the catastrophic
disruption threshold found in previous work are the result of
differences in impact velocity and mass ratio (few studies varied
impact parameter). 

Here, we have derived a general formulation for the catastrophic
disruption criteria that accounts for material properties, impact
velocity, mass ratio, and impact angle. The forward calculation of
$Q^{\prime *}_{RD}$ for a specific impact scenario between bodies with
material parameters $c^*$ and $\bar \mu$ is described in the Appendix
and in the companion paper \citep{Stewart:2011}.

\subsection{Fragment size distribution} \label{sec:fragsize}

\begin{figure*}
 \includegraphics[scale=0.8]{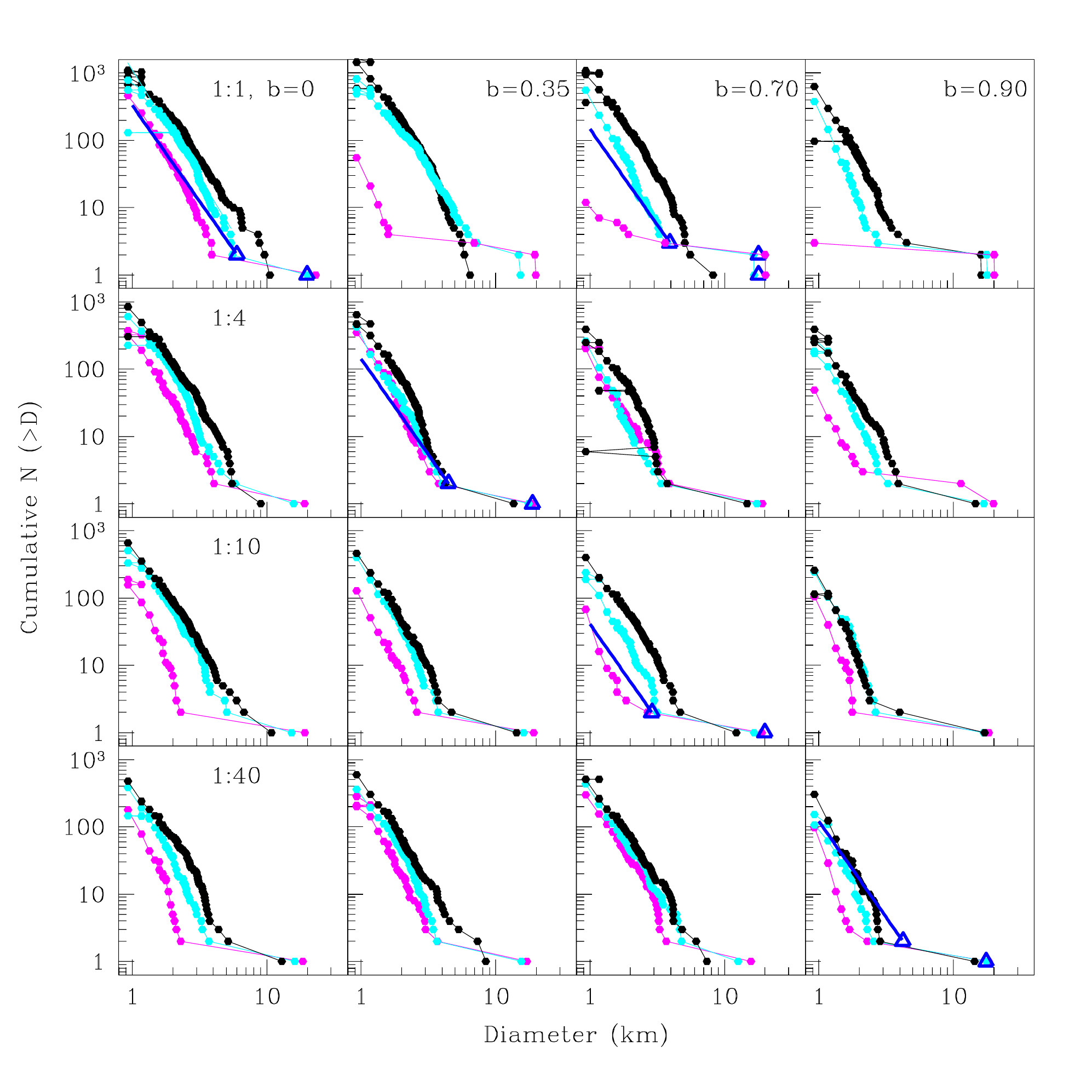}
 \caption{Cumulative size distribution versus fragment diameter. For
   each mass ratio $\gamma$ and impact parameter $b$, size
   distributions are shown for three different impact energies. Table \ref{tab:pkd} provides the details for these
   simulations. The
   colors are an aid for the eye: magenta is the lowest energy impact
   in each block of three in Table \ref{tab:pkd}, black is the highest energy, and cyan is
   in-between.  In five panels, the fragment size distribution scaling
   law (blue line and triangles) is compared to the data. The impact
   parameters used for the model comparison are indicated in Table
   \ref{tab:pkd} by a * in the first column.  \label{fig:cumr}}
\end{figure*}

In the disruption regime, our new simulations resolve the size
distribution of fragments over a decade in size (Figure~\ref{fig:cumr}).
In general, the post-collision fragments smaller than the largest
remnant form a smooth tail that can be fit well by a single power
law. The second-largest remnant forms the base of this tail. For most
collisions there is a significant separation in size between the
largest and second largest remnant. However, if the collision is very
energetic, the largest remnant joins the power-law distribution (e.g.,
in $\gamma=1, b=0.35$). In addition, for the hit-and-run impacts with
$\gamma = 1$, the two largest remnants are comparable in size. Only
the most energetic scenarios with $\gamma=1$ fall in the disruption
regime (e.g., black lines in $b=0.35$ and 0.7). In all disruption
regime collisions, the slope of the cumulative power-law tail,
$-\beta$ (see Table \ref{tab:pkd}), is effectively independent of the
impact conditions ($b$, $V_i$, $\gamma$).

Using the method from \citet{Wyatt:2002} and Paardekooper, Leinhardt,
and Thebault (in preparation), the mass of the second largest remnant,
$M_{\rm slr}$, is fully constrained by knowledge of the mass/size of the
largest remnant and the power-law slope for the size distribution of
the smaller fragments. Let us consider a differential size
distribution
\begin{equation}\label{eqn:diff}
n(D)dD = CD^{-(\beta+1)} dD,
\end{equation}
where $n(D)$ is the number of objects with a radius between $D$ and $D + dD$, $-(\beta + 1)$ is the slope of the differential size distribution, and $C$ is the proportionality constant. Integrating equation~\ref{eqn:diff}, the number of bodies between $D_{\rm lr}$ and $D_{\rm slr}$ is 
\begin{equation}
N(D_{\rm lr},D_{\rm slr}) = -\frac{C}{\beta} \left( D_{\rm slr}^{-\beta} - D_{\rm lr}^{-\beta} \right).
\end{equation}
Therefore, the number of objects larger than the second largest remnant ($D_{\rm slr}$) is $N_{\rm slr} = N(D_{\rm slr},\infty)$. Assuming that $\beta > 0$, 
\begin{equation}\label{eqn:slr}
D_{\rm slr} =  \bigg[ \frac{N_{\rm slr}}{C}\beta  \bigg]^{\frac{-1}{\beta}}.
\end{equation}
For spherical bodies with bulk density $\rho$, the mass of material between $D_{\rm lr}$ and $D_{\rm slr}$ is
\begin{equation}
M(D_{\rm lr},D_{\rm slr}) = \frac{4}{3} \pi \rho C \frac{D_{\rm slr}^{3-\beta} - D_{\rm lr}^{3-\beta}}{3-\beta}.
\end{equation}
In order to enforce a negative slope of the remnants, $\beta$ must be less than 3. Mass is conserved in the impact; thus, the mass in the remnant tail must equal the total mass minus the mass in the largest remnant(s), $M(0,D_{\rm slr}) = M_{\rm tot} - N_{\rm lr}M_{\rm lr}$, where $N_{\rm lr}$ is the number of objects with mass equal to the largest remnant (here, we allow for multiple largest remnants). Substituting for $D_{\rm slr}$ from equation~\ref{eqn:slr}, $C$ is given by
\begin{equation}
\frac{C}{N_{\rm slr}\beta} = \left[ \frac{(3 - \beta)(M_{\rm tot} - N_{\rm lr} M_{\rm lr})}{(4/3) \pi \rho N_{\rm slr} \beta} \right ]^\frac{\beta}{3}.
\end{equation}
Substituting this expression for $C$ into equation \ref{eqn:slr} and assuming that all of the objects are spherical, the size and mass of the second largest remnant is expressed in terms of the total mass by
\begin{equation}\label{eqn:dd}
\frac{D_{\rm slr}}{D_{\rm tot}} = \left[ \frac{(3 -\beta) (1 - N_{\rm lr} \frac{M_{\rm lr}}{M})}{N_{\rm slr} \beta} \right]^\frac{1}{3},
\end{equation}
where $D_{\rm tot} = 2((3 M_{\rm tot})/(4 \pi \rho))^{1/3}$. In the
simulations presented here, the calculated diameter of the fragments
is not informative because most PKDGRAV particles merge with other
particles in gravitationally bound clumps; in these cases, the bulk
density is assumed for the size of the merged particle. The mass of
the remnants is accurate, however. Rewriting equation \ref{eqn:dd} in
terms of mass,
\begin{equation}\label{eqn:mslr}
\frac{M_{\rm slr}}{M_{\rm tot}} = \frac{(3 - \beta) (1 - N_{\rm lr} \frac{M_{\rm lr}}{M_{\rm tot}})}{N_{\rm slr} \beta},
\end{equation}
where $M_{\rm lr}$ is given by the universal law and the catastrophic
disruption criteria $Q_{RD}^{\prime*}$ (equation \ref{eqn:univlawangle}).

In the last column of Table \ref{tab:pkd}, the predicted mass of the
second largest remnant (equation \ref{eqn:mslr}) is compared to the
numerical simulations using the empirically fit $Q^{\prime *}_{RD}$,
$\beta = 2.85$, $N_{\rm lr} =1$, and $N_{\rm slr} = 2$. Since the
analytic method presented here assumes an infinite size distribution
in the fragment tail, we selected the value of $\beta$ to optimise the
fit to the value of $M_{\rm slr}$ in the simulations. This simple
method of predicting $M_{\rm slr}$ works well for all impact
conditions.\footnote{Because the analytic model for the fragment size
  distribution assumes an infinite range of sizes in the tail, $\beta$
  is constrained to be less than 3, which is slightly smaller than the
  slope of power laws that are best fit to the data (Table
  \ref{tab:pkd}).  The fragment size distribution may be modeled under
  different assumptions: e.g., choosing a minimum diameter in the
  integral of equation \ref{eqn:diff}, which would represent the
  smallest constituent particles or grain size. We have chosen not the
  impose any assumptions about material properties in the model
  presented here, but there may be situations were more is known about
  the colliding bodies and the model for determining $M_{\rm slr}$ and
  $\beta$ may be modified.} To illustrate the model, the predicted
size distribution (blue line and triangles) is compared to selected
numerical simulation results in Figure~\ref{fig:cumr}. In order to
predict the size distribution of fragments in the hit-and-run regime
impacts between comparable mass bodies ($\gamma=1$ and $b>0$), the
model needs to be modified slightly.  In this special case, we suggest
adopting $N_{\rm lr} = 2$ and $N_{\rm slr} = 4$ because the target and
projectile each have a nearly identical size distribution of
fragments.  In this example, the model is calculated for the same
impact conditions as the cyan data set with $\gamma = 1$, $b = 0.7$,
and $V_i = 80$ m s$^{-1}$ (see section \ref{sec:hitandrun} for more
detailed discussion of the hit-and-run regime).

The fragment size distributions calculated using our subsonic $N$-body
simulations are consistent with shock code calculations investigating
asteroid family formation via catastrophic impact events
\citep{Nesvorny:2006, Jutzi:2010}. All of the asteroid family-forming
simulations used a hybridized numerical technique, combining an SPH
code with PKDGRAV in order to model the propagation of the initial
shock wave and the subsequent gravitational reaccumulation of the
collision remnants. The asteroid family-forming collisions have
significantly different impact parameters compared to our simulations:
$V_i$ was orders of magnitude larger, $\gamma$ was an order of
magnitude smaller than our smallest $\gamma$, and targets were larger
(10's km in diameter). These differences notwithstanding, we find that
the range in the values of $\beta$ for the tail of the size
distribution is very similar to our $N$-body results (note that some
published values for $\beta$ include the largest remnant in the fit,
whereas we do not). Qualitatively, we also find a general trend in
curvature of the size distribution consistent with \citet{Durda:2007},
with slightly convex size distributions for super-catastrophic impact
events (section \ref{sec:super}).

\subsection{Fragment velocity distribution} \label{sec:fragvel}

\begin{figure}
 \includegraphics[scale=0.4]{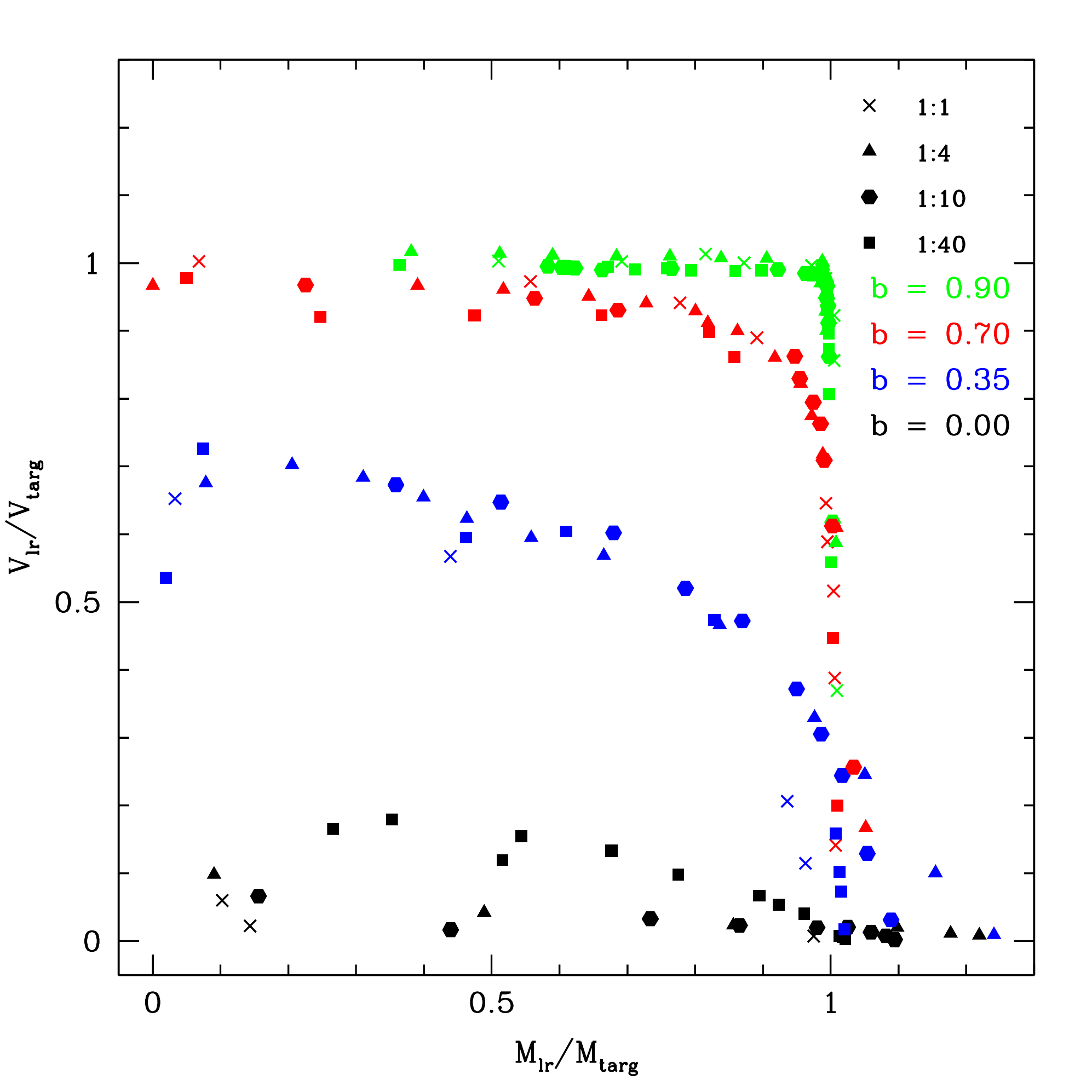}
 \caption{Velocity of largest remnant with respect to the initial center
   of mass target velocity versus the mass of largest remnant normalized
   by the mass of the target. Impact angle is indicated by color;
   mass ratio is indicated by symbol. \label{fig:vej}}
\end{figure}

\begin{figure*}
 \includegraphics[scale=0.8]{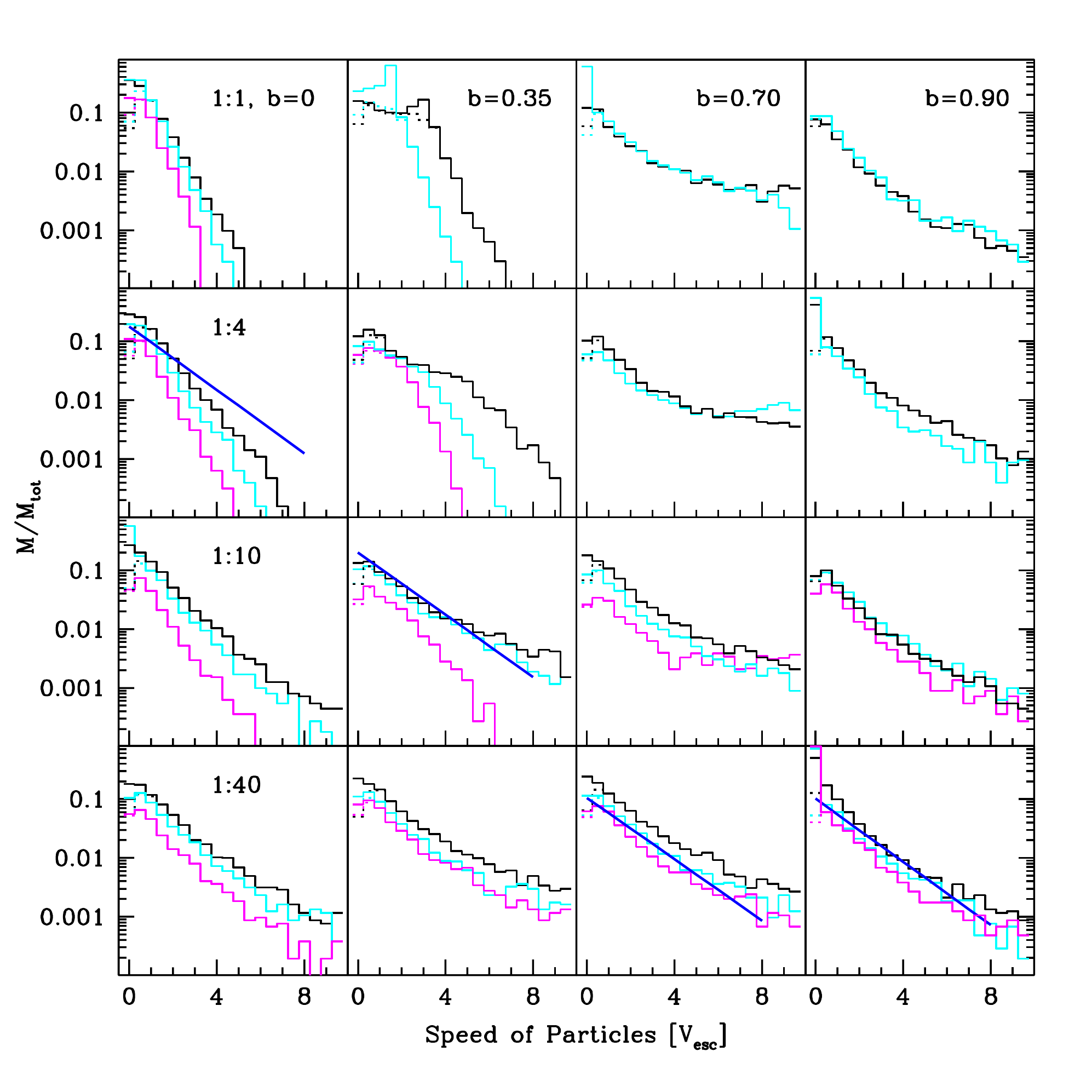}
 \caption{Fragment mass--velocity histograms for simulations in Figure
   \ref{fig:cumr} and Table \ref{tab:pkd}. The fragment velocities are
   relative to the largest remnant in units of the escape velocity
   from the combined mass of the target and projectile, $V_{\rm esc} =
   (2GM_{\rm tot}/R_{C1})^{1/2}$. The color coding is the same as in
   Figure \ref{fig:cumr}. The scaling law predictions are shown in
   blue. \label{fig:vwrtlr}}
\end{figure*}

Next, we consider the velocity of the collision remnants. The results
are easier to interpret by separating the velocities of the largest
remnant from the rest of the collision remnants. 

We first consider the speed of the largest remnant with respect to the
center of mass of the collision (Figure~\ref{fig:vej}).  For erosive
events ($M_{\rm lr}<M_{\rm targ}$), there is almost no change in the amplitude
of the target velocity for impacts with $b=0.9$. Even at $b=0.7$, the
velocity reduction is minimal for all fractions of mass lost. Because
$b=0.7$ is the center of the probability distribution of impact
angles, fully half of all erosive impacts have $<10\%$ change in
the target velocity amplitude. After head-on collisions ($b=0$), the
largest remnant moves with the center of mass velocity. Note there is
significant scatter in the data from $\gamma = 0.025$, which is due to
the fact that there was a small number of particles delivering the
impact energy to a localized region of the target; thus, the
organization of the surface features on both objects become
important. For disruptive impacts at $b=0.35$, there is partial
velocity reduction of the largest remnant. From these data, we cannot
define a unique function for the dependence of $V_{\rm lr}$ on $b$, and we
suggest that a quasi-linear relationship for $0<b<0.7$ is a reasonable
approximation.  We stress that the specific dependence of $V_{\rm lr}$ on
$b$ in the disruptive regime is likely to be sensitive to internal
structure and composition, so extrapolation of these results beyond
weak, constant density objects should be done with caution.

In complete merging events, of course, the post-impact velocity is
zero with respect to the center of mass. The $b=0.35$ data with
$M_{\rm lr}>M_{\rm targ}$ steadily approach the center of mass velocity with
more mass accreted.  The $b=0.7$ and 0.9 data points plotted near
$M_{\rm lr}/M_{\rm targ}=1$ are primarily hit-and-run events, which will be
discussed in \S \ref{sec:hitandrun}.

The smaller remnants of disruptive collisions have a more complex
behavior. Figure~\ref{fig:vwrtlr} presents mass histograms of fragments
versus velocity with respect to the largest remnant from the
simulations summarized in Table \ref{tab:pkd}. The slowest simulations
are not plotted for the $\gamma = 0.25$ and 1.0 grazing impacts because
there are only a small number of fragments. A significant number of
the fragments consist of 10 PKDGRAV particles or less; in
Figure~\ref{fig:vwrtlr}, mass associated 10 or less particles is shown
as dotted histograms.  The dotted histograms overlay the total mass
histograms for all but the lowest velocity bins; thus, within most of
the velocity bins, the simulations do not have the resolution to
robustly constrain the size-frequncy distribution of the mass in the
bin. The smallest (poorly resolved) fragments are found in all
velocity bins, while the largest fragments tend to move slowly with
respect to the largest remnant. For example, the second largest
remnant falls in one of the lowest velocity bins, but that bin is also
occupied by smaller fragments.

Hence, to describe the velocity field after a collision, we fit the
velocity-binned mass of the collision remnants. The binned mass versus
velocity is a fairly well defined exponential function for most of the
simulations. In general, the lowest velocity bin in
Figure~\ref{fig:vwrtlr} is of order $0.1M_{\rm tot}$. Using a
least-squares fit of the subset of simulations in Table \ref{tab:pkd},
we find the mass fraction in the lowest velocity bin is proportional
to the largest remnant mass:
\begin{equation}
A = -0.3 M_{\rm lr}/M_{\rm tot} + 0.3.
\end{equation}
To determine the slope, $S$, of the binned mass versus velocity
exponential function, we integrate the differential mass function,
\begin{eqnarray}
\log \left(\Delta v \frac{dm}{dv}\right) &=& (A - S v), \label{eqn:velscaling}\\
\Delta v\frac{dm}{dv} &=& 10^{A-Sv},\\
\frac{dm}{dv} &=& \frac{10^{A-Sv}}{\Delta v}\\
\frac{M_{\rm rem}}{M_{\rm tot}} &=& \int^\infty_0 \frac{10^{A-Sv}}{\Delta v} dv, \\
S &=&\frac{10^A}{\ln(10)\, \Delta v \,(M_{\rm rem}/M_{\rm tot})}, 
\end{eqnarray}
where $m = M/M_{\rm tot}$, $v = V/V_{\rm esc}$, $\Delta v$ is the bin width,
and the total mass in the histogram is the total mass in the remaining
remnants, $M_{\rm rem} = M_{\rm tot} - M_{\rm lr}$. 

The fragment velocity scaling law (equation \ref{eqn:velscaling}) is
shown in blue in Figure~\ref{fig:vwrtlr} for selected cases indicated by
$^{\dag *}$ in Table \ref{tab:pkd}. The velocity distributions of the
remnants agree qualitatively with those found in hypervelocity
simulations of asteroid family forming events, although previous
workers have not fit any function to the velocity distribution of the
fragments \citep[e.g.,][]{Michel:2002,Nesvorny:2006}.

\section{Other Collision Regimes} \label{sec:resultsother}

\subsection{Super-catastrophic regime}\label{sec:super}

\begin{figure}
 \includegraphics[scale=0.4]{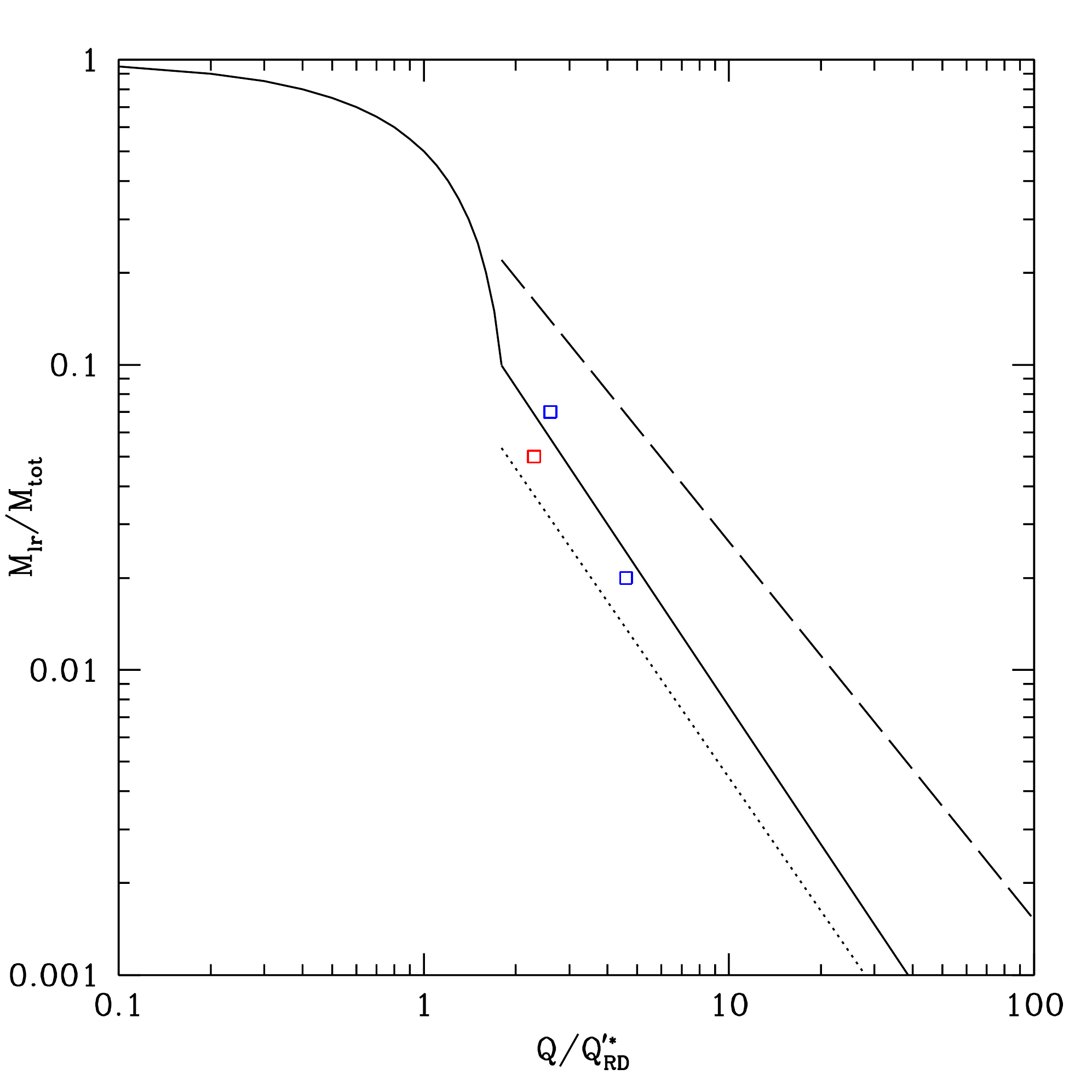}
 \caption{The mass of the largest remnant in the catastrophic and
   super-catastrophic disruption regimes. The solid line shows the
   combined universal law (equation~\ref{eqn:univlawangle}) and
   recommended power law relation for $M_{\rm lr}/M_{\rm tot}<0.1$
   (equation \ref{eqn:mlrsuper}). The symbols are new gravity regime
   simulations and the dotted and dashed lines represent the range of
   super-catastrophic disruption data in laboratory experiments in the
   strength regime. The shape and color of the symbols are the same
   as in Figure~\ref{fig:univ}. \label{fig:univsuper}}
\end{figure}

In both laboratory experiments in the strength regime
\citep[e.g.,][]{Kato:1995,Matsui:1982} and the few high resolution
disruption simulations in the gravity regime
\citep[e.g.,][]{Korycansky:2009}, the relationship between the mass of
the largest remnant and the specific impact energy $Q_R$ shows a
marked change in slope at around $M_{\rm lr}/M_{\rm tot} \sim 0.1$. We define
the super-catastrophic regime when $M_{\rm lr}/M_{\rm tot}<0.1$ (e.g., when
$Q_R/Q^{\prime *}_{RD} > 1.8$ by the universal law,
equation~\ref{eqn:univlawangle}).  In the super-catastrophic regime,
the mass of the largest remnant follows a power law with $Q_R$ rather
than the linear universal law.

The slope of the power law for the largest remnant mass vs.\ impact
energy shows some scatter in laboratory data, primarily in the range
of -1.2 to -1.5. In Figure~\ref{fig:univsuper}, our few simulations of
super-catastrophic collisions (symbols) are compared to the range of
outcomes from laboratory experiments (dotted and dashed lines). Based
on the simulations in the gravity regime and laboratory experiments in
the strength regime, we recommend a power law in the
super-catastrophic regime,
\begin{equation} \label{eqn:mlrsuper}
M_{\rm lr}/M_{\rm tot} = \frac{0.1}{1.8^{\eta}} (Q_R/Q^{'*}_{RD})^{\eta},
\end{equation}
where $\eta \sim -1.5$ and the coefficient is chosen for continuity
with the universal law (equation \ref{eqn:univlawangle}).  The slope of the
power law, about -1.5, is consistent with our gravity regime
simulations and a wide range of laboratory experiments summarized in
Figure 1 in \citet{Holsapple:2002}.

In Figure~\ref{fig:univsuper}, the solid line is the combined universal
law and the recommended super-catastrophic power law (equations
\ref{eqn:univlawangle} and \ref{eqn:mlrsuper}). The dotted line is our fit
to disruption data on solid ice from \citet{Kato:1992},
$M_{\rm lr}/M_{\rm tot} = 0.125 (Q_R/Q^{\prime *}_{RD})^{-1.45}$. The dashed line is
our fit to disruption data on basalt from \citet{Fujiwara:1977},
$M_{\rm lr}/M_{\rm tot} = 0.457 (Q_R/Q^{\prime *}_{RD})^{-1.24}$. Note that lab data
are available up to very high values of $Q_R/Q^{\prime *}_{RD} \sim 100$. The
lab data spanning very weak to very strong geologic materials can be
considered lower and upper bounds for the parameters in equation
\ref{eqn:mlrsuper}.

The general agreement between the gravity and strength regimes suggests
that gravitational reaccumulation of fragments has a negligible effect
in the super-catastrophic regime. In other words, the mass of the
largest fragment is primarily controlled by the shattering process. 

Based on the similarity of the size distribution of fragments in
laboratory experiments to the gravity regime data presented here
(Figure~\ref{fig:cumr}), we suggest that the dynamical properties of the
smaller fragments in super-catastrophic collisions are similar to the
disruption regime. Therefore, the size and velocity distributions
described in \S \ref{sec:fragsize} and \S \ref{sec:fragvel} can be
applied.

\subsection{Hit-and-run regime} \label{sec:hitandrun}

\begin{figure*}
\includegraphics[scale=0.7]{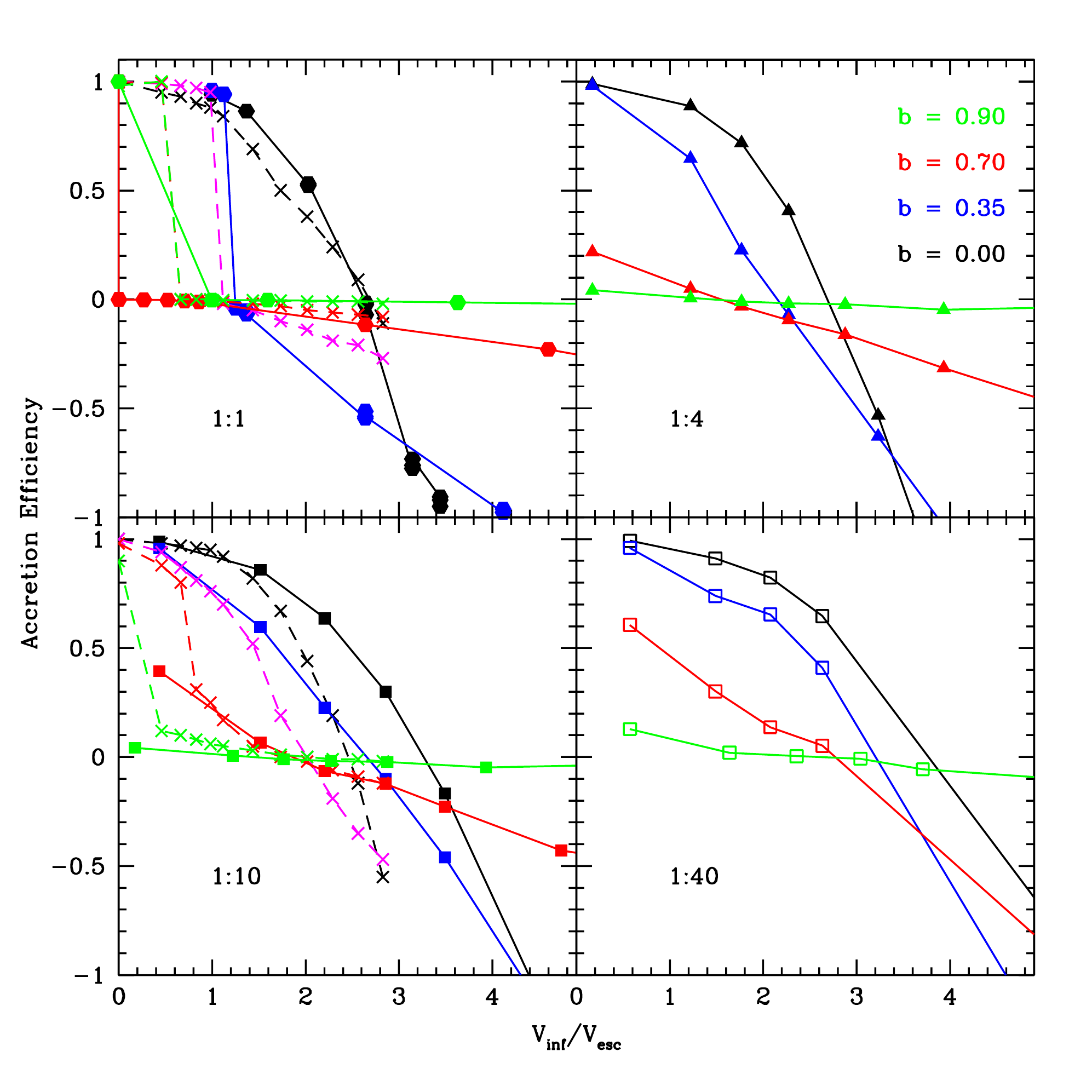}
\caption{Accretion efficiency (equation \ref{eqn:acceff}) versus
  velocity at infinity normalized by mutual escape velocity for
  different projectile-to-target mass ratios and impact
  parameters. Note that the impact velocity $V_{i}=\sqrt{V^{2}_{\rm
      inf}+V^{2}_{\rm esc}}$. Results from this work are connected by
  solid lines; previous results for supersonic impacts between
  protoplanets are connected by dashed lines
  \citep{Agnor:2004,Agnor:2004a} and symbols are an aid to
  differentiate simulation groups. Magenta lines are for
  $b=0.5$. \label{fig:acceff}}
\end{figure*}

Non-grazing impacts in the gravity regime transition from perfect
merging to the disruption regime with increasing impact velocity.
However, for impact angles greater than a critical value, an
intermediate outcome may occur: hit-and-run
\citep{Agnor:2004,Asphaug:2006,Marcus:2009,Marcus:2010,Asphaug:2010,Leinhardt:2010}. In
a hit-and-run collision, the projectile hits the target at an oblique
angle but separates again, leaving the target almost intact. Some
material from the topmost layers of the two bodies may be transferred
or dispersed. Depending on the exact impact conditions, the projectile
may escape largely intact or may sustain significant damage and
deformation \citep[e.g., Figure~7 in][]{Asphaug:2010}.
 
The hit-and-run regime is clearly identified by considering the
accretion efficiency of a collision, defined by \citet{Asphaug:2009}
as
\begin{equation} \label{eqn:acceff}
\xi = \frac{M_{\rm lr} - M_{\rm targ}}{M_{\rm p}}.
\end{equation}
In a perfect hit-and-run event ($M_{\rm lr} = M_{\rm targ}$), $\xi =
0$. For a perfect accretion event ($M_{\rm lr} = M_{\rm targ} + M_{\rm
  p}$), $\xi = 1$. An erosive event in which $M_{\rm lr} < M_{\rm
  targ}$ leads to $\xi < 0$. Note that the negative value of $\xi$
that corresponds to catastrophic disruption ($M_{\rm lr}=0.5M_{\rm
  tot}$) depends on the specific mass ratio of the two bodies
($\xi^*=0.5-0.5/\gamma$).
 
There is remarkably good agreement in the accretion efficiency and
transitions from merging to hit-and-run and from hit-and-run to
disruption between this work and previous simulations of higher
velocity impacts between large planetary bodies
\citep{Agnor:2004,Agnor:2004a,Marcus:2009,Marcus:2010}. Figure~\ref{fig:acceff}
shows the accretion efficiency from our simulations in solid colored
lines for four different projectile-to-target mass ratios and impact
parameters. Data for collisions between protoplanets at supersonic
velocities from \citet{Agnor:2004,Agnor:2004a} \citep[and plotted
in][]{Asphaug:2009} are shown in dashed lines for the common mass
ratios (1:1 and 1:10). Hit-and-run collisions are indicated by a
sudden drop from merging outcomes ($\xi=1$) to a nearly constant value
of $\xi \sim 0$ for a range of impact velocities. Note that the drop
in $\xi$ is sharpest for equal-mass bodies. For smaller mass ratios,
the transition is not as sharp, and partial accretion of the
projectile occurs at energies just above perfect merging ($\xi$ just
above 0).

Outcomes that are defined by the disruption regime have steep negative
sloped accretion efficiencies. The disruption regime equations apply
for partial accretion ($0<\xi<1$) and for erosion of the target ($\xi
< 0$).  Note that for high impact parameters (e.g., $b=0.7$), there
exists an intermediate regime where the accretion efficiency has a
very shallow negative slope and values of $\xi$ just below 0. These
impact events, termed erosive hit-and-run, lead to some erosion of the
target and more severe deformation of the projectile. The erosive
hit-and-run regime is eventually followed by a disruptive style
erosive regime at sufficiently high impact velocities. The
post-hit-and-run disruptive regime may be identified by finding the
impact energy that leads to a linear relationship that satisfies the
universal law. The required impact velocity increases substantially
with increasing impact parameter; see \S \ref{sec:transitions} and
\citet{Marcus:2010} for an example disruption regime after a hit-and
run regime ($\gamma=0.5$ and $b=0.5$).

In an ideal hit-and-run event, the target is almost unaffected by the
collision, and the velocity of the largest remnant (the target) is
about equal to the initial speed of the target with respect to the
center of mass. More commonly, there is a small velocity change in
both bodies which increases the probability of merging in subsequent
encounters \citep{Kokubo:2010}.  \citet{Agnor:2004} referred to this
collision outcome as inelastic bouncing.  In our hit-and-run
simulations with $b=0.9$ (green cluster of points in
Figure~\ref{fig:vej} at $M_{\rm lr}=M_{\rm targ}$), the targets
typically lose about 10\% of their pre-impact velocity. For $b=0.7$
(red points), there is more significant slowing of the target. Our
data does not provide a robust description of the dependence of the
post-impact velocity on the impact parameter and impact velocity.

The projectile may be significantly deformed and disrupted
during a hit-and-run event. The level of disruption of the projectile
may be approximated by considering the reverse impact scenario: a
fraction of the larger body impacts the smaller body. In this case, we
estimate the interacting mass from the larger body with a simple
geometric approximation. For the example geometry given in
Figure~\ref{fig:cartoon}, the cross-sectional area of the circular
projectile interacting with the target is calculated. The apothem is
given by $l-r$, and the central angle is
$\phi=2\cos^{-1}((l-r)/r)$. Then, the projectile collision cross
section is
\begin{equation}
A_{\rm interact}=r^2 (\pi - (\phi-\sin\phi)/2)
\end{equation}
The interacting length through the target is approximated by the chord at $l/2$,
\begin{equation}
L_{\rm interact}=2\sqrt{R^2-(R-l/2)^2}.
\end{equation}
And the interacting mass from the target is of order
\begin{equation}
M_{\rm interact}=A_{\rm interact}L_{\rm interact}.
\end{equation}
Note that the interacting mass depends on the impact angle (through
$l$).

To estimate the disruption of the projectile, we consider an idealized
hit-and-run scenario between gravity-dominated bodies: the
fraction of the target that does not intersect the projectile is
sheared off with negligible change in momentum and gravitationally
escapes the interacting mass. Hence, we ignore the escaping target mass
and consider only the impact between $M_{\rm interact}$ and the projectile
mass, $M_{\rm p}$.

The reverse impact is thus defined by $M^{\dag}_{\rm p}=M_{\rm interact}$ and
$M^{\dag}_{\rm targ}=M_{\rm p}$, and the $^{\dag}$ denotes the reverse impact
variables. For each impact angle, calculate $R^\dag_{C1}$ for
$M^{\dag}_{\rm tot}=M^{\dag}_{\rm p}+M^\dag_{\rm targ}$, $Q^{\dag *}_{RD, \gamma=1}$
from the principal disruption curve (equation \ref{eqn:principalq}),
and $V^{\dag *}_{\gamma=1}$ from equation \ref{eqn:vequal}. The
reverse variables are
\begin{eqnarray}
  \mu^{\dag}  & = & M^{\dag}_{\rm p} M^{\dag}_{\rm targ} / (M^{\dag}_{\rm p}+M^{\dag}_{\rm targ}),\\
 \gamma^{\dag} & = & M^{\dag}_{\rm p}/M^{\dag}_{\rm targ}.
\end{eqnarray}
The mass ratio correction from the principal disruption curve is
\begin{eqnarray}
  V^{\dag *}  &=& \left [ \frac{1}{4} \frac{(\gamma^{\dag}+1)^2}{\gamma^{\dag}} \right ]^{1/(3 \bar  \mu)}  V^{\dag *}_{\gamma=1}, \\
  Q^{\dag *}_{RD} &=& Q^{\dag *}_{RD,\gamma=1} \left ( \frac{1}{4} \frac{(\gamma^{\dag}+1)^2}{\gamma^{\dag}} \right )^{2/(3\bar \mu) - 1}.
\end{eqnarray}  
Once the reverse impact disruption criteria is calculated, we use the
universal law for the mass of the largest remnant to determine the
collision regime for the projectile.  If the projectile disrupts, then
the size distribution of the projectile fragments may be estimated to
first order from the disruption regime scaling laws.

\section{Transitions between Collision Regimes} \label{sec:transitions}

\subsection{Empirical transitions between accretion, erosion, and hit-and-run} \label{sec:transitionssim}

\begin{figure*}
\includegraphics[scale=0.7]{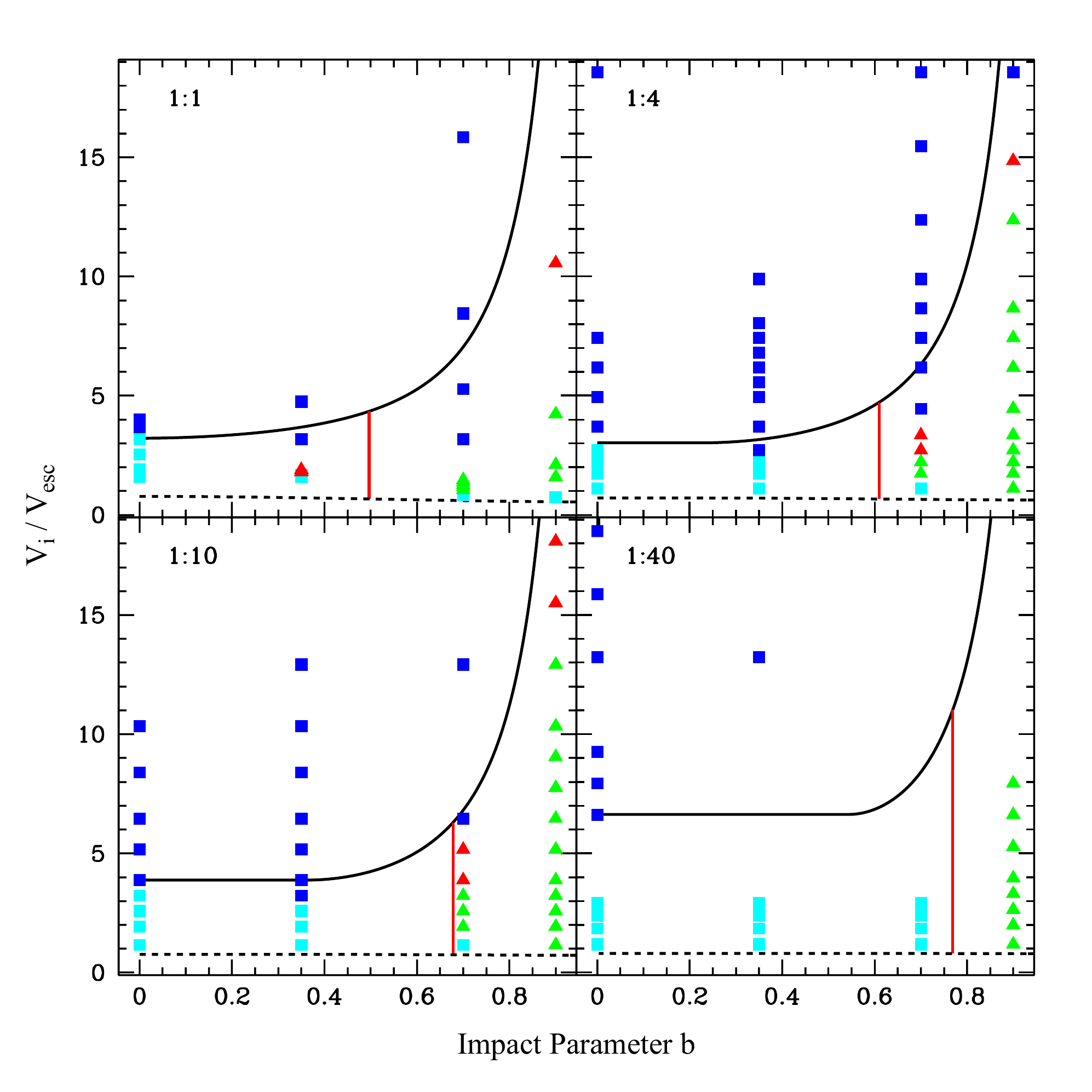}
\caption{Map of the major collision regimes as a function of mass
  ratio, impact parameter, and impact velocity normalized by the
  escape velocity from the combined mass with radius $R_{C1}$. Cyan
  squares --- a full or partial accretion event, $M_{\rm lr} > M_{\rm
    targ}$; blue squares --- target is eroded, $M_{\rm lr} < M_{\rm
    targ}$; green triangles --- ideal hit-and-run event, $M_{\rm lr} =
  M_{\rm targ}$; red triangles --- erosive hit-and-run event, $M_{\rm
    lr}$ slightly less than $M_{\rm targ}$.  Red vertical line
  corresponds to $b_{\rm crit}$ for the given mass ratio (equation
  \ref{eqn:bcrit}).  Black curve is onset of erosion predicted from
  the catastrophic disruption model (\S \ref{sec:univ})
  with $c^*=4.3$ and $\bar \mu=0.35$; dashed black curve is predicted
  transition from perfect merging to hit-and-run (equation
  \ref{eqn:vescmod}). \label{fig:transition}}
\end{figure*}

We have classified the collision outcome regime for all of our new
simulations. The outcome is sensitive to the mass ratio of the two
bodies, the impact parameter, and the impact velocity.  Four regimes
are mapped in Figure~\ref{fig:transition}: 
\begin{enumerate}
\item{Accretion of some or all of the projectile onto the target
    ($M_{\rm lr}>M_{\rm targ}$ and $\xi > 1$, light blue squares),}
\item{Partial erosion of the target ($M_{\rm lr}<M_{\rm targ}$ and
    $\xi < 1$, dark blue squares),}
\item{Pure hit-and-run ($M_{\rm lr}=M_{\rm targ}$ and $\xi = 0$, green
    triangles), and}
\item{Erosive hit-and-run ($M_{\rm lr}$ slightly less
than $M_{\rm targ}$ and $\xi$ slightly less than 0, red triangles).}
\end{enumerate}
Note that the 1:40 mass ratio simulations reach impact velocities that
exceed the physics included in PKDGRAV; impact velocities greater than
about 1~km~s$^{-1}$ should use a shock physics code. Hence, the
transition to the erosive regime at high impact parameters could not
be derived directly.

For impacts at small impact parameters (more head-on), the collision
outcomes transition from accretion to erosion with increasing impact
velocity. For more oblique impacts, the collision outcomes transition
from merging to hit-and-run to erosion with increasing impact
velocity. As suggested by \citet{Asphaug:2010}, $b_{\rm crit}$ (red
vertical line in Figure~\ref{fig:transition}) is indeed a good indicator
of the minimum impact parameter necessary to enter the hit-and-run
regime.  However, for $\gamma =1$, we find a small region of erosive
hit-and-run events when $b < b_{\rm crit}$.  The use of $b_{\rm crit}$
to define grazing and non-grazing impacts makes the very simplifying
assumption that the velocity vector of the center of mass of the
projectile remains constant during the event. In reality, the
projectile center of mass will be deflected to some extent during the
encounter, and the true interactive mass will be larger than assumed
here. The deflection is greatest for more equal-mass bodies, and a
narrow region of erosive hit-and-run events is observed for $b = 0.35$
and $\gamma=1$. Note that the transition between erosion and
hit-and-run occurs near $b_{\rm crit}$ for all size bodies studied to
date, from 1~km rubble pile planetesimals \citep{Leinhardt:2000} to
super-earths \citep{Marcus:2009}.

For grazing collisions, the hit-and-run regime is bounded by perfect
merging at low impact velocities the onset of disruption at high
velocities. The projectile merges with the target when the impact
velocity is less than the mutual escape velocity (in other words, the
velocity at infinity $V_{\rm inf}$ is zero). Since only a fraction of
the projectile may interact in oblique impacts, the appropriate mutual
escape velocity for perfect merging is slightly less than the mutual
escape velocity from the total mass. Then the appropriate measure for
merging is
\begin{equation} 
  V^{\prime}_{\rm esc} = \sqrt{(2 G M^{\prime}/R^{\prime})}, \label{eqn:vescmod} 
\end{equation}
where $M^{\prime} = M_{\rm targ} + m_{\rm interact}$ and $R^{\prime} =
((3 M^{\prime})/(4 \pi \rho))^{1/3}$, assuming that the projectile and
target have the same bulk density $\rho$. The boundary between merging
and hit-and-run is well matched by equation \ref{eqn:vescmod} in
Figure \ref{fig:transition} (dashed black line).

Of course, the concept of an interacting mass is a simplistic limit
because it assumes that the part of the projectile that impacts the
target can separate from the rest of the projectile without loss of
momentum.  In Figure~\ref{fig:transition}, the only set of simulations
that did not show a sharp transition from merging to hit-and-run is
$\gamma=0.1$ and $b=0.7$. In this case, the impact parameter is very
close to $b_{\rm crit}=0.66$, and the outcomes include partial
accretion of the projectile, erosive hit-and-run, and fully erosive
collisions with increasing impact velocity.

Loss of momentum by the projectile in grazing collisions does lead to
merging when $V_{\rm inf}>0$; in Figure~\ref{fig:acceff}, note the
nearly complete merging in the $\gamma=1$ simulations for small values
of $V_{\rm inf}$ with $b=0.5$ and 0.9 \citep[data
from][]{Agnor:2004}. For $V_{\rm inf}$ slightly above zero, merging
occurs in graze-and-merge events \citep[e.g.,][]{Leinhardt:2010}. In
such cases, the two bodies hit, separate as nearly intact bodies with
decreased velocity, and then merge upon a second collision. The impact
velocity range for graze-and-merge outcomes is quite narrow; previous
studies have demonstrated the small velocity increase needed to
transition from perfect merging to graze-and-merge to hit-and-run
\citep[e.g.,][]{Canup:2004,Leinhardt:2010}.  Concurrent with this
work, the graze-and-merge regime has been explored in more detail
using hydrodynamic SPH simulations by \citet{Genda:2011b}.

Grazing collisions transition out of hit-and-run to erosion of the
target when the impact velocities reach the disruption regime. The
transition to the disruption regime is a strong function of the impact
parameter because of the rapidly shrinking projectile interaction mass
and the dependence of the disruption criteria on the mass ratio and
impact velocity.

Our general model for the catastrophic disruption criteria combined
with the universal law for the mass of the largest remnant is used to
derive the impact velocity needed to begin eroding the target mass
($M_{\rm lr}=M_{\rm targ}$, black line in Figure
\ref{fig:transition}). Our new simulation data are best fit
with a value of $c^*=4.3$ and $\bar \mu=0.35$. Our model for
the disruption regime provides an excellent estimate for the
transition to erosion of the target for the wide range of impact
parameters considered here. In particular, the analytic model captures
the sharp increase in the upper bound to the hit-and-run regime
between $b=0.7$ ($45^{\circ}$) and 0.9 ($64^{\circ}$).

\subsection{Predicted transitions between accretion, erosion, and hit-and-run} \label{sec:transitionspredicted}

\begin{figure*}
\includegraphics[scale=0.33]{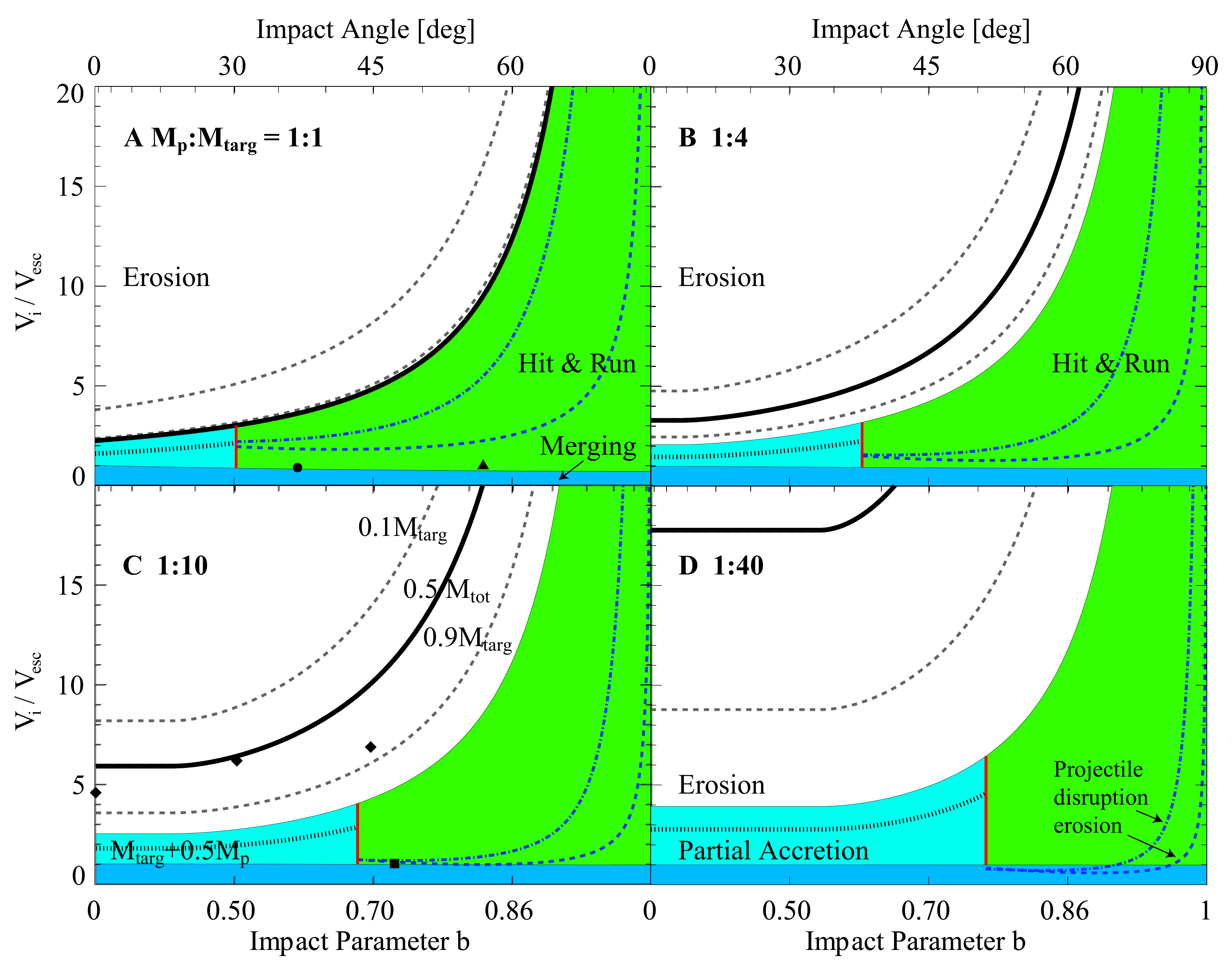}
\caption{Predicted collision outcome maps using the analytic model for
  strengthless planets ($c^*=1.9$ and $\bar \mu=0.36$) for selected
  projectile-to-target mass ratios. Impact velocity is normalized by
  the mutual surface escape velocity assuming a bulk density of
  3000~kg~m$^{-3}$; impact parameter is spaced according to equal
  probability. Colored regions denote perfect merging (dark blue),
  partial accretion (light blue), net erosion to the target (white)
  and hit-and-run (green). Vertical red line denotes the onset of
  hit-and-run events at $b_{\rm crit}$. Thick black curve -- critical
  disruption velocity for half the total mass remaining; grey dashed
  curves -- 10\% and 90\% of target mass in largest remnant; dotted
  curve -- 50\% of projectile accreted; dot-dashed blue curve --
  catastrophic disruption of the projectile; dashed blue curve --
  erosion of the projectile.  Example proposed giant impact events:
  $\bullet$ -- Haumea system \citep{Leinhardt:2010}; $\blacktriangle$
  -- Pluto-Charon \citep{Canup:2005}; $\blacklozenge$ -- Mercury
  \citep{Benz:2007}; $\blacksquare$ -- Earth-Moon
  \citep{Canup:2004}. \label{fig:planetmap}}
\end{figure*}

Using our analytic model, we derive example collision outcome maps for
collisions between protoplanets. We fit values of $\bar \mu=0.36$ and
$c^*=1.9$ to the data from collisions between planet-sized bodies
using SPH codes (Figure \ref{fig:qsrd}B). Collision maps, which are
color-coded for outcome regime, are shown in
Figure~\ref{fig:planetmap} for four mass ratios.

The details of the forward calculation of the collision regimes are
given in the Appendix\footnote{A code to generate collision outcome
  maps and to calculate specific impact scenarios is available from
  the authors.}.  In Figure~\ref{fig:planetmap}, the impact parameter
axis is scaled by the probability of an impact at that angle. The
probability of an impact within an interval $(\theta, \theta + {\rm
  d}\theta)$ is proportional to $\sin(\theta)\cos(\theta) {\rm
  d}\theta$ \cite{Shoemaker:1962}. The corresponding impact angle is
shown on the top axis with $5^{\circ}$ tick intervals. The model
assumes an abrupt transition between grazing and non-grazing impacts,
which is certainly artificial. Near the critical impact parameter,
collision outcomes will have elements from both the disruption and
hit-and-run regimes.

Contours of impact velocities that correspond to a constant mass of
the largest remnant are calculated using the general model for
catastrophic disruption and the universal law or power law for the
mass of the largest remnant (equations \ref{eqn:univlawangle} or
\ref{eqn:mlrsuper}). In Figure~\ref{fig:planetmap}, the thick black
curve corresponds to the critical velocity for catastrophic
disruption, where the largest remnant contains half the total
mass. Note that this curve corresponds to the target erosion boundary
for 1:1 scenarios (the transition from partial accretion (light blue)
or hit-and-run (green) to erosion (white) regions). The grey dashed
curves correspond to the impact velocity needed to disperse 10\% and
90\% of the target mass.

Between perfect merging and erosion of the target, there is a region
of partial accretion of the projectile. For non-grazing impacts, the
dotted curve corresponds to accretion of 50\% of the projectile
mass. Grazing impacts transition rapidly between perfect merging and
hit-and-run with increasing impact velocity.

Most hit-and-run collisions with $M_{\rm p} \le 0.1M_{\rm targ}$ result in
significant disruption of the projectile. In the collision outcome
maps, the onset of projectile erosion in a hit-and-run event is given
by
\begin{equation}
 V^{\dag}_{i,\rm lr=M^{\dag}_{\rm targ}} = \sqrt{ 2 Q_{R, \rm lr=M^{\dag}_{\rm targ}} M^{\dag}_{\rm tot} / \mu^{\dag} }.
\end{equation}
Note that for impact parameters near $b_{\rm crit}$, $M^{\dag}_{\rm p} \sim
M^{\dag}_{\rm targ}$ for projectile-to-target mass ratios less than about
0.1. Thus the velocity contours of constant remnant mass intersect for
catastrophic disruption ($M^{\dag}_{\rm lr}=0.5M^{\dag}_{\rm tot}$) and onset
of projectile erosion ($M^{\dag}_{\rm lr}=M^{\dag}_{\rm targ}$). Futhermore,
there is a minima in the projectile erosion curve at an optimal
fraction of total interacting mass from the target (in other words,
the reverse projectile-to-target mass ratio is varying with impact
parameter). The two velocity contours diverge at higher impact
parameter as $M^{\dag}_{\rm p}$ becomes much less than $M^{\dag}_{\rm targ}$.

Collision maps for planetesimals are presented in \S
\ref{sec:strength}, and the implications of the diversity of collision
outcomes for planet formation are discussed in \S
\ref{sec:implications}.


\section{Discussion} \label{sec:discussion}

\subsection{Scaling of collision outcomes in the gravity regime}

For all gravity-regime bodies studied to date, collision outcomes are
strikingly similar for a tremendous range of target composition and
size.  Furthermore, the transitions between the major collision
regimes (merging, hit-and-run, disruption, and super-catastrophic
disruption) occurs under similarly scaled conditions.  The types of
bodies studied, ranging from km to several 1000's km in size, included
rubble-pile and porous planetesimals \citep[this
work,][]{Stewart:2009,Benz:2000,Jutzi:2010,Korycansky:2009}, pure rock
or pure ice planetesimals with strength
\citep{Benz:1999,Leinhardt:2009,Jutzi:2010}, strengthless
differentiated rock and iron planets
\citep{Benz:1988,Benz:2007,Agnor:2004,Marcus:2009,Genda:2011b},
strengthless differentiated water and rock planets
\citep{Marcus:2010}, and strengthless pure rock planets
\citep{Marcus:2009}. The studies focused on a variety of stages during
planet formation, from accretion of planetesimals to destruction of
planets; thus, the impact velocities ranged from $\sim 1$ m~s$^{-1}$
to over 100 km~s$^{-1}$. The computational methods included three
different shock physics codes and two $N$-body codes. Our analysis of
the results from these studies suggest that the same scaling laws may
be applied over an incredibly broad range of impact scenarios during
planet formation.

As stressed by \citet{Asphaug:2010}, similarity of outcome is not the
same as true scale invariance. He notes that scale invariance applies
only for idealized incompressible, self-gravitating inviscid fluid
planets. In reality, many aspects of collision outcomes will not scale
simply with size: e.g., the mass of collision-produced melt depends on
the specifics of impact velocity, target composition, and the internal
temperature and pressure history. Here, we investigated the similarity
of the dynamics of collision outcomes for a variety of non-ideal
gravity-regime bodies, from icy planetesimals to differentiated
super-earths. Specifically, we developed scaling laws to define the
mass and velocity distribution of bodies after any gravity-regime
collision.

Why do the dynamics of collision outcomes appear to scale similarly
with size in the gravity regime? At impact velocities just above the
escape velocity, momentum dominates the outcome at all scales. Hence,
the transition from merging to hit-and-run depends primarily on the
geometric cross section of the collision for all size bodies. As
impact velocities increase, the energy required for disruption is
dominated by the gravitational dispersal of fragments rather than the
energy required to shatter an intact body into small pieces
\citep{Melosh:1997}. As a result, erosive outcomes require that the
velocity of the fragments exceed a critical value that relies
primarily upon the gravitational potential of the total colliding
mass.

For small bodies, the critical fragment velocity may be reached with
impact velocities that impart negligible irreversible work on the
materials (Figure~\ref{fig:qsrd}). For larger bodies, the critical
velocity requires sufficiently high impact velocities that strong
shock waves are formed. The shock wave permanently deforms the
materials and, in the process, reduces the total energy available for
the final velocity distribution of fragments. The energy of
deformation is often referred to as ``waste heat''; for a fixed impact
energy, a larger fraction of waste heat is generated with increasing
impact velocity (primarily due to the onset of shock-induced melting
and vaporization at high shock pressures). As a result, the
catastrophic disruption criteria increases with increasing impact
velocity (equation~\ref{eqn:qstarred}).

Based on currently available data we argue that in the disruption regime the dynamics of the outcome is
similar over the entire gravity regime when scaled by the catastrophic
disruption criteria. The post-collision size distribution is similar,
as it is controlled by the largest remnant and the gravitationally
accreted clumps from the shattered parent bodies. The general
catastrophic disruption law accounts for both the increasing
gravitational potential with total mass of the colliding bodies and
the increase in waste heat at higher impact velocities
\citep{Housen:1990}.

The development of equation~\ref{eqn:qstarred} relied upon the concept
of a coupling parameter, $\Lambda \propto R_{\rm p}V_i^{\bar \mu}$, a
point source approximation of the coupling of the projectile's energy
and momentum into the target \citep[c.f.][]{Holsapple:1987}. The
velocity exponent $\bar \mu$ is bounded by pure momentum coupling
($\bar \mu=1/3$) and pure energy coupling ($\bar \mu=2/3$). In the
gravity regime, the coupling parameter distills the physical response
of the geologic material into the variable $\bar \mu$. Some
constraints on $\bar \mu$ are available from laboratory cratering
experiments: e.g., $\bar \mu=0.4$ for sand and $\bar \mu=0.55$ for
water \citep{Holsapple:1987}. Here, we fit the coupling parameter to
the numerical simulation results for disruption of a wide variety of
materials. The derived best fit range of $0.33\le \bar \mu \le 0.37$
is close to pure momentum scaling.

Why does the concept of a point source approximation apply to
collisions between comparably sized planetary bodies? The point source
approximation was developed for impact cratering by a finite size
projectile onto a half space target. \citet{Holsapple:1987} show that
the concept of a point source is equivalent to a variety of models
that describe a similar material velocity field far from the impact
point. In the case of catastrophic disruption, the late-time far-field
criteria is a fragment size-velocity distribution where half the mass
is escaping the gravitational potential of the largest remnant.  The
principal dynamical factors governing the collision are incorporated
into the $Q^{\prime *}_{RD}$ formulation: relative velocity, mass
ratio, impact parameter, and bulk density. The similar outcomes of
collisions with similar $Q_R/Q^{\prime *}_{RD}$ indicate that the
remaining details of how the energy and momentum is distributed into
the target and projectile during the initial stage of the collision are
negligible in determining the late time dynamics following a
catastrophic disruption event.

In summary, the primary factors that bound the different collision
outcomes regimes scale similarly with size in the gravity regime:
momentum, geometric cross section, and normalized critical impact
energy ($Q^{\prime *}_{RD}$). Other factors that lead to second order
perturbations to the dynamics of the collision outcomes are discussed
in the next section.

\subsection{Scaling laws limits of applicability}
Planet formation involves a vast range of bodies with distinct
physical characteristics, including dust aggregates, rubble-pile
planetesimals, differentiated molten and solid protoplanets, solid
planets with extended atmospheres, and gas-dominated planets. The
constituent materials (iron-alloys, silicates, ices, and gases) span
orders of magnitude in density and material strength. The complex and
time-varying physical properties of planetary bodies significantly
limits the application of any single equation to all bodies over the
course of planet formation. And yet, judicious simplification is
necessary for planet formation simulations to be both physically
robust and computationally tractable.

We have focused on developing scaling laws that describe the dynamical
outcome of collisions between any two gravity-dominated bodies.  The
dynamical outcome from collisions seems to be rather insensitive to
the internal composition, when the results are scaled by the
appropriate value for $Q^*_{RD}$.  However, the types of bodies
studied to date do not contain any gas mass fraction \citep[see for example,][]{Kobayashi:2011}, and so the
scaling laws may need modification for a planet with a significantly
different internal structure than the differentiated and homogeneous
planets included in this study.  One area that warrants further
investigation is the sensitivity of the velocity of the largest
remnant to the internal structure/composition (Figure~\ref{fig:vej}).

The role of tidal effects during collisions or in close encounters may
be important factors during the fragmentation of planetary bodies
\citep{Asphaug:2006, Asphaug:2010}. In this work, all bodies are
assumed to be approximately spherical at the time of impact. Tidal
affects will change the interacting mass and contribute to the
fragmentation process in ways that lead to different size and
velocity distributions than found here. 

Similarly, the role of pre-impact spin during collisions has received
modest attention \citep{Leinhardt:2000,Canup:2008}. While the net spin
of a growing body may essentially average near zero during the rapid
growth phase where collisions are frequent, the effect of pre-impact
spin and the collision angular momentum may be very important in the
final giant impact phase of planet formation \citep{Agnor:1999}.
While, we did not consider any pre-impact spin in this study, a few
simulations with high collision angular momentum are notable. In 1:1
collisions with $0<b<b_{\rm crit}$ \citep[Table \ref{tab:pkd} in this
work and Table 1 in ][]{Leinhardt:2000}, the catastrophic disruption
criteria is less than the value at $b=0$ (e.g., closed and open stars
in Figure~\ref{fig:qsrd}). We interpret the lower disruption criteria
with pre-impact spin to arise from the significant collision angular
momentum. As a result, the gravitational potential is reduced and
dispersal requires slightly less energy. We suggest that future work
investigate the possibility of using the reduction in the
gravitational potential due to pre-impact and collision angular
momentum to account for the first order affects of spin. Specifically,
the spin-modified catastrophic disruption criteria may remain at a
constant offset ($c^*$) from the spin-modified gravitational
potential.

\subsubsection{Strength and porosity in the gravity regime} \label{sec:strength}

\begin{figure}
\includegraphics[scale=0.4]{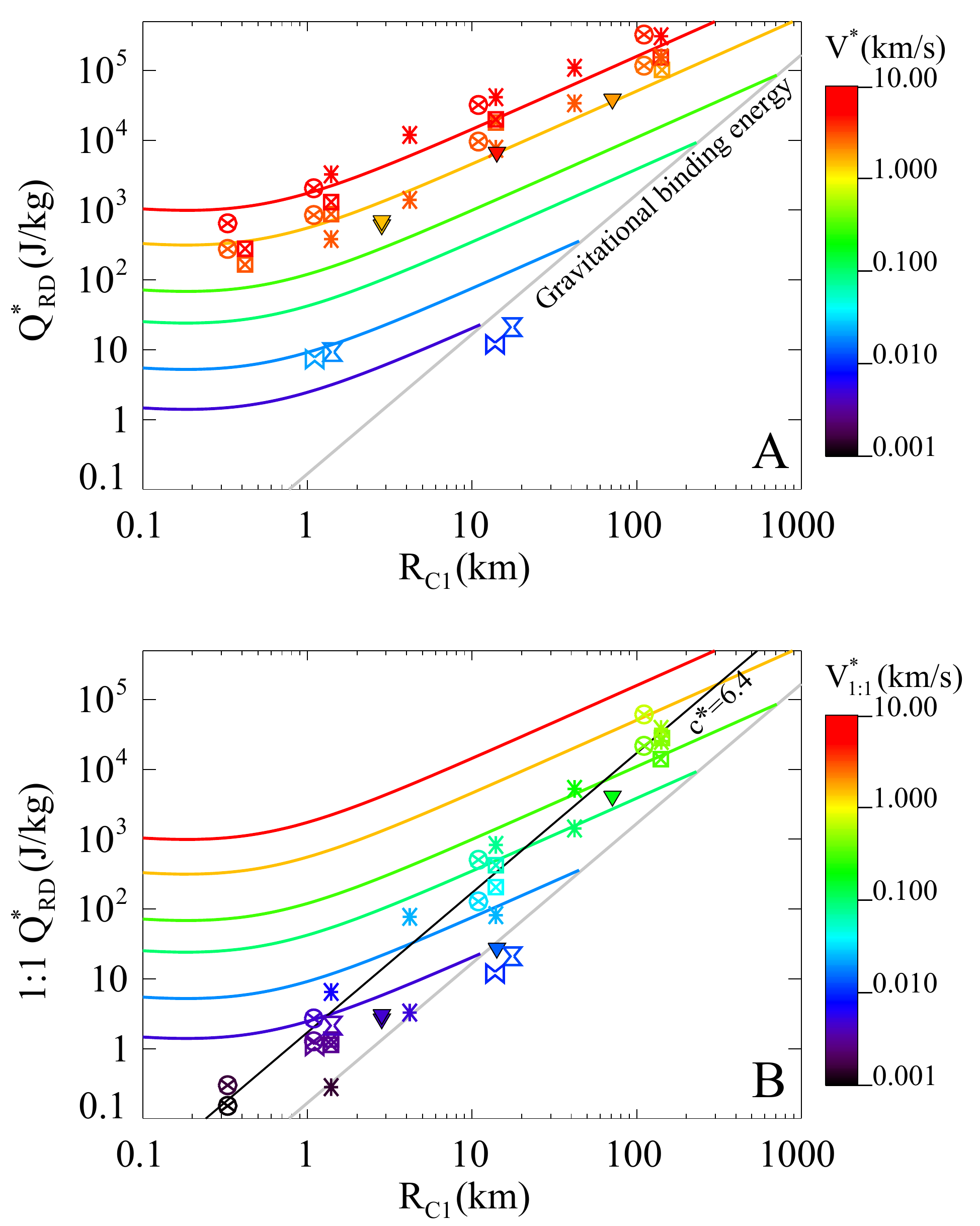}
\caption{Catastrophic disruption simulation results for strong rock
  targets (porous and nonporous). Same notation as in
  Figure~\ref{fig:qsrd} and Table \ref{tab:datasources}.
  A. Simulation data corrected to an equivalent head-on
  impact. B. Simulation data converted to an equivalent equal-mass
  disruption criteria.  The results for critical velocities from
  m~s$^{-1}$ to 5 km~s$^{-1}$ demonstrate that energy scaling is
  incorrect. Best fit $Q^*_{RD}$ curves with $\bar{\mu}=0.35$ and
  $c^*=6.4$ for $V^*=.005$, .02, .1, .3, 1.5, and 5
  km~s$^{-1}$. \label{fig:strongqsrd}}
\end{figure}

The study of catastrophic disruption of strong rock targets has been
motivated by collisional evolution studies of the asteroid and Kuiper
belts. The strength models were tested by fitting laboratory
quasi-static strength measurements and fragment size distributions
from head-on disruption experiments. Particular attention was paid to
the development of the model for tensile fracture \citep{Benz:1994},
as the tensile strength dominates the catastrophic disruption criteria
for head-on impacts in the strength regime.

Results from several numerical simulations of catastrophic disruption
of strong rock targets in the gravity regime are shown in
Figure~\ref{fig:strongqsrd}. The head-on basalt disruption data at
impact velocities of 3 and 5 km~s$^{-1}$ ($\ast$) are shown from the
canonical study by \citet{Benz:1999} using the SPH code with the
detailed tensile strength model. Using the same code,
\citet{Benz:2000} studied the disruption of strong nonporous basalt
(hourglass) and a macroporous target, composed of overlapping
clusters of SPH particles representing strong interconnected boulders
($\bowtie$), at very low impact velocities (5-40 m~s$^{-1}$) and
$b=0.7$.  In Figure~\ref{fig:strongqsrd}, the 10-km target data, which
fall below the specific gravitational binding energy, are derived from
the equal-mass collisions presented in Benz's Figure~5 and will be
discussed below. The \citet{Benz:2000} 1-km data are less certain
using our catastrophic disruption variables because both impact
velocity and mass ratio were varied and the specific values were not
reported. Nevertheless, the significant offset in the disruption
criteria is irrefutable evidence that pure energy scaling does not
apply. In fact, the total dispersion in the specific impact energy is
slightly larger than can be accommodated by the momentum scaling limit
of $\bar \mu=1/3$, which is likely a result of differences in the
details of the strength models.

More recent simulations ($\otimes, \boxtimes$) by \citet{Jutzi:2010}
with critical velocities of 3 and 5~km~s$^{-1}$ fall in-between the
data from \citet{Benz:1999}. Their work uses the same SPH code with an
updated strength model that includes the extra dissipation of
microporosity. Simulations using the CTH shock physics code with
different shear and tensile strength models yield similar results as
found for nonporous basalt targets using the SPH code
\citep[$\blacktriangledown$,][]{Leinhardt:2009}.

In Figure~\ref{fig:strongqsrd}, the strong target data, with mass
ratios from 1:1 to almost 1:45,000 and impact velocities from 0.001 to
5 km~s$^{-1}$, are best fit by $\bar \mu=0.35$ and $c^*=6.4$. The
equivalent equal-mass disruption data have $c^*$ values from 1 to
20. For comparison, the best fit to only the PKDGRAV rubble pile data
is $c^*=5.5$ and $\bar \mu=0.365$. We note that the data from
\citet{Jutzi:2010} and the 1-km targets from \citet{Benz:2000} nicely
cluster around the best fit principal disruption
curve. \citet{Jutzi:2010} fit their 3 and 5-km~s$^{-1}$ data with
$\bar \mu=0.43$; however, such a high value for $\bar \mu$ cannot
simultaneously fit the data at lower velocities. The two-dimensional
simulations from \citet{Leinhardt:2009} fall systematically below the
best fit curve. The 3 and 5-km~s$^{-1}$ head-on data from
\citet{Benz:1999} have a dispersion greater than can be explained with
our model; the low and high-velocity data fall below and above the
best fit curve, respectively. The data from \citet{Benz:1999} and the
10-km data from \citet{Benz:2000} were excluded from the global fits
presented in \S \ref{sec:massratio}.

The 10-km equal-mass data from \citet{Benz:2000} ($\bowtie$ and
hourglass) require closer examination. At an impact angle of 45
degrees, all 1:1 data on weaker bodies pass through the hit-and-run
regime. However, both the nonporous and porous data show disruption
results similar to the non-grazing regime. We interpret the
non-grazing outcome to be due to the high shear strength of the target
preventing a hit-and-run outcome. We hypothesize that the disruptive
outcome and disruption energy below the gravitational binding energy
are related to the strength and angular momentum of the event. A
collision between two equal-size strong bodies has a larger
interacting mass than assumed in our model, so the adjustment from the
oblique to equivalent head-on collision disruption energy is
overestimated. In addition, the collision generates significant spin
angular momentum. The angular momentum reduces the effective
gravitional binding energy and, similarly, the required disruption
energy. These data illustrate the need to better understand the
physical properties of strong targets in oblique impacts and the role
of angular momentum.

In the strong rock target simulations, the typical limiting shear
strength is 3.5 GPa, comparable to the quasi-static shear strength in
laboratory rock under high confining pressure.  In the SPH
simulations, the shear strength was fixed throughout the
simulation. In the CTH simulations, the shear strength was dependent
on the confining pressure and the accumulation of damage (e.g.,
fractures). \citet{Leinhardt:2009} demonstrated that shear strength is
important for the size bodies considered here, which are usually
considered to be purely in the gravity regime. Higher shear strength
leads to greater dissipation of the shock energy into material
deformation; hence, higher specific energies are required to disrupt
stronger targets.  None of the published work has investigated the
role of strain rate on zones of shear localization in catastrophic
disruption simulations, which leads to significant reduction of shear
strength during impact cratering events
\citep[e.g.,][]{Senft:2009}. More work is needed to develop more
sophisticated shear strength models for strong rock targets and to
validate model calculations for oblique impacts.

There has been some recent work on the catastrophic disruption of
porous planetesimals. Porosity has been modeled in three different
ways: as hard sphere rubble piles with various bulk densities in
studies using PKDGRAV (see references in Figure~\ref{fig:qsrd}),
macroporous overlapping clusters of SPH particles representing intact
boulders \citep{Benz:2000}, and microporous bodies using a
constitutive model for porosity in an SPH code \citep{Jutzi:2010}. The
SPH simulations found significant effects of porosity in the strength
regime; however, porosity was a second order effect in the gravity
regime, and the catastrophic disruption criteria agreed with the
nonporous simulations when the data were normalized by the difference
in bulk density \citep{Jutzi:2010}. The low-velocity macroporous SPH
simulation results in the gravity regime overlap with the PKDGRAV
rubble pile results. Finally, \citet{Jutzi:2010} found similar
fragment size and velocity distributions between their porous and
nonporous gravity regime results.

We note that the transition between the gravity and the strength
regime should be handled carefully and appropriate coefficients should
be chosen for different material composition and strength. There
appears to be significantly more variation in the disruption criteria
in the strength regime compared to the gravity regime; however, future
work should consider whether or not a scaling analysis similar to the
one presented here may capture most of the variance.

\subsubsection{Other collision outcomes}

In cases where the impact velocity is above the escape velocity but
the mass of the projectile is too small to lead to disruption, some
material will escape the target in the form of crater ejecta. In
recent work, \citet{Housen:2011} has conducted a detailed study of the
scaling of ejecta from impact craters. Based on many laboratory
experiments, \citet{Housen:2011} find that approximately $0.01M_{\rm
  p}$ of material achieves escape velocity in cratering events at $V_i
\sim V_{\rm esc}$ (see their Figure~16). Empirical fits to the
material eroded during cratering events onto self-gravitating bodies
has also been studied numerically by \citet{Svetov:2011}.

The bulk composition of a body may change during planet formation by
either preferentially accreting material of a certain composition
(e.g., ice fragments chipped off smaller bodies) or by stripping of
mantle material. The loss of a mantle during catastrophic disruption
has been studied for rock/iron and water/rock differentiated planets
by \citet{Marcus:2009, Marcus:2010}. They developed two models to
calculate the resulting change in the mantle mass fraction that bound
the simulation results. Their method for calculating the change in
composition is described in the Appendix and may be incorporated into
planet formation studies that track the composition of growing and
eroding planets \citep{Stewart:2011}.

Previous work has addressed collision outcomes in the strength regime
to various levels of generality. We refer the reader to
\citet{Beauge:1990} and \citet{Kenyon:2008} and references therein.

\subsection{Implications for planet formation} \label{sec:implications}

\subsubsection{Giant impact events} 

Even with limited understanding of the full dynamics of collision
outcomes, the significant role of giant impact events in determining
the final physical properties of rocky/icy planets has been recognized
\citep[e.g.,][]{Agnor:1999,Ida:2010}. Any event between similar
sized bodies ($\gamma \gsim 0.1$) may be considered a giant impact
event, although the outcomes are more dramatic for larger mass bodies
\citep{Asphaug:2006,Asphaug:2010}. 

\citet{Agnor:1999} found that the most common collision events at the
end stage of terrestrial planet formation (under our solar system
conditions) have $\gamma \sim 0.01-0.2$ and $V_i$ from about 1 to
$4V_{\rm esc}$. Over this range of mass ratios and impact velocities,
collision outcomes span all the regimes: accretion, erosion, and
hit-and-run. In Figure~\ref{fig:planetmap}C, note that $b_{\rm crit}=0.66$
for $\gamma=0.1$; hence, about half of all impacts fall in the regime
that transitions from accretion to erosion and half transition through
a hit-and-run regime.  Hence, the implementation of self-consistent
scaling laws to describe collision outcomes is crucial to the accurate
treatment of the giant impact phase of planet formation. Although
\citet{Agnor:1999} typically found that impact velocities fell in the
range of 1 to $2V_{\rm esc}$, temporary dynamical excitation by migrating
giant planets may further increase the impact velocities in our solar
system and in exoplanetary systems
\citep{Nagasawa:2005,Morbidelli:2010,Walsh:2011}. Therefore, robust
characterization of all collision outcomes is necessary for any planet
formation calculation.

With the strong dependence of collision outcome on the mass ratio, the
final stage of planet formation is likely to produce more diverse
outcomes than previously anticipated. As argued by
\citet{Asphaug:2010}, the increased frequency of hit-and-run events
during the giant impact stage may routinely lead to compositional
modification of the second-largest body. As shown in
Figure~\ref{fig:planetmap}, the escaping projectile is nearly always
eroded in hit-and-run events. Consequently, the atmosphere,
hydrosphere, and even the mantle of these bodies may be stripped
away. Such interesting details may now be explored directly in planet
formation simulations. \citet{Asphaug:2010} suggested that the growth
of large rocky planets occurs often by a series of hit-and-run events
followed by an eventual merger. Under these circumstances, each
accreting protoplanet could have been partly devolatilized before
merging. In this manner, the final composition of planets may be
altered significantly compared to the initial protoplanets during
accretion into the final planets.

Note that our analytic calculation of collision outcomes agrees very
well with the proposed giant impact scenarios for the formation of the
Haumea system \citep{Leinhardt:2010}, the formation of Pluto and
Charon \citep{Canup:2005}, the formation of Earth's moon
\citep{Canup:2004}, and the increased density of Mercury
\citep{Benz:2007} (Figure~\ref{fig:planetmap}). The formation models
plotted for Haumea and Pluto-Charon are the result of graze-and-merge
events, where two equal-mass bodies collide and separate nearly
intact. The loss of velocity by the first collision leads to a merging
upon a second collision, creating a final body with enough angular
momentum to spin off a disk of material. In contrast, the canonical
formation of the Moon involves a collision where the projectile is
disrupted upon the first impact. The analytic calculation for
disruption of the projectile agrees very well with these moon-forming
studies. Because the giant impact phase of planet formation is
dominated by collisions slightly above the mutual escape velocity, the
probability scaled axis in Figure~\ref{fig:planetmap} emphasizes the
high likelihood that the giant impacts will be on the boundary of the
merging and hit-and-run regimes (see also \citet{Stewart:2011}).

Given the range of impact velocities found by \citet{Agnor:1999} in
the giant impact stage (up to about $6V_{\rm esc}$), stripping the mantle
from Mercury by a catastrophic disruption event is reasonably 
probable. Recently, collision outcomes alone have been used to define
a limit to the possible density of super-earth mass exoplanets
(1--10$M_{\otimes}$). Based on the criteria to strip off mantle
material during catastrophic disruption, \citet{Marcus:2010a} used the
extremely high impact velocities required to disrupt 1 to 10 earth
mass planets to place an empirical limit to the iron fraction of a
planet that has suffered a single catastrophic impact event. The
prediction is in very good agreement with observations of rocky
exoplanets \citep[e.g., Kepler 10b and 55 Cnc
e,][]{Winn:2011,Batalha:2011}. Consideration should be given to the
potential for stripping mantles off the planets by erosive hit-and-run
events: e.g., the smaller projectile has its mantle stripped but it is
never incorporated into a larger body.

\subsubsection{Collisional evolution of small body populations}

\begin{figure}
\includegraphics[scale=0.32]{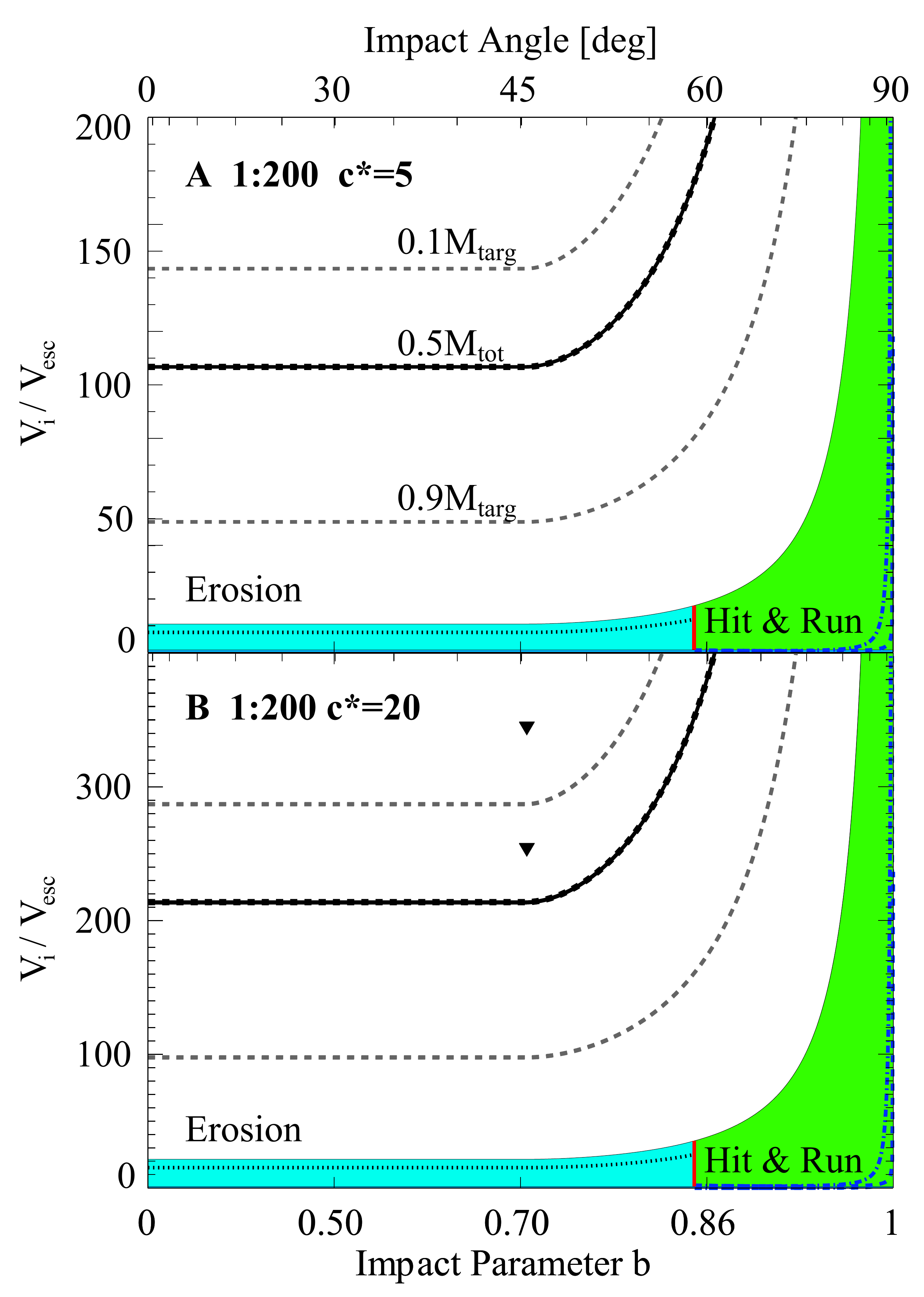}
\caption{Predicted collision outcome maps for small projectiles
  ($M_{\rm p}:M_{\rm targ}=1:200$) using the analytic model for (A) a
  nominal planetesimal with $c^*=5$ and (B) best fit to asteroid
  family formation simulations with $c^*=20$. Axes, colors and line
  notations are the same as defined in
  Figure~\ref{fig:planetmap}. Proposed asteroid family-forming events
  with $M_{\rm lr} \sim 0.1-0.2M_{\rm targ}$: $\blacktriangledown$ --
  Karin \citep{Nesvorny:2006}. \label{fig:planetesimalmap}}
\end{figure}

The asteroid and Kuiper belts contain a valuable record of the
dynamics of the solar system. The collisional evolution and dynamical
clearing of these reservoirs of small bodies has been modeled
extensively
\citep[e.g.][]{Nesvorny:2002,Bottke:2002,Morbidelli:2008,Kenyon:2008a}. Observations
of asteroid families, in particular, have been important in the study
of planetary dynamics and impact processes. Asteroid families and
their associated dust bands are believed to have formed in recent
catastrophic impact events \citep{Nesvorny:2003}. Simulations of
asteroid disruption have been compared to observations of the size and
velocity distribution of asteroid families to derive possible impact
scenarios.  For example, \citet{Nesvorny:2006} simulated the formation
and dynamical evolution of the Karin family.  Using the same SPH code
and strength model used by \citet{Benz:1999}, their best fit scenarios
for the Karin family involve a 5 to 7 km~s$^{-1}$ impact onto a 15-km
target with a mass ratio of 1:200 and $M_{\rm lr} \sim 0.1-0.2 M_{\rm
  targ}$ (Figure~\ref{fig:planetesimalmap}).  In order to match this
combination of impact energy and largest remnant mass with our
analytic model, a value of $c^*$ of approximately 20 is required,
which is significantly higher than the best fit value of 6.4 derived
here for strong targets. Figure~\ref{fig:planetesimalmap} presents
collision outcome maps for the best fit for all small bodies ($c^*=5$)
and the very strong bodies inferred from the asteroid-family formation
simulations ($c^*=20$). Catastrophic disruption begins at impact
velocities of $107V_{\rm esc}$ for the generic small body, whereas
values about 2 times higher are required for the strong targets
simulated by \citet{Nesvorny:2006}.

In addition to studying individual collisions, the collisional
evolution of the asteroid belt size distribution has been studied in
detail \citep[e.g.,][]{Davis:1979,Bottke:2005,Morbidelli:2009}. Such
studies seek to understand the relative weight of collisional versus
dynamical clearing of the belt and the initial size distribution of
planetesimals. From our discussion of the role of strength during the
evolution of planetesimals from weak aggregates to protoplanets, it is
clear that a single catastrophic disruption criteria cannot apply at
all times.

Also, the influence of mass ratio on the disruption criteria will be
important if the first planetesimals were born
big. \citet{Morbidelli:2009} argue that the observed size-frequency
distribution of asteroids is best fit by an initial population of
planetesimals that were 100's km in size. (Note this suggestion has
not been fully accepted as a requirement for the observed the size
distribution of asteroids \citep[see][] {Weidenschilling:2011}.) Two
processes have been proposed to form km to 100 km-scale initial
planetesimals: turbulent concentration \citep{Cuzzi:2008} and
streaming instabilities \citep{Johansen:2007}. A mechanism to form
km-scale planetesimals is attractive because it could help bypass the
so-called meter-size barrier, the size particle that radially drifts
in toward the sun faster than it can grow. In the collision evolution
model by \citet{Morbidelli:2009}, the catastrophic disruption
threshold is the angle-averaged 5 km~s$^{-1}$ constant velocity curve
for strong basalt from \citet{Benz:1999}. Note, however, that the
proposed mechanisms to form km-scale planetesimals would generate weak
aggregates of small (e.g., cm to m-size) particles. These aggregates
will be comparable to simulations of weaker
materials. \citet{Morbidelli:2009} considered a collisional evolution
simulation using a catastrophic disruption criteria that was a factor
of 8 lower than the basalt criteria. They note that the size-frequency
distribution was not significantly affected by the larger amount of
collisional grinding in the one test case; however, they could not
form the lunar to mars-size embryos expected in the early asteroid
belt. They reject the lower disruption criteria as being inconsistent
with observations (and their initial assumed population of 1~km-radius
bodies).

Here, we stress that a population of comparably sized bodies will be
subject to the lowest possible disruption criteria,
$Q^{*}_{RD,\gamma=1}$. For example, two colliding bodies with
individual radii of 1 km and density of rock have
$R_{C1}\sim2$~km. Using equation~\ref{eqn:vequal} with $c^*=5$ and
$\bar \mu=0.37$ for small bodies, $Q^*_{RD,\gamma=1}=5.3$~J~kg$^{-1}$,
and the corresponding $V^*_{\gamma=1}$ is 6.5~m~s$^{-1}$. For a
45-degree impact, the value for $Q^{\prime *}_{RD}$ rises by almost a
factor of 5 (equation \ref{eqn:qgamma}) and $V^*=14$~m~s$^{-1}$. Note
that this value of $Q^{\prime *}_{RD}$ is more than a factor of 100
lower than the 5 km~s$^{-1}$ strong basalt disruption curve (Figure
\ref{fig:strongqsrd}). This example illustrates the need to
incorporate self-consistent disruption criteria that account for
impact velocity and mass ratio in order to infer the magnitude of
collisional evolution in a given population of bodies.

\subsubsection{Application of collision scaling laws in planet formation}

To date, all numerical implementations of fragmentation during
collisional growth of planets assume pure energy scaling. That is,
$\bar \mu$ is assumed to be 2/3 and, thus, there is no velocity
dependence in the catastrophic disruption criteria (equation
\ref{eqn:qstarred}). In contrast, the vast collection of data in the
gravity regime indicate that catastrophic disruption is best fit by
nearly pure momentum scaling. With nearly linear dependence on the
critical velocity, the criteria for fragmentation may vary by orders
of magnitude during planet formation \citep{Stewart:2009}. Several
studies have investigated shifting a single reference size-dependent
disruption curve by a constant value that is fixed over the course of
the simulation to represent stronger or weaker bodies, but none have
considered a fragmentation criteria that may be variable in time and
space.

Furthermore, planet formation models have not included the dependence
on the mass ratio of the bodies on the disruption criteria. A recent
simple treatment of the collision parameters required for hit-and-run
versus merging indicated that planet formation was only slightly
delayed compared to simulations with only merging outcomes
\citep{Kokubo:2010}. However, this work did not include any treatment
of fragmentation. Based on our calculation of the region of partial
accretion for non-grazing impacts and projectile disruption in
hit-and-run events (Figure~\ref{fig:planetmap}), fragmentation is a
critical component of the end stage of planet formation. Numerical
simulations cannot assume pure merging or pure hit-and-run. The
diversity of collision outcomes during the end stage of planet
formation is described in detail in the companion paper, which uses
the impact parameters from recent $N$-body simulations that assumed
perfect merging to calculate the range of collision outcomes predicted
by our model \citep{Stewart:2011}.

In this work, the general catastrophic disruption law relies upon only
two independent material constants ($\bar \mu$ and $c^*$; $q_g$ is
related through equation \ref{eqn:qg}) and the impact conditions
(mass, mass ratio, impact angle and velocity).  The material
parameters are well constrained. The coupling parameter $\bar \mu$ is
tightly constrained by a large dataset (Figure~\ref{fig:qsrd}) to fall
close to pure momentum scaling (0.33 to 0.37).

The energy dissipation parameter $c^*$ is a measure of the physical
characteristics of the body. For small bodies with varying composition
and strength, we find $c^*=5\pm2$. As bodies grow into protoplanets
($\sim1000$~km), they heat internally from the heat of accretion and
radioactive decay. Then, the energy dissipation parameter for
hydrodynamic planets and planetesimals, $c^*=1.9\pm0.3$, is
appropriate. After molten planetesimals recrystallize, they will be
temporarily stronger until they experience sub-catastrophic shattering
impact events. Based on the strong rock simulations, collisional
evolution models should test for sensitivity to values of $c^*$ up to
about 20.

Two additional parameters describe the distribution of fragments
during erosive collisions. $\beta$ is the exponent to the size
distribution of small fragments and $\eta$ is the slope of the power
law size distribution for the largest fragment in the
super-catastrophic regime. The values for $\beta$ and $\eta$ are also
tightly constrained by simulations and laboratory experiments,
respectively, and may be considered, to first order, constant. 

\section{Conclusions} \label{sec:conclusions}

We present a completely self-consistent set of scaling laws to
describe the dynamical outcome of a collision between any two bodies
in the gravity regime. The scaling laws rely upon the concept of the
interacting mass, the fraction of the projectile involved in the
collision. Using the interacting mass, we derive a general
catastrophic disruption scaling law. The general forms include two
independent material parameters: the coupling parameter $\bar \mu$ and
the energy dissipation parameter $c^*$. The catastrophic disruption
criteria is used to bound the transitions between collision outcome
regimes.  The subsequent description for the size and velocity of
collision fragments are derived in closed-form analytic equations with
two well-constrained parameters.

With this powerful new tool to describe the dynamical outcome of
collisions, the physics of collisions in planet formation models will
have much greater fidelity. Planet formation models now have a very
small number of free parameters to describe collision outcomes
(primarily the energy dissipation parameter, $c^*$). With a more
robust physical model for collisions, more detailed factors may be
studied, such as the evolution of composition during planet
formation. Alternatively, other fundamental aspects of planet
formation may be investigated more deeply (e.g., migration) now that
the collision model is tightly constrained.

Future work should now bring greater scrutiny to the scaling laws used
in the strength regime. Although this regime has historically been
better constrained by the abundance of laboratory experiments,
self-consistent scaling laws also require attention to the dependence
of fragmentation on the impact velocity, mass ratio, and impact angle.

By fully constraining the dynamics of collisions in the gravity
regime, this work represents a major advancement in the robustness of
simulations of planet formation and the collisional evolution of
populations of planets.

 {\it Acknowledgements.} The $N$-body calculations were run using the University of Cambridge, Astrophysical Fluids Research Group computational facilities.  ZML is supported by an Advanced STFC fellowship; STS by NASA grant \# NNX09AP27G.

\bibliographystyle{apj}
\bibliography{GeneralBib,stsmnew}

\section{Appendix}
\setcounter{table}{0}
\renewcommand{\thetable}{A.\arabic{table}}

A description of variables and annotations is given in Table
\ref{tab:vars}.  The compilation of simulation data on catastrophic
disruption presented in Figures \ref{fig:qsrd} and
\ref{fig:strongqsrd} is summarized in Table \ref{tab:datasources}.  A
summary of all the PKDGRAV simulations conducted in this study is
presented in Table \ref{tab:pkdall}.

\begin{table*} 
  \caption{Summary of primary variables and annotations. \label{tab:vars}}
\begin{tabular}{ll}
  Symbol & Definition \\
  \hline\hline
  Material parameters & \\
  $c^*$ & Head-on equal-mass disruption energy in units of specific gravitational binding energy \\
  $\bar \mu$ & Velocity exponent in coupling parameter \\
  $\beta$    & Slope of fragment size distribution \\
  $\eta$     & Exponent in power-law fragment distribution in super-catastrophic regime \\
  Principal variables & \\
  $V, V_i$   & Impact velocity \\
  $V_{\rm esc}, V_{\rm inf}$   & Escape velocity, velocity at infinity \\
  $Q$          & Specific impact energy \\
  $Q_R$        & Specific impact energy for the collision in center of mass frame \\
  $Q^*_{RD}$  & Catastrophic disruption criteria -- specific impact
  energy to disperse half the total mass \\
  $M, m$        & Mass \\
  $R_{C1}$    & Radius of total mass in a body with density 1000~kg~m$^{-3}$\\
  $\mu$      & Reduced mass \\
  $\gamma$   & Projectile-to-target mass ratio \\
  $\alpha$    & Mass fraction of the projectile that intersects the  target \\
  $q_g$    & Coefficient of gravity term in general formula for  $Q^*_{RD}$ \\  
  $\xi$    & Accretion efficiency \\ 
  $v$      & Velocity of fragments \\
  $\rho$  & Density \\
 Geometric variables & \\
  $\theta$   & Impact angle (0 is head-on) \\
  $b$        & Impact parameter sin($\theta$) \\
  $b_{\rm crit}$  & Critical impact parameter denoting transition from non-grazing to grazing collision \\
  $R$        & Target radius \\
  $r$        & Projectile radius \\
  $D$        & Diameter \\
  $l/(2r)$   & Fraction of projectile diameter that overlaps with  target \\
 Superscripts & \\
  $^{*}$       & Value at the catastrophic disruption threshold \\
  $^{'}$       & Oblique impact \\
  $^{\dag}$   & Reverse impact onto the projectile in the hit-and-run regime \\
  Subscripts & \\
  $_{\rm targ}$ & Target \\
  $_{\rm p}$  & Projectile \\
  $_{\rm tot}$ & Target + projectile  \\
  $_{\rm interact}$ & Interacting fraction \\
  $_{\alpha}$ & Interacting projectile fraction \\
  $_{\gamma=1}$ & Equal-mass collision \\
  $_{\rm lr}$ & Largest remnant \\
  $_{\rm slr}$ & Second largest remnant \\
  $_{\rm rem}$ & Fragments smaller than the largest remnant \\
  $_{\rm core}$ & Core fraction of a differentiated body \\
  Constants & \\
  $\rho_1$   & Density of 1000~kg~m$^{-3}$ \\
  $G$          & Gravitational constant \\
  \hline
\end{tabular}
\end{table*}

\begin{table*} 
  \caption{Sources and description of catastrophic disruption data
    presented in Figures
    \ref{fig:qsrd} and  \ref{fig:strongqsrd}. Filled
    and line symbols
    indicate head-on collisions ($b=0$). Open symbols indicate oblique
    impacts: $b=0.15,0.3$ for open star,
    $0.35<b<0.9$ for $\bigtriangleup$, $b=0.5$ for $\Box$ and
    $\lozenge$, and $b=0.71$ for $\bowtie$, hourglass, hexagon,
    $\otimes$, and $\boxtimes$. \label{tab:datasources}}
\begin{tabular}{lll}
  Symbol & Target Description & Source \\
  \hline\hline
  Weak targets & & \\
  $\blacktriangle$,$\bigtriangleup$ & 10-km PKDGRAV rubble piles & This work \\
  $+$ & 1 to 50-km PKDGRAV rubble piles & \citet{Stewart:2009} \\
  $\bigstar$, open star & 1-km PKDGRAV rubble piles & \citet{Leinhardt:2000} \\
  $\blacktriangleleft$ & weak 2 to 50-km basalt & \citet{Leinhardt:2009} \\
  $\bullet$ & hydrodynamic 2 to 50-km basalt & \citet{Leinhardt:2009} \\
  hexagon & 50-km ice & \citet{Leinhardt:2009} \\

   & & \\
  Strong targets & & \\
  $\ast$ & 1 to 100-km radius basalt & \citet{Benz:1999} \\
  $\blacktriangledown$ & 2 to 50-km basalt & \citet{Leinhardt:2009} \\
  $\otimes$ & 0.3 to 100-km microporous rock (pumice) & \citet{Jutzi:2010} \\
  $\boxtimes$ & 0.3 to 100-km basalt & \citet{Jutzi:2010} \\
  $\bowtie$ & 10-km macroporous basalt & \citet{Benz:2000} \\
  hourglass & 10-km basalt & \citet{Benz:2000} \\

  & & \\
  Hydrodynamic planets & & \\
  $\blacklozenge$, $\lozenge$ & 2.2 Mercury-mass bodies, differentiated rock and iron & \citet{Benz:2007} \\
  $\blacktriangleright$ & 0.4 and 7 Earth-mass bodies, pure rock & \citet{Marcus:2009} \\
  $\times$ & 1 to 10 Earth-mass bodies, differentiated rock and iron & \citet{Marcus:2009} \\
  $\blacksquare$, $\Box$ & 0.5 to 5 Earth-mass bodies, differentiated water and rock & \citet{Marcus:2010} \\

  \hline
\end{tabular}
\end{table*}

\subsection{A general formulation for collision outcomes} \label{sec:decisiontree}

We summarize the sequence of logic that should be applied to determine
the dynamical outcome of any collision in the gravity regime using our
analytic model. First, we identify the boundaries of the major
collision regimes:
\begin{enumerate}
\item{For a given collision scenario ($M_{\rm p}$, $M_{\rm targ}$, $b$, $V_i$,
    and $R_{\rm p}$ and $R_{\rm targ}$ from the bulk densities of the bodies),
    calculate the interacting mass fraction of the projectile,
    $m_{\rm interact}=\alpha M_{\rm p}$ (equation \ref{eqn:alpha}).}
\item{If $V_i < V_{\rm esc}^{\prime}$ (equation \ref{eqn:vescmod}), then
    the impact is in the {\it perfect merging regime}.}
\item{Calculate the critical impact parameter $b_{\rm crit}$ for the
    collision (equation \ref{eqn:bcrit}). If $b<b_{\rm crit}$, then it is
    a non-grazing impact, else it is a grazing impact.}
\item{Calculate the catastrophic disruption criteria, $Q^{\prime *}_{RD}$, and corresponding
    critical impact velocity, $V^{'*}$, for the specific impact scenario:
   \begin{enumerate}
    \item{Calculate $R_{C1}$ from the total mass and density of
        1000~kg~m$^{-3}$.}
    \item{Calculate the principal disruption value for an equivalent
        equal-mass collision at $R_{C1}$, $Q^*_{RD,\gamma=1}$
        (equation \ref{eqn:principalq}), and its corresponding
        critical impact velocity, $V^*_{\gamma=1}$ (equation
        \ref{eqn:vequal}), using the material parameter $c^*$.}
    \item{Calculate the reduced mass, $\mu$, and the reduced mass using
        the interacting mass, $\mu_{\alpha}$ (equation
        \ref{eqn:mualpha}).}
    \item{Calculate the disruption criteria, $Q^*_{RD}$, and critical
        impact velocity, $V^*$, for a head-on impact with the desired
        mass ratio, $\gamma$ using equations \ref{eqn:qgamma} and
        \ref{eqn:vgamma} and the material parameter $\bar \mu$.}
    \item{The value for the disruption energy, $Q^{\prime *}_{RD}$,
        and critical impact velocity, $V^{\prime *}_{RD}$, for the
        desired impact angle are found using equations
        \ref{eqn:qangle} and \ref{eqn:vstarprime}.}
   \end{enumerate}
}
\item{Calculate the value for $Q_R$ required for onset of erosion,
    $M_{\rm lr}=M_{\rm targ}$, using the value of $Q^{\prime *}_{RD}$
    and the universal law for the mass of the largest remnant
    (equation \ref{eqn:univlawangle}). From this $Q_R$, derive the
    impact velocity for the onset of target erosion, $V_{\rm erosion}$,
    from equation \ref{eqn:qr}.}
\item{For grazing impacts ($b>b_{\rm crit}$), the {\it hit-and-run regime}
    is bounded by $V^{\prime}_{\rm esc}<V_i<V_{\rm erosion}$. Note that
    the graze-and-merge regime is a subset of this range, but it is
    not explicitly defined in this work \citep[see][]{Genda:2011b}.}
\item{Calculate the value for $Q_R$ required for the onset of
    super-catastrophic disruption, $M_{\rm lr}=0.1M_{\rm tot}$, using
    the value of $Q^{\prime *}_{RD}$ and the universal law for the
    mass of the largest remnant (equation
    \ref{eqn:univlawangle}). From this $Q_R$, derive the impact
    velocity for the onset of super-catastrophic disruption,
    $V_{\rm supercat}$, from equation \ref{eqn:qr}.}
\item{For all impact angles, the target is eroded when
    $V_i>V_{\rm erosion}$ and the impact is in the {\it erosion
      regime}.}
\item{For all impact angles, the impact is in the {\it
      super-catastrophic disruption regime} when
    $V_i>V_{\rm supercat}$.}
\item{For non-grazing events and $V^{\prime}_{\rm esc}<V_i<V_{\rm supercat}$,
    the impact is in the {\it disruption regime} and the universal law
    for the mass of the largest remnant applies. The partial {\it accretion
      regime} is bounded by $V^{\prime}_{\rm esc}<V_i<V_{\rm erosion}$.}
\item{For grazing events and $V_{\rm erosion}<V_i<V_{\rm supercat}$,
    the impact is in the {\it disruption regime} and the universal law
    for the mass of the largest remnant applies only for $M_{\rm
      lr}<M_{\rm targ}$.}
\item{In the hit-and-run regime, calculate the critical disruption
    energy for the reverse impact onto the projectile, $Q^{\dag
      \prime *}_{RD}$, as described in \S \ref{sec:hitandrun}, and its
    corresponding $V^{\dag '*}$. Use the equation
    \ref{eqn:univlawangle} or \ref{eqn:mlrsuper} to determine the
    largest remnant after disruption of the total mass involved in the
    reverse impact, $M_{\rm interact}+M_{\rm p}$. }
\end{enumerate}

In the {\it merging regime}, mass and momentum are conserved. 

In the {\it disruption regime}:
\begin{enumerate}
\item{Determine the mass of the largest remnant $M_{\rm lr}$ from the
    universal law (equation \ref{eqn:univlawangle}) using $Q_R$ and
    $Q_{RD}^{\prime *}$.}
\item{Determine the mass of the second largest remnant $M_{\rm slr}$ using
    equation \ref{eqn:mslr} with $\beta = 2.85$, $N_1 = 1$, and $N_2 =
    2$. The size distribution of the tail of smaller fragments is
    described by equation \ref{eqn:diff}.}
\item{For $b=0$, assume that the largest remnant obtains the velocity
    of the center of mass; for $b>0.7$ assume that the largest remnant
    maintains $V_{\rm targ}$. For $0<b<0.7$, the largest remnant velocity
    is some quasi-linear function of $b$.}
\item{The mass-velocity distribution of the smaller fragments with
    respect to the largest remnant is given by equation
    \ref{eqn:velscaling}.}
\end{enumerate}

In the {\it super-catastrophic disruption regime}:
\begin{enumerate}
\item{Determine the mass of the largest remnant $M_{\rm lr}$ from the
    power law (equation \ref{eqn:mlrsuper}) using $Q_R$ and
    $Q_{RD}^{\prime *}$ (equation \ref{eqn:qangle}).}
\item{The size and velocity distribution of the fragments with respect
    to the largest remnant are the same as in the disruption regime.}
\end{enumerate}

In the {\it hit-and-run regime}:
\begin{enumerate}
\item{The mass of the largest remnant $M_{\rm lr}$ is approximately equal to the target
    mass $M_{\rm targ}$.}
\item{The mass of the second largest remnant $M_{\rm slr}$ is estimated
    using the universal law and disruption criteria for the reverse
    impact on the projectile, $Q^{\dag \prime *}_{RD}$.}
\item{When the projectile is disrupted, the size and velocity
    distribution of the fragments are described as in the disruption
    regime with respect to the largest remnant from the projectile. }
\item{In the special case of $\gamma \sim 1$, the onset of erosion occurs
    simultaneously in both bodies and $M_{\rm lr} \sim M_{\rm slr}$.  Then, use
    $N_1=2$ and $N_2=4$ to calculate the size distribution. One can
    assume that the fragments from both the projectile and target have
    identical size and velocity distributions with respect to their
    body of origin.}
\end{enumerate}

In the disruption and super-catastrophic disruption regimes, a
differentiated target may change its bulk composition by stripping off
a portion of the mantle. Bulk compositional changes may be tracked
using the results from \citet{Marcus:2010}. They found that the core
mass fraction after a disruptive collision falls between two idealized
models, and we suggest using an average of these two results:
\begin{enumerate}
\item{Model 1 -- Cores always merge: Given the original
    $M_{\rm core,targ}$ and $M_{\rm core,p}$, the post-impact core is
    $M_{\rm core}={\rm min}(M_{\rm lr},M_{\rm core,targ}+M_{\rm core,p})$.}
\item{Model 2 -- Cores only merge on accretion: When
    $M_{\rm lr}>M_{\rm targ}$, $M_{\rm core}=M_{\rm core,targ}+{\rm
      min}(M_{\rm core,p},M_{\rm lr}-M_{\rm targ}$). When $M_{\rm lr}<M_{\rm targ}$,
    assume that none of the projectile accretes and the mantle is
    stripped first. Then, $M_{\rm core}={\rm min}(M_{\rm core,targ},M_{\rm lr})$.}
\end{enumerate}
In hit-and-run events with projectile disruption, the same relations
may be used to estimate the bulk changes in composition for the
projectile.

Finally, \citet{Housen:2011} provide scaling laws for the
gravitationally escaping ejecta from the target in the impact
cratering regime. The impact cratering regime occurs at the low
velocity end of the disruption regime, when $V_i>V_{\rm esc}^{'}$,
$M_{\rm p} << M_{\rm targ}$, and $M_{\rm lr} \sim M_{\rm targ}$. Based
on many laboratory experiments, \citet{Housen:2011} find that
approximately $0.01M_{\rm p}$ achieves escape velocity in cratering
events of $V_i \sim V_{\rm esc}$ (see their Figure~16). In addition,
\citet{Svetov:2011} provides empirical equations for ejected material
in cratering collisions on self-gravitating bodies.

\begin{table}
\caption{Summary of all simulation parameters and results.}
\label{tab:pkdall}
\begin{tabular}{lcccccc}
  $\underline{M_{\rm p}}$ & $b$ &  $V_i$ & $\underline{M_{\rm lr}}$ & $\underline{M_{\rm slr}}$ & $Q_{R}$ \\
$M_{\rm targ}$ & -- & m/s & $M_{\rm tot}$ & $M_{\rm tot}$ & J/kg\\
\hline\hline
0.025 & 0.00 & 9 & 1.00 & 9.64e-05 & 9.69e-01 & \\ 
0.025 & 0.00 & 14 & 1.00 & 9.64e-05 & 2.34e+00 & \\ 
0.025 & 0.00 & 18 & 1.00 & 9.64e-05 & 3.88e+00 & \\ 
0.025 & 0.00 & 22 & 0.99 & 1.93e-04 & 5.79e+00 & \\ 
0.025 & 0.00 & 50 & 0.94 & 3.86e-04 & 2.99e+01 & \\ 
0.025 & 0.00 & 60 & 0.90 & 5.78e-04 & 4.31e+01 & \\ 
0.025 & 0.00 & 70 & 0.88 & 7.71e-04 & 5.86e+01 & \\ 
0.025 & 0.00 & 100 & 0.77 & 1.45e-03 & 1.20e+02 & \\ 
0.025 & 0.00 & 120 & 0.67 & 2.41e-03 & 1.72e+02 & \\ 
0.025 & 0.00 & 140 & 0.55 & 1.19e-02 & 2.34e+02 & \\ 
0.025 & 0.00 & 160 & 0.51 & 6.17e-03 & 3.06e+02 & \\ 
0.025 & 0.00 & 180 & 0.35 & 1.84e-02 & 3.88e+02 & \\ 
0.025 & 0.00 & 200 & 0.27 & 1.65e-02 & 4.79e+02 & \\ 
0.025 & 0.00 & 400 & 0.00 & 2.99e-03 & 1.91e+03 & \\ 
\hline
0.025 & 0.35 & 9 & 1.00 & 9.64e-05 & 9.69e-01 & \\ 
0.025 & 0.35 & 14 & 0.99 & 1.93e-04 & 2.34e+00 & \\ 
0.025 & 0.35 & 18 & 0.99 & 1.93e-04 & 3.88e+00 & \\ 
0.025 & 0.35 & 22 & 0.99 & 1.93e-04 & 5.79e+00 & \\ 
0.025 & 0.35 & 100 & 0.81 & 7.71e-04 & 1.20e+02 & \\ 
0.025 & 0.35 & 160 & 0.60 & 6.07e-03 & 3.06e+02 & \\ 
0.025 & 0.35 & 200 & 0.45 & 5.88e-03 & 4.79e+02 & \\ 
0.025 & 0.35 & 300 & 0.07 & 4.76e-02 & 1.08e+03 & \\ 
0.025 & 0.35 & 400 & 0.02 & 1.13e-02 & 1.91e+03 & \\ 
\hline
0.025 & 0.70 & 9 & 0.99 & 1.45e-03 & 9.69e-01 & \\ 
0.025 & 0.70 & 14 & 0.98 & 9.64e-04 & 2.34e+00 & \\ 
0.025 & 0.70 & 18 & 0.98 & 1.93e-04 & 3.88e+00 & \\ 
0.025 & 0.70 & 22 & 0.98 & 1.93e-04 & 5.79e+00 & \\ 
0.025 & 0.70 & 160 & 0.84 & 6.75e-04 & 3.06e+02 & \\ 
0.025 & 0.70 & 200 & 0.80 & 8.68e-04 & 4.79e+02 & \\ 
0.025 & 0.70 & 300 & 0.65 & 2.41e-03 & 1.08e+03 & \\ 
0.025 & 0.70 & 400 & 0.47 & 6.07e-03 & 1.91e+03 & \\ 
0.025 & 0.70 & 500 & 0.26 & 1.33e-02 & 2.99e+03 & \\ 
0.025 & 0.70 & 600 & 0.05 & 2.80e-02 & 4.31e+03 & \\ 
\hline
0.025 & 0.90 & 9 & 0.98 & 1.88e-02 & 9.69e-01 & \\ 
0.025 & 0.90 & 15 & 0.98 & 1.58e-02 & 2.69e+00 & \\ 
0.025 & 0.90 & 20 & 0.98 & 1.27e-02 & 4.79e+00 & \\ 
0.025 & 0.90 & 25 & 0.98 & 9.26e-03 & 7.48e+00 & \\ 
0.025 & 0.90 & 30 & 0.97 & 6.27e-03 & 1.08e+01 & \\ 
0.025 & 0.90 & 40 & 0.97 & 1.93e-03 & 1.91e+01 & \\ 
0.025 & 0.90 & 50 & 0.97 & 6.75e-04 & 2.99e+01 & \\ 
0.025 & 0.90 & 60 & 0.97 & 2.12e-03 & 4.31e+01 & \\ 
0.025 & 0.90 & 400 & 0.88 & 2.89e-04 & 1.91e+03 & \\ 
0.025 & 0.90 & 500 & 0.84 & 4.82e-04 & 2.99e+03 & \\ 
0.025 & 0.90 & 600 & 0.78 & 7.71e-04 & 4.31e+03 & \\ 
0.025 & 0.90 & 700 & 0.70 & 2.70e-03 & 5.86e+03 & \\ 
0.025 & 0.90 & 800 & 0.74 & 1.45e-03 & 7.66e+03 & \\ 
0.025 & 0.90 & 900 & 0.66 & 2.02e-03 & 9.69e+03 & \\ 
0.025 & 0.90 & 1000 & 0.36 & 2.80e-03 & 1.20e+04 & \\ 
\hline\hline
0.10 & 0.00 & 9 & 1.00 & 8.99e-05 & 3.37e+00 & \\
0.10 & 0.00 & 15 & 0.99 & 1.80e-05 & 9.35e+00 & \\ 
0.10 & 0.00 & 20 & 0.97 & 1.80e-04 & 1.66e+01 & \\ 
0.10 & 0.00 & 25 & 0.94 & 8.09e-04& 2.60e+01 & \\ 
0.10 & 0.00 & 30 & 0.90 & 7.19e-04 & 3.74e+01 & \\ 
\end{tabular}
\end{table}

\begin{table}
\begin{tabular}{lcccccc}
0.10 & 0.00 & 40 & 0.79 & 1.35e-03 & 6.65e+01 & \\ 
0.10 & 0.00 & 50 & 0.67 & 2.61e-03 & 1.04e+02 & \\ 
0.10 & 0.00 & 65 & 0.41 & 1.44e-02 & 1.76e+02 & \\ 
0.10 & 0.00 & 80 & 0.14 & 3.44e-02 & 2.66e+02 & \\ 
\hline
0.10 & 0.35 & 9 & 1.00 & 1.71e-03 & 3.37e+00 & \\ 
0.10 & 0.35 & 15 & 0.96 & 1.08e-03 & 9.35e+00 & \\ 
0.10 & 0.35 & 20 & 0.93 & 1.89e-03 & 1.66e+01 & \\ 
0.10 & 0.35 & 25 & 0.90 & 1.44e-03 & 2.60e+01 & \\ 
0.10 & 0.35 & 30 & 0.87 & 1.44e-03 & 3.74e+01 & \\ 
0.10 & 0.35 & 40 & 0.79 & 1.98e-03 & 6.65e+01 & \\ 
0.10 & 0.35 & 50 & 0.72 & 1.89e-03 & 1.04e+02 & \\ 
0.10 & 0.35 & 65 & 0.62 & 5.13e-03 & 1.76e+02 & \\ 
0.10 & 0.35 & 80 & 0.47 & 5.85e-03 & 2.66e+02 & \\ 
0.10 & 0.35 & 100 & 0.33 & 1.16e-02 & 4.15e+02 & \\ 
\hline
0.10 & 0.70 & 9 & 0.95 & 4.76e-02 & 3.37e+00 & \\ 
0.10 & 0.70 & 15 & 0.92 & 3.72e-02 & 9.35e+00 & \\ 
0.10 & 0.70 & 20 & 0.90 & 3.43e-02 & 1.66e+01 & \\ 
0.10 & 0.70 & 25 & 0.90 & 1.42e-02 & 2.60e+01 & \\
0.10 & 0.70 & 30 & 0.89 & 8.63e-03 & 3.74e+01 & \\ 
0.10 & 0.70 & 40 & 0.87 & 2.07e-03 & 6.65e+01 & \\ 
0.10 & 0.70 & 50 & 0.86 & 1.53e-03 & 1.04e+02 & \\ 
0.10 & 0.70 & 100 & 0.77 & 2.16e-03 & 4.15e+02 & \\ 
0.10 & 0.70 & 150 & 0.63 & 1.53e-03 & 9.35e+02 & \\ 
0.10 & 0.70 & 200 & 0.52 & 3.51e-03 & 1.66e+03 & \\ 
0.10 & 0.70 & 300 & 0.21 & 1.15e-02 & 3.74e+03 & \\ 
\hline
0.10 & 0.90 & 9 & 0.92 & 8.21e-02 & 3.37e+00 & \\ 
0.10 & 0.90 & 15 & 0.91 & 8.44e-02 & 9.35e+00 & \\ 
0.10 & 0.90 & 20 & 0.91 & 8.21e-02 & 1.66e+01 & \\ 
0.10 & 0.90 & 25 & 0.91 & 8.21e-02 & 2.60e+01 & \\ 
0.10 & 0.90 & 30 & 0.91 & 7.85e-02 & 3.74e+01 & \\ 
0.10 & 0.90 & 40 & 0.91 & 7.68e-02 & 6.65e+01 & \\ 
0.10 & 0.90 & 50 & 0.91 & 6.83e-02 & 1.04e+02 & \\ 
0.10 & 0.90 & 60 & 0.90 & 6.43e-02 & 1.50e+02 & \\ 
0.10 & 0.90 & 70 & 0.90 & 6.09e-02 & 2.04e+02 & \\ 
0.10 & 0.90 & 80 & 0.90 & 5.96e-02 & 2.66e+02 & \\ 
0.10 & 0.90 & 100 & 0.90 & 1.71e-02 & 4.15e+02 & \\ 
0.10 & 0.90 & 120 & 0.89 & 2.88e-03 & 5.98e+02 & \\ 
0.10 & 0.90 & 140 & 0.88 & 2.07e-03 & 8.14e+02 & \\ 
0.10 & 0.90 & 300 & 0.84 & 4.50e-04 & 3.74e+03 & \\ 
0.10 & 0.90 & 400 & 0.70 & 6.29e-04 & 6.65e+03 & \\ 
0.10 & 0.90 & 500 & 0.61 & 1.80e-03 & 1.04e+04 & \\ 
0.10 & 0.90 & 600 & 0.57 & 4.77e-03 & 1.50e+04 & \\ 
0.10 & 0.90 & 700 & 0.53 & 2.07e-03 & 2.04e+04 & \\ 
0.10 & 0.90 & 800 & 0.55 & 2.16e-03 & 2.66e+04 & \\ 
0.10 & 0.90 & 900 & 0.57 & 7.19e-03 & 3.37e+04 & \\ 
\hline\hline
0.25 & 0.00 & 9 & 1.00 & 7.89e-05 & 6.52e+00 & \\ 
0.25 & 0.00 & 14 & 0.98 & 1.58e-04 & 1.58e+01 & \\ 
0.25 & 0.00 & 18 & 0.94 & 3.95e-04 & 2.61e+01 & \\ 
0.25 & 0.00 & 22 & 0.88 & 1.66e-03 & 3.89e+01 & \\ 
0.25 & 0.00 & 30 & 0.69 & 6.55e-03 & 7.24e+01 & \\ 
0.25 & 0.00 & 40 & 0.40 & 1.95e-02 & 1.29e+02 & \\ 
0.25 & 0.00 & 50 & 0.09 & 1.66e-02 & 2.01e+02 & \\ 
0.25 & 0.00 & 60 & 0.01 & 1.07e-02 & 2.90e+02 & \\ 
\hline
0.25 & 0.35 & 14 & 0.93 & 1.40e-02 & 1.58e+01 & \\ 
0.25 & 0.35 & 18 & 0.84 & 1.89e-02 & 2.61e+01 & \\ 
0.25 & 0.35 & 22 & 0.78 & 9.00e-03 & 3.89e+01 & \\ 
0.25 & 0.35 & 30 & 0.67 & 5.29e-03 & 7.24e+01 & \\ 
0.25 & 0.35 & 40 & 0.53 & 6.31e-03 & 1.29e+02 & \\ 
0.25 & 0.35 & 45 & 0.46 & 5.52e-03 & 1.63e+02 & \\ 
0.25 & 0.35 & 50 & 0.37 & 4.97e-03 & 2.01e+02 & \\ 
0.25 & 0.35 & 55 & 0.33 & 1.71e-02 & 2.43e+02 & \\ 
0.25 & 0.35 & 60 & 0.25 & 7.42e-03 & 2.90e+02 & \\ 
0.25 & 0.35 & 65 & 0.17 & 2.36e-02 & 3.40e+02 & \\ 
0.25 & 0.35 & 80 & 0.07 & 1.12e-02 & 5.15e+02 & \\ 
0.25 & 0.35 & 9 & 1.00 & 1.89e-03 & 6.52e+00 & \\ 
\hline
\end{tabular}
\end{table}

\begin{table}
\begin{tabular}{lcccccc}
0.25 & 0.70 & 9 & 0.84 & 1.55e-01 & 6.52e+00 & \\ 
0.25 & 0.70 & 14 & 0.81 & 1.54e-01 & 1.58e+01 & \\ 
0.25 & 0.70 & 18 & 0.79 & 1.36e-01 & 2.61e+01 & \\ 
0.25 & 0.70 & 22 & 0.78 & 1.18e-01 & 3.89e+01 & \\ 
0.25 & 0.70 & 27 & 0.77 & 1.02e-01 & 5.86e+01 & \\ 
0.25 & 0.70 & 36 & 0.74 & 5.44e-02 & 1.04e+02 & \\ 
0.25 & 0.70 & 50 & 0.69 & 5.84e-03 & 2.01e+02 & \\ 
0.25 & 0.70 & 60 & 0.66 & 6.16e-03 & 2.90e+02 & \\
0.25 & 0.70 & 70 & 0.64 & 2.05e-03 & 3.94e+02 & \\ 
0.25 & 0.70 & 80 & 0.58 & 1.42e-03 & 5.15e+02 & \\
0.25 & 0.70 & 100 & 0.52 & 3.79e-03 & 8.04e+02 & \\ 
0.25 & 0.70 & 125 & 0.42 & 3.79e-03 & 1.26e+03 & \\ 
0.25 & 0.70 & 150 & 0.32 & 5.21e-03 & 1.81e+03 & \\ 
0.25 & 0.70 & 175 & 0.10 & 7.89e-05 & 2.46e+03 & \\ 
\hline
0.25 & 0.90 & 9 & 0.81 & 1.91e-01 & 6.52e+00 & \\ 
0.25 & 0.90 & 14 & 0.80 & 1.93e-01 & 1.58e+01 & \\ 
0.25 & 0.90 & 18 & 0.80 & 1.90e-01 & 2.61e+01 & \\ 
0.25 & 0.90 & 22 & 0.80 & 1.88e-01 & 3.89e+01 & \\ 
0.25 & 0.90 & 27 & 0.79 & 1.87e-01 & 5.86e+01 & \\ 
0.25 & 0.90 & 36 & 0.79 & 1.74e-01 & 1.04e+02 & \\ 
0.25 & 0.90 & 50 & 0.79 & 1.87e-01 & 2.01e+02 & \\ 
0.25 & 0.90 & 60 & 0.79 & 1.87e-01 & 2.90e+02 & \\
0.25 & 0.90 & 70 & 0.79 & 1.83e-01 & 3.94e+02 & \\ 
0.25 & 0.90 & 100 & 0.79 & 1.62e-01 & 8.04e+02 & \\ 
0.25 & 0.90 & 120 & 0.77 & 1.44e-01 & 1.16e+03 & \\ 
0.25 & 0.90 & 150 & 0.73 & 5.26e-02 & 1.81e+03 & \\ 
0.25 & 0.90 & 200 & 0.67 & 3.63e-03 & 3.22e+03 & \\ 
0.25 & 0.90 & 250 & 0.61 & 1.34e-03 & 5.03e+03 & \\ 
0.25 & 0.90 & 300 & 0.55 & 1.10e-03 & 7.24e+03 & \\ 
0.25 & 0.90 & 350 & 0.47 & 3.39e-03 & 9.85e+03 & \\ 
0.25 & 0.90 & 400 & 0.42 & 4.02e-03 & 1.29e+04 & \\ 
0.25 & 0.90 & 450 & 0.31 & 5.84e-03 & 1.63e+04 & \\ 
\hline\hline
1.00 & 0.00 & 15 & 0.97 & 2.46e-04 & 2.83e+01 & \\ 
1.00 & 0.00 & 18 & 0.93 & 4.42e-04 & 4.07e+01 & \\ 
1.00 & 0.00 & 24 & 0.76 & 3.66e-03 & 7.24e+01 & \\ 
1.00 & 0.00 & 24 & 0.76 & 2.72e-03 & 7.24e+01 & \\ 
1.00 & 0.00 & 24 & 0.77 & 3.06e-03 & 7.24e+01 & \\ 
1.00 & 0.00 & 30 & 0.50 & 1.06e-02 & 1.13e+02 & \\ 
1.00 & 0.00 & 30 & 0.49 & 1.04e-02 & 1.13e+02 & \\ 
1.00 & 0.00 & 30 & 0.47 & 1.23e-02 & 1.13e+02 & \\ 
1.00 & 0.00 & 35 & 0.12 & 5.47e-02 & 1.54e+02 & \\ 
1.00 & 0.00 & 35 & 0.12 & 3.76e-02 & 1.54e+02 & \\ 
1.00 & 0.00 & 35 & 0.14 & 3.01e-02 & 1.54e+02 & \\ 
1.00 & 0.00 & 38 & 0.05 & 2.77e-02 & 1.81e+02 & \\ 
1.00 & 0.00 & 38 & 0.04 & 3.86e-02 & 1.81e+02 & \\ 
1.00 & 0.00 & 38 & 0.03 & 2.07e-02 & 1.81e+02 & \\ 
\hline
1.00 & 0.35 & 15 & 0.98 & 3.93e-04 & 2.83e+01 & \\ 
1.00 & 0.35 & 16 & 0.97 & 2.95e-04 & 3.22e+01 & \\ 
1.00 & 0.35 & 17 & 0.48 & 4.56e-01 & 3.63e+01 & \\ 
1.00 & 0.35 & 18 & 0.47 & 4.46e-01 & 4.07e+01 & \\ 
1.00 & 0.35 & 30 & 0.23 & 1.98e-01 & 1.13e+02 & \\ 
1.00 & 0.35 & 45 & 0.02 & 1.22e-02 & 2.55e+02 & \\
\hline
1.00 & 0.70 & 8 & 1.00 & 0.00e+00 & 8.04e+00 & \\ 
1.00 & 0.70 & 10 & 0.50 & 4.97e-01 & 1.26e+01 & \\ 
1.00 & 0.70 & 11 & 0.50 & 4.97e-01 & 1.52e+01 & \\ 
1.00 & 0.70 & 12 & 0.50 & 4.93e-01 & 1.81e+01 & \\ 
1.00 & 0.70 & 13 & 0.50 & 4.95e-01 & 2.12e+01 & \\ 
1.00 & 0.70 & 14 & 0.50 & 4.89e-01 & 2.46e+01 & \\ 
1.00 & 0.70 & 30 & 0.44 & 4.41e-01 & 1.13e+02 & \\ 
1.00 & 0.70 & 50 & 0.39 & 3.82e-01 & 3.14e+02 & \\ 
1.00 & 0.70 & 80 & 0.28 & 2.74e-01 & 8.04e+02 & \\ 
1.00 & 0.70 & 150 & 0.04 & 1.05e-02 & 2.83e+03 & \\ 
\hline
1.00 & 0.90 & 7 & 1.00 & 4.97e-01 & 6.16e+00 & \\ 
1.00 & 0.90 & 15 & 0.50 & 4.95e-01 & 2.83e+01 & \\ 
1.00 & 0.90 & 20 & 0.50 & 4.95e-01 & 5.03e+01 & \\ 
1.00 & 0.90 & 40 & 0.50 & 4.90e-01 & 2.01e+02 & \\ 
\end{tabular}
\end{table}

\begin{table}
\begin{tabular}{lcccccc}
1.00 & 0.90 & 100 & 0.48 & 4.80e-01 & 1.26e+03 & \\ 
1.00 & 0.90 & 200 & 0.44 & 4.30e-01 & 5.03e+03 & \\ 
1.00 & 0.90 & 300 & 0.41 & 4.06e-01 & 1.13e+04 & \\ 
1.00 & 0.90 & 400 & 0.35 & 3.39e-01 & 2.01e+04 & \\ 
1.00 & 0.90 & 600 & 0.26 & 2.51e-01 & 4.52e+04 & \\ 
\hline
\end{tabular}
\end{table}

\end{document}